\documentclass[12pt]{article}
\usepackage[utf8]{inputenc}
\usepackage[margin=1.5in]{geometry}
\newgeometry{top=.5in,bottom=1in,left=1in,right=1in}
\usepackage{mwe}
\usepackage{float}
\usepackage{mathtools}
\usepackage{booktabs}
\usepackage{multicol}
\usepackage{amsmath}
\usepackage{bm}
\usepackage[table,xcdraw]{xcolor}
\usepackage{url}
\usepackage{subcaption}
\usepackage{graphicx}
\usepackage{comment} 
\usepackage[affil-it]{authblk} 
\usepackage{etoolbox}
\usepackage{lmodern}
\usepackage{todonotes}
\usepackage{econometrics}
\usepackage[page,header]{appendix}
\usepackage{titletoc}
\usepackage{array}
\usepackage{makecell}
\usepackage{lscape}
\usepackage{ amssymb }
\usepackage{xfrac}
\usepackage{multirow}
\usepackage{rotating}
\usepackage{threeparttable}
\usepackage{nicefrac}

\newtheorem{proposition}{Proposition}
\newtheorem{theorem}{Theorem}
\newtheorem{lemma}{Lemma}
\newtheorem{corollary}{Corollary}

\newtheorem{remark}{Remark}
\newtheorem{assumption}{Assumption}

\usepackage[colorlinks=true, citecolor=blue]{hyperref}
\hypersetup{
    colorlinks=true,
    linkcolor=blue,
    citecolor=blue,
    filecolor=blue,
    urlcolor=blue,
}

\makeatletter
\renewcommand{\todo}[2][]{%
    \@todo[caption={#2}, #1]{\begin{spacing}{0.85}#2\end{spacing}}%
} 
\makeatother 

\usepackage{pifont}
\newcommand{\xmark}{\ding{55}}%

\makeatletter
\newcommand{\distas}[1]{\mathbin{\overset{#1}{\kern\z@\sim}}}%
\newsavebox{\mybox}\newsavebox{\mysim}
\newcommand{\distras}[1]{%
  \savebox{\mybox}{\hbox{\kern3pt$\scriptstyle#1$\kern3pt}}%
  \savebox{\mysim}{\hbox{$\sim$}}%
  \mathbin{\overset{#1}{\kern\z@\resizebox{\wd\mybox}{\ht\mysim}{$\sim$}}}%
}
\makeatother

\usepackage{natbib,bibunits}
\usepackage{setspace}
\title{\textbf{Implicit score-driven filters \\
for time-varying parameter models}\footnote{We thank the co-editor Michael Jansson for helpful editorial guidance, and the anonymous AE and two
referees for their valuable comments and suggestions.  We also thank the following individuals for their feedback: Christian Brownlees, Janneke van Brummelen, Leopoldo Catania, Guillaume Chevillon, Simon Donker van Heel, Gustavo Freire, Ana Galvao, Peter Hansen, Andrew Harvey, Onno Kleen, Erik Kole, Andrew Patton, Phyllis Wan and participants of the NESG 2022, CEF 2022, SoFiE 2022, Aarhus Econometrics Workshop 2022, UvA Econometrics Seminar 2022, SNDE Workshop for Young Scholars 2022, Barcelona Workshop in Financial Econometrics 2023, CREST 2023, and NBER-NSF 2023. Remaining errors are our own.}}
\author{ \textsc{Rutger-Jan Lange}\footnote{Econometric Institute, Erasmus University Rotterdam (lange@ese.eur.nl)},\; \textsc{Bram van Os}\footnote{Econometrics and Data Science Department, Vrije Universiteit Amsterdam (b.van.os2@vu.nl)}  \,  and\,  \textsc{Dick van Dijk}\footnote{Econometric Institute, Erasmus University Rotterdam (djvandijk@ese.eur.nl)}\\ }

\usepackage[vpos=1cm]{draftwatermark}
\SetWatermarkText{\textnormal{Forthcoming in} \textit{Journal of Econometrics}}
\SetWatermarkScale{0.06}
\SetWatermarkColor[gray]{0.5}
	\SetWatermarkAngle{0}

\begin{document}
\maketitle

\thispagestyle{empty}
\small
\begin{spacing}{1.3}

{\centering \textbf{Abstract}\par} \noindent 
We propose an observation-driven modeling framework that allows model parameters to vary over time through an implicit score-driven (ISD) update. The ISD update maximizes the logarithmic observation density with respect to the parameter vector while penalizing the weighted \(\ell_2\) norm relative to a one-step-ahead predicted parameter. This yields an implicit stochastic-gradient update. We show that the popular class of explicit score-driven (ESD) models arises when the observation log density is linearly approximated around the prediction. By preserving the full density, the ISD update extends the favorable local properties of the ESD update to a global setting. For log-concave observation densities, whether correctly specified or not, the ISD filter is stable for all learning rates, and its updates are contractive in mean squared error toward the (pseudo-)true parameter at every time step. We demonstrate the usefulness of ISD filters in simulations and empirical applications in finance and macroeconomics.
\end{spacing}
\vspace{0.1in} \noindent \textit{Keywords}: Implicit gradient, proximal-point method, stochastic-gradient descent, observation-driven models
\clearpage
\setcounter{page}{1}
\newgeometry{top=1in,bottom=1in,left=1in,right=1in}

\normalsize

\clearpage
\begin{bibunit}[chicago]

\section{Introduction}

Empirical evidence shows that assuming constant model parameters over prolonged periods of time is often too restrictive. In economics and finance, parameters are often regime dependent or subject to structural breaks (e.g., \citealp{stock1996evidence}). They may also evolve gradually without an obvious pattern, making it unclear how to update them after observing new data. In some cases, ex-post estimators can be constructed; for example, in ARCH-type models (see \citealp{terasvirta2009introduction}), the squared shock is an unbiased proxy for the true variance. In general, however, such proxies may be difficult to derive, inefficient, or nonexistent.

We propose a new framework that makes model parameters time-varying in an observation-driven setting via an implicit score-driven (ISD) filter. Analogous to \citeauthor{kalman1960new}'s (\citeyear{kalman1960new}) filter, the ISD filter alternates between prediction and  update steps. The update solves an optimization problem that maximizes the current observation's log-likelihood contribution subject to a weighted $\ell_2$ penalty centered at the one-step-ahead prediction. The penalty weights are determined by a positive-definite matrix, whose inverse can be interpreted as a learning-rate matrix. Maximizing the log-likelihood contribution allows new information to be efficiently incorporated, while the penalty regularizes the parameter adjustment. The ISD setup also enables automatic coordination of multiple interacting parameters and the incorporation of parameter constraints without necessitating parameter transformations (e.g., link functions).

The first-order condition of the ISD update can be written as an implicit stochastic-gradient step: \emph{implicit} because the gradient is evaluated at the updated rather than predicted parameter, and \emph{stochastic} because it uses noisy data. In the optimization literature, implicit updates are known as proximal-point methods and recognized as inherently more stable than their explicit counterparts, which arise as first-order approximations. 
For example, ISD updates guarantee log-likelihood improvements at each observation---unlike explicit score-driven (ESD) updates (Section~\ref{subsec:lit}).

The ISD filter enjoys several attractive theoretical properties that are typically sought in observation-driven models, but rarely combined in a single framework. First, the ISD filter is invertible under easily verifiable conditions: concavity of the logarithmic postulated density is typically sufficient, even when misspecified (Theorem~\ref{Thm_Invertability}). As such, this invertibility result requires no knowledge of the data-generating process (DGP). Hence, any differences due to the initialization vanish almost surely and exponentially fast, a key property in the filtering literature (e.g., \citealp{bougerol1993kalman}; \citealp{straumann2006quasi}).

Second, for concave logarithmic densities (again, even when misspecified), the ISD update is contractive in  mean squared error toward a small region around the \mbox{(pseudo-)}true parameter (Theorem~\ref{Thm_GlobOpt}). On average, therefore, updates improve on predictions. In fact, the largest gains are expected for the worst predictions. This result requires the existence of a pseudo-true parameter, but is otherwise largely independent of the DGP. Only when the prediction is very close to the pseudo-true parameter may the update be less accurate, as is unavoidable for any stochastic-optimization method (e.g., \citealp{patrascu2018nonasymptotic}).

Beyond being largely DGP agnostic, Theorems~\ref{Thm_Invertability}--\ref{Thm_GlobOpt} are also robust in that they hold for an arbitrary (positive-definite) learning-rate matrix. In contrast, ESD filters require additional Lipschitz-type conditions and sufficiently small learning rates to avoid instability or 
divergence, especially in misspecified settings. Theorems~\ref{Thm_Invertability}--\ref{Thm_GlobOpt} assume  concavity of the researcher-postulated log density to obtain strong theoretical guarantees. While this reliance on concavity/convexity is standard in the optimization literature (e.g., \citealp{boyd2004convex}; \citealp{nesterov2018lectures}), throughout the article we highlight extensions indicating that the proposed method is broadly applicable (see Remark~\ref{remark1}).

We illustrate the properties of ISD filters through simulation experiments and empirical applications. In simulations, we show that under correct specification the static \mbox{(hyper-)}parameters are accurately recovered (Section~\ref{Sec_Est}). Under misspecification, ISD filters reliably track the pseudo-true parameter path even when the ESD filter diverges (Section~\ref{Sec_Sim}), which can arise even in simple settings if the gradient is not Lipschitz continuous. Although this risk of divergence is well known in optimization, it appears to have been overlooked in the score-driven filtering literature (Section~\ref{subsec:lit}). In a high-dimensional network-flow design, we further demonstrate that, as dimensionality increases, ISD filtering outperforms particle filtering in both estimation accuracy and computational efficiency (Section~\ref{sec: Network}).

Finally, we present three empirical applications highlighting advantages of ISD over ESD filters. First, we estimate a linear regression of Microsoft returns on market returns with a time-varying beta. Second, we analyze growth-at-risk via lower quantiles of U.S. GDP growth, where the ISD filter yields an implicit version of \citeauthor{engle2004caviar}'s (\citeyear{engle2004caviar}) adaptive CAViaR model. Because ISD quantile updates cannot cross the observation on which they are based, simple restrictions on the prediction step ensure that jointly modeled quantiles remain properly ordered. Third, in a study of T-bill rate spreads, the ISD filter outperforms its explicit counterpart even when our concavity assumptions are violated.

The paper is structured as follows. Section~\ref{Sec_Method} introduces ISD filters and compares them with their conventional ESD counterparts. Sections~\ref{Sec_Stability} and \ref{Sec_Glob} develop our main theoretical results on filter stability and accuracy, respectively. Section~\ref{Sec_Est} discusses maximum-likelihood estimation of the static parameters, showing accurate parameter recovery in simulations. Section~\ref{Sec_Sim}  contains simulations investigating the filtering performance under misspecification. Section~\ref{Sec_Emp} presents the empirical applications, while Section~\ref{Sec_Con} concludes. The appendix contains proofs of the main results (Appendix~\ref{appendix A}), additional theoretical results (Appendix~\ref{appendix B}), and further details for the simulations and empirical analyses (Appendix~\ref{Appendix C}).

\subsection{Positioning in the literature}
\label{subsec:lit}

This paper intersects with two strands of literature, summarized along two axes in Table~\ref{Tab_refs}, differentiated by (a) the methodology (explicit vs.\ implicit gradient methods), and (b) the goal (learning vs.\ tracking). To our knowledge, this paper is unique in the fourth quadrant: it uses an \emph{implicit} gradient method to \emph{track} a dynamic parameter.\footnote{The overview in Table~\ref{Tab_refs} is not exhaustive; for instance, we omit simulation-based approaches such as particle filters (e.g., \citealp{chopin2020introduction}).}

Along the methodology axis (implicit vs.\ explicit gradient methods), the ISD filter aligns with implicit methods for static optimization, notably \citeauthor{rockafellar1976monotone}'s (\citeyear{rockafellar1976monotone}) proximal-point algorithm, which couples a static target with a quadratic penalty around the previous iterate. Because our log likelihood uses random observations from the true density, the ISD filter at each step can be viewed as a stochastic proximal-point method (e.g., \citealp{bauschke2003bregman}; \citealp{ryu2016stochastic}; \citealp{bianchi2016ergodic}; \citealp{patrascu2018nonasymptotic}; \citealp{asi2019stochastic}). Proximal optimization is equivalent to an implicit stochastic-gradient step (e.g., \citealp{toulis2015scalable}; \citealp{toulis2016towards}; \citealp{toulis2017asymptotic}; \citealp{toulis2021proximal}). Our approach also ties in with online-learning methods that process data sequentially (e.g.,\ \citealp{orabona2019modern}; \citealp{cesa2021online}), including machine-learning applications (e.g.,\ \citealp{kulis2010implicit}). What sets our work apart from implicit gradient methods in the (stochastic) optimization literature is that the parameter to be estimated is dynamic.

\begin{table}[t!]
\caption{Overview of related methods. \label{Tab_refs}}
\begin{threeparttable}
\begin{tabular}{l|ll}
         & \footnotesize Explicit gradient method & \footnotesize  Implicit gradient method    \\ \hline
\footnotesize  Learning & \scriptsize SGD (e.g., \citealp{robbins1951stochastic};    & \scriptsize ISGD (e.g., \citealp{patrascu2018nonasymptotic};  \\
\footnotesize (static target) &\scriptsize\citealp{amari1993backpropagation}; \citealp{kushner2010stochastic}; \citealp{bottou2012stochastic})& \scriptsize\citealp{asi2019stochastic}; \citealp{toulis2021proximal})\\
\hline 
\footnotesize  Tracking & \scriptsize Constant-gain tracking (e.g., \citealp{benveniste2012adaptive}) & \scriptsize {\textbf{ISD filter}} (this article)   
\\
\footnotesize (dynamic target) & \scriptsize ESD filter (e.g., \citealp{creal2013generalized}; \citealp{harvey2013dynamic};
\\
 &\scriptsize for examples, see \href{https://www.gasmodel.com}{www.gasmodel.com})
\end{tabular}
\begin{tablenotes}
\item \footnotesize Note: (I)SGD = (implicit) stochastic gradient descent. (I/E)SD = (implicit/explicit) score driven. 
\end{tablenotes}

\end{threeparttable}
\end{table}

Along the goal-related axis (learning vs.\ tracking), our work aligns with engineering and optimization approaches that, following the seminal work of \citet{robbins1951stochastic}, use explicit gradient methods with constant (i.e., non-vanishing) step sizes to track dynamic targets (e.g., \citealp{ljung1977analysis}; \citealp{benveniste2003measure}). The resulting procedures are usually referred to as constant-gain tracking (for an overview, see \citealp[Ch.~4]{benveniste2012adaptive}), although extensions with adaptive step sizes have also been considered (e.g., \citealp{kushner2002analysis}). In econometrics, dynamic conditional score (DCS; \citealp{harvey2013dynamic}) models and generalized autoregressive score (GAS; \citealp{creal2013generalized}) models likewise use the (explicit) gradient of the log-likelihood function, known as the \emph{score}, to update time-varying parameters. This framework encompasses many established models (e.g., GARCH) and is popular for its ease of use and strong forecasting performance (e.g., \citealp{creal2014observation}; \citealp{harvey2014filtering}; \citealp{koopman2016predicting}; \citealp{harvey2017volatility}; \citealp{harvey2018modeling}; \citealp{opschoor2018new}; \citealp{gorgi2020beta}). It has been used in ${\sim}400$ published articles; for a near-exhaustive list, see \href{https://www.gasmodel.com}{www.gasmodel.com}. Recent surveys (\citealp{artemova2022score1,artemova2022score2}; \citealp{harvey2022score}) have converged on the terminology of score-driven (SD) models. To align with this nomenclature while distinguishing our approach, we refer to this model class as using \emph{explicit} score-driven (ESD) filters. As this article demonstrates, ESD filters can be obtained within the ISD framework by locally linearizing the logarithmic observation density around the prediction at each time step, an apparently new insight in the econometrics literature. We will show that avoiding this local-linear approximation has both theoretical and practical benefits.

\section{Implicit score-driven filters}
\label{Sec_Method}

We consider an $N\times 1$ variable of interest $y_t$, observed at times $t=1,\dots,T$, drawn from a data-generating process (DGP) characterized by a time-varying density, which we denote by $p^0_t(\cdot)$.
We refer to $p^0_t(\cdot)$ as a density, though it could also be a mass function (i.e., we allow for discrete observations $y_t$). The dynamics and functional form of the true density $p_t^0(\cdot)$ are left, for the most part, unspecified. It could be parametric, in which case $p^0_t(\cdot)=p^0(\cdot |\theta^0_t)$, where $\theta^0_t$ is a $K_0 \times 1$ time-varying parameter vector taking values in some (non-empty) parameter space $\Theta^0$. Alternatively, $p^0_t(\cdot)$ could be a non-parametric time-varying density.

The aim of this paper is to construct a modeling framework that tracks the true density $p^0_t(\cdot)$ using filters that alternate between prediction and update steps. Let $p(\cdot | \theta_{t}, x_{t}, \psi)$ denote the researcher-postulated observation density, which may be misspecified. Here, $\theta_t$ is a $K \times 1$ vector of time-varying parameters taking values in a (non-empty) convex parameter space $\Theta \subseteq \mathbb{R}^K$, $x_t$ is an $\mathcal{F}_{t-1}$-measurable variable, where $\mathcal{F}_{t-1}$ is the information set at time $t-1$, and $\psi$ is a vector of static parameters. The inclusion of $x_{t}$ allows the density to depend on exogenous variables and/or lags of $y_t$. For readability, we write $p(\cdot | \theta_t)$, suppressing the dependence on $x_t$ and $\psi$. We denote the researcher's predicted and updated parameter vectors by $\theta_{t\mid t-1} \in \Theta$ and $\theta_{t\mid t} \in \Theta$, which are based on $\mathcal{F}_{t-1}$ and $\mathcal{F}_t$, respectively.

While the postulated observation density $p(\cdot|\theta_t)$ may be misspecified relative to the true density $p_t^0(\cdot)$, model selection remains relevant. This can be illustrated using a trivial case: if $\Theta$ is a singleton, filtering is useless because $\theta_{t|t}$ cannot vary, so the filter is severely misspecified whenever the true density $p^0_t(\cdot)$ is time varying. In practice, we aim to narrow the misspecification gap by postulating a sufficiently flexible density $p(\cdot|\theta_t)$ that can closely approximate $p_t^0(\cdot)$. For our theory development, however, we treat $p(\cdot|\theta_t)$ as generic but fixed. We impose conditions relating only to the postulated density $p(\cdot|\theta_t)$ (i.e., Assumptions~\ref{A1_RL_Existence}--\ref{A5_Concavity}), thereby remaining agnostic about the true density $p^0_t(\cdot)$, or minimal conditions relating $p(\cdot|\theta_t)$ to $p^0_t(\cdot)$, such as the existence of a pseudo-true parameter (i.e., Assumptions~\ref{A7_Uniqueness}--\ref{A8_BoundedInformation}).

Given some postulated density $p(\cdot|\theta_t)$, the main difficulty in designing filtering algorithms lies in specifying how the update $\theta_{t|t}$ should be constructed from the prediction $\theta_{t|t-1}$ and the observation $y_t$. Here we argue that a sound updating scheme should satisfy at least two criteria. First, the update should yield an improved fit of the observed data $y_t$ in terms of the postulated density; i.e., we want $p(y_t|\theta_{t|t}) \geq p(y_t|\theta_{t|t-1})$. As we shall see, explicit score-driven (ESD) filters, despite their popularity (Section~\ref{subsec:lit}), generally fail to meet this requirement. Second, as each observation $y_t$ is inherently noisy, it is desirable to regularize the extent to which the update $\theta_{t|t}$ deviates from the prediction $\theta_{t|t-1}$. Penalizing the magnitude of $\theta_{t|t}-\theta_{t|t-1}$ prohibits the filtered parameter path from becoming excessively volatile. 

To satisfy both criteria, we propose the new class of implicit score-driven (ISD) filters. These filters perform the parameter update at time $t$ by maximizing the researcher-postulated logarithmic observation density $\log p(y_t|\cdot)$ subject to a weighted $\ell_2$ penalty centered at the prediction $\theta_{t|t-1}$. That is, the parameter update is defined as
\begin{equation}
  \label{ProParUpdate}
    \theta_{t|t} \; :=\; \underset{\theta \in \Theta}{\mathrm{argmax}} \; f(\theta|y_t, \theta_{t|t-1}, P_t),   
\end{equation}
where
\begin{equation}
    \label{RegLike}
    f(\theta|y_t, \theta_{t|t-1}, P_t) \; :=\; \log p(y_t|\theta) \, -\,  \frac{1}{2}\big\|\theta - \theta_{t|t-1}\big\|_{P_t}^2.
\end{equation}
\noindent
Here,  $f(\theta|y_t, \theta_{t|t-1}, P_t)$ denotes the ``regularized'' log-likelihood contribution and $\|z\|_{P_t}^2 = z' P_t z$ is the squared $\ell_2$ norm with respect to a $K\times K$ positive-definite penalty matrix $P_t$.

Formulating the parameter update as a maximization problem yields several benefits: (a) it uses the full conditional density, not just moments, (b) the elements of $\theta_{t|t}$ are interdependent because they jointly solve the multivariate problem~\eqref{ProParUpdate}, and (c) as an optimizer, $\theta_{t|t}$ automatically lies in the correct space $\Theta$ without necessitating link functions. For example, if $\theta_t$ contains positive shape parameters (as in Section~\ref{sec: dynamic gamma}), update~\eqref{ProParUpdate} keeps them positive; likewise, the ordering of multiple quantiles (as in Section~\ref{sec: GAR}) is preserved. More generally, the optimization framework easily accommodates additional constraints.

The $\ell_2$ penalty yields tractable updates and can be 
interpreted as a second-order Taylor expansion around $\theta_{t|t-1}$ of a loss function, where $P_t$ acts as the Hessian. The ISD update \eqref{ProParUpdate}–\eqref{RegLike} mirrors \citeauthor{rockafellar1976monotone}'s (\citeyear{rockafellar1976monotone}) classic proximal-point algorithm, which similarly considers the optimization of a target function---in our case, the log-likelihood contribution of the (\emph{a priori} random) observation~$y_t$---subject to a quadratic penalty. Hence, for a fixed time step, the approach is a stochastic proximal-point algorithm (e.g., \citealp{bauschke2003bregman}; \citealp{asi2019stochastic}); the difference is that we consider a moving target (Section~\ref{subsec:lit}).

Update~\eqref{ProParUpdate} can also be viewed as the posterior mode in a (possibly misspecified) Bayesian setup, with the quadratic penalty acting as a Gaussian prior. This resembles Laplace approximations in both the Bayesian (e.g., \citealp{rue2009approximate}) and frequentist approaches (e.g., \citealp{koyama2010approximate}). Update~\eqref{ProParUpdate} reduces to Kalman's level update when $p(\cdot|\theta)$ is Gaussian, the mean is linear in $\theta$, and the penalty matrix $P_t$ is the inverse predicted covariance (\citealp{lange2024short}). For a Gaussian linear regression with time-varying slopes, update~\eqref{ProParUpdate} specializes to the normalized least mean squares (NLMS) filter (\citealp{nagumo1967learning}); we return to this case in an empirical illustration (Section~\ref{Sec_Emp:linreg}). Although the link between (least-squares) optimization and the Kalman filter has been known since \citet{bierman1977factorization} and \citet{bertsekas1996incremental}, it has recently attracted renewed interest in signal processing (e.g., \citealp{akyildiz2019probabilistic}), control (e.g., \citealp{simonetto2024nonlinear}), and econometrics (e.g., \citealp{lange2024bellman,lange2024short}).

These links justify investigation of the proximal method~\eqref{ProParUpdate} in a more \emph{general} setting: the observation density may be non-Gaussian, $\theta$ need not relate to the mean, and the quadratic penalty need not represent a Gaussian prior. We aim to remain agnostic about the true density sequence $\{p_t^0(\cdot)\}$; i.e., we do not presume latent states with linear Gaussian dynamics. Accordingly, we postulate \eqref{ProParUpdate} as part of a \emph{filter} or \emph{algorithm}, a conceptually distinct approach from imposing conditions on the DGP. We then investigate the algorithm's performance, especially when some or all of the classic assumptions fail. Despite its simplicity and close connection with existing methods, the proposed method is---at this level of generality---new.

Whereas the DGP is generally unknown, the postulated density $p(\cdot|\theta)$ is under the researcher's control; typically, it is known in closed form. Assumptions~\ref{A1_RL_Existence}--\ref{A5_Concavity} relate only to this postulated density and are, therefore, practically verifiable. Assumptions~\ref{A1_RL_Existence} and~\ref{A2_RL_StrictConcave} are standard in optimization, ensuring existence and uniqueness of the maximizer in~(\ref{ProParUpdate}). Assumptions~\ref{A3_Boundary} and~\ref{A4_Differentiability} allow its characterization using standard first-order conditions; while not strictly necessary (e.g., subgradients could be used), this aids clarity and tractability.

\begin{assumption} \label{A1_RL_Existence} \textbf{\emph{(Existence)}}
The solution set of $\smash{\underset{\theta \in \Theta}{\mathrm{argmax}} \: f(\theta|y_t,\theta_{t|t-1}, P_t)}$ is non-empty with probability one. 
\end{assumption}
\begin{assumption} \label{A2_RL_StrictConcave} \textbf{\emph{(Strictly concave regularized log likelihood)}}
$f(\theta|y_t,\theta_{t|t-1}, P_t)$ is proper concave and strictly concave in $\theta$, $\forall \theta \in \Theta$ with probability one.
\end{assumption}
\begin{assumption} \label{A3_Boundary} \textbf{\emph{(Interior solution)}}
$\theta_{t|t} \in \mathrm{Int}(\Theta)$ with probability one.
\end{assumption}
\begin{assumption} \label{A4_Differentiability} \textbf{\emph{(Differentiability)}}
$\log p(y_t|\theta)$ is at least (a) once or (b) twice continuously differentiable in $\theta$, $\forall \theta \in \mathrm{Int}(\Theta)$ with probability one. When left unspecified, (b) holds.
\end{assumption}

Assumptions \ref{A1_RL_Existence} and \ref{A2_RL_StrictConcave} can typically be satisfied by choosing a sufficiently large penalty $P_t$. In particular, even when the postulated log density is poorly behaved (e.g.,\ non-concave or multimodal), a sufficiently strong penalty ensures that update~\eqref{ProParUpdate} remains well-behaved. 

Under Assumptions \ref{A1_RL_Existence} through \ref{A4_Differentiability}a, the first-order condition for the parameter update $\theta_{t|t}$ in the maximization problem~(\ref{ProParUpdate}), i.e.,\ $0=\nabla( y_t | \theta_{t|t})-P_t(\theta_{t|t}-\theta_{t|t-1})$, can be rearranged as  
\begin{equation}
  \theta_{t|t}\; =\; \theta_{t|t-1}\, +\, H_t \, \nabla( y_t | \theta_{t|t}),
  \label{ProximalFOC}
\end{equation}
where the inverse penalty $H_t := P_t^{-1}$ is referred to as the learning-rate matrix at time $t$ and $\nabla(y_t|\theta_{t|t}) := (\partial  \log p(y_t|\theta) /\partial \theta)|_{\substack{\theta = \theta_{t|t}}}$ is the \emph{score} evaluated at the updated parameter $\theta_{t|t}$.

Representation~\eqref{ProximalFOC} shows that the ISD update yields a gradient-type update, in which the learning-rate matrix $H_t$ controls the step size. Because the score is evaluated at the update $\theta_{t|t}$ instead of the prediction $\theta_{t|t-1}$, it is an \emph{implicit} method. Indeed, the update $\theta_{t|t}$ appears on both sides of equation~\eqref{ProximalFOC}; hence, it is not immediately computable. As the update also depends on the (\emph{a priori} random) observation $y_t$, our method is also closely related to implicit stochastic-gradient methods (Section~\ref{subsec:lit}). 

While ISD update~(\ref{ProximalFOC}) may not allow a closed-form solution, Assumptions~\ref{A2_RL_StrictConcave} and~\ref{A4_Differentiability}a ensure the global solution to~\eqref{ProParUpdate} can be found numerically using standard techniques. Specifically, for a scalar parameter ($K=1$), root-finding (e.g., bisection) applies. More generally, quasi-Newton methods with line search can be used, with projected or interior-point variants to handle constraints. For a toolbox of proximal schemes, see \citet[Ch.~6]{parikh2014proximal}.

In the optimization literature, $H_t$ is often set to decrease over time (e.g., $H_t=\mathcal{O}(t^{-1})$), so the estimated parameter converges to a constant vector. Here, we aim to track a moving target, so the parameter path must remain responsive, even asymptotically. We may therefore keep the learning rate constant, as is standard in tracking applications (e.g., \citealp[p.\ 93]{kushner2010stochastic}; \citealp[Ch.\ 4]{benveniste2012adaptive}). Indeed, we typically set $H_t=H$ for all $t$, where $H$ may contain static parameters to be estimated (Section~\ref{Sec_Est}). The ISD filter can thus be viewed as an implicit stochastic gradient method with a non-vanishing learning rate. It is more flexible in that $H_t$ can be a time-varying positive-definite matrix (rather than  a scalar, as is typical in the optimization literature) and includes a separate prediction step, described below, which is crucial in time-series modeling.

The prediction step generates one-step-ahead forecasts of the time-varying parameters. For simplicity, we consider a linear first-order specification as follows:
\begin{equation}
    \theta_{t+1|t} \;=\; \omega\, + \, \Phi\, \theta_{t|t},
    \label{GASPred}
\end{equation}
where $\omega$ is a $K\times 1$ vector of constants and $\Phi$ is a $K\times K$ autoregressive matrix. Conditions ensuring stable recursions are discussed in the next section. Prediction step~\eqref{GASPred}  could be generalized to allow for non-linear and/or higher order dynamics if these were found to be relevant for a particular application. In economics and statistics, however, mean reversion is often critical, while no additional information is available at the time of the prediction. For these reasons, a more complicated structure may not yield immediate benefits.

Throughout, we require that the prediction step~\eqref{GASPred}  maps $\Theta$ to itself. For $\Theta=\mathbb{R}^K$, this requirement imposes no restrictions beyond $\omega$ and $\Phi$ having real-valued elements. For bounded parameter spaces, it can often be achieved by restricting $\omega$ and $\Phi$. For a scalar variable in $\Theta=\mathbb{R}_{>0}$ (e.g., a variance), it suffices to take $\omega,\Phi>0$. For a scalar variable in $\Theta=(-1,1)$ (e.g., a correlation), we can impose $\Phi\in[0,1)$ and $\tilde{\omega}:=\omega/(1-\Phi)\in(-1,1)$, so that
$
\theta_{t+1|t}=(1-\Phi)\tilde{\omega}+\Phi\,\theta_{t|t},
$ is a convex combination of two elements in the convex set $\Theta=(-1,1)$. The same logic extends to higher dimensions; for example, if $\Theta$ is the positive orthant, then a diagonal matrix $\Phi$ with entries $\Phi_{ii} > 0$ and intercepts $\omega_i > 0$ for $i=1,\ldots,K$ suffices. Finally, when the prediction step is the identity mapping (i.e., $\omega=0_K$, $\Phi=I_K$), $\theta_{t+1|t}\in \Theta$ is immediate.

As an alternative to imposing constraints on $\omega$ and $\Phi$, we can take $\Theta=\mathbb{R}^K$ and employ link functions mapping $\mathbb{R}^K$ into an appropriate domain. To illustrate, for $K=1$ we can use
the exponential map from $\mathbb{R}$ to $\mathbb{R}_{>0}$ to enforce positivity. Although widely used, such non-linear transformations may complicate the analysis of the filter's theoretical properties.

\subsection{Relationship with existing (explicit) score-driven filters}
\label{Sec:ESDvsISD}

The implicit gradient update~\eqref{ProximalFOC} suggests a close link with existing (i.e., explicit) updating schemes. Here we show that linearizing the logarithmic observation density in the ISD optimization problem~\eqref{ProParUpdate} produces the familiar ESD update. To explain the relationship, we may approximate~(\ref{RegLike}) via a first-order Taylor expansion at $\theta_{t|t-1}$, i.e., using $\log p(y_t|\theta)\approx \log p(y_t|\theta_{t|t-1})+\langle\theta-\theta_{t|t-1},\nabla(y_t|\theta_{t|t-1})\rangle$, where $\langle x_1,x_2\rangle:=x_1'x_2$. To avoid boundary solutions, we may assume $\Theta=\mathbb{R}^K$. The regularized log-likelihood $f(\cdot|\cdot,\cdot,\cdot)$ in \eqref{ProParUpdate} then becomes a linear objective with a quadratic penalty, so the optimization can be solved in closed form. Indeed, the resulting linearized version of (\ref{ProParUpdate}) and its associated closed-form solution read
\begin{equation}
    \theta_{t|t}^\textnormal{ex} := \underset{\theta \in \mathbb{R}^K}{\mathrm{argmax}}\left\{ \underbrace{\log p(y_t|\theta_{t|t-1}) + \langle \theta - \theta_{t|t-1},\nabla( y_t | \theta_{t|t-1})\rangle }_{\textnormal{linear approximation of $\log p(y_t|\theta)$ at $\theta_{t|t-1}$}}- \frac{1}{2}\|\theta - \theta_{t|t-1}\|_{P_t}^2\right\},
    \label{ExplicitUpdate}
\end{equation}
\noindent
\begin{equation}
  \theta_{t|t}^\textnormal{ex} \;=\; \theta_{t|t-1}\, +\, H_t \, \nabla( y_t | \theta_{t|t-1}),
  \label{ExplicitFOC}
\end{equation}
where the explicit update is denoted $\theta_{t|t}^\textnormal{ex}$ 
to differentiate it from the ISD update~\eqref{ProximalFOC}.

Combining the explicit gradient update~\eqref{ExplicitFOC} with the linear prediction step \eqref{GASPred} reproduces the well-known class of dynamic conditional score (DCS; \citealp{harvey2013dynamic}) models or generalized autoregressive score (GAS; \citealp{creal2013generalized}) models. These equivalent model classes are collectively known as \emph{score driven} (Section~\ref{subsec:lit}). To align with this standard terminology, while emphasizing the difference with our approach, we refer to this existing model class as using \emph{explicit} score-driven (ESD) filters. Indeed, combining the explicit gradient step~\eqref{ExplicitFOC} with the linear prediction step~\eqref{GASPred} gives
\begin{equation}
    \theta_{t+1|t}^\textnormal{ex} \;=\; \omega \,+\, 
    \Phi\, H_t \,\nabla( y_t | \theta_{t|t-1})   \,+\, \Phi\, \theta_{t|t-1}.
\end{equation}
In this literature, it is standard to take $H_t = H S_t$, where $H$ is static, while $S_t$ is known as a \emph{scaling} matrix, which is often based on the Fisher information of the postulated density (e.g.,\ \citealp{artemova2022score1}). Subsequently writing $A := \Phi H$ yields the canonical prediction-to-prediction recursion in the ESD literature (e.g., \citealp[eq.\ 2]{creal2013generalized}).
While the relationship between explicit and implicit gradient steps is well-known in the optimization literature (e.g.,\ \citealp{rockafellar1976monotone}), it has not yet been recognized in the econometrics literature.

The inherent locality of the first-order Taylor expansion suggests that ESD updates may perform satisfactorily only for small (or, technically, infinitesimal) parameter adjustments $\theta_{t|t}^\textnormal{ex}-\theta_{t|t-1}$. Whenever the parameter adjustment is sizable, no likelihood improvement $p(y_t|\theta_{t|t}^\textnormal{ex})>p(y_t|\theta_{t|t-1})$ is guaranteed. Indeed, the likelihood may regularly deteriorate.

In contrast, for the ISD update with an arbitrary (positive-definite) learning-rate matrix $H_t$, the typical case $\theta_{t|t}\neq \theta_{t|t-1}$ automatically implies $p(y_t|\theta_{t|t})>p(y_t|\theta_{t|t-1})$. This property is due to the formulation of the ISD update as an optimization problem, where the value of the objective $f(\cdot|y_t,\theta_{t|t-1},P_t)$ at the optimizer $\theta_{t|t}$ must weakly exceed the value at the prediction $\theta_{t|t-1}$. Simple algebra yields the inequality $\log p(y_t|\theta_{t|t})-\log p(y_t|\theta_{t|t-1})\geq \sfrac{1}{2}\|\theta_{t|t}-\theta_{t|t-1}\|^2_{P_t}$. Hence, the ISD update cannot lead to a worse fit, and in fact produces a strictly better one whenever $\theta_{t|t} \neq \theta_{t|t-1}$.
By the same inequality, $\|\theta_{t|t}-\theta_{t|t-1}\|^2_{P_t}$ is bounded if the likelihood improvement is bounded, which can be viewed as a robustness property of ISD updates.

Generally, implicit and explicit gradients can point in opposite directions. When the maximization problem~\eqref{RegLike} is concave, however, both recommend adjustments in roughly the same direction; i.e., the angle between $\theta_{t|t}-\theta_{t|t-1}$ and $\nabla(y_t|\theta_{t|t-1})$ cannot exceed $90^\circ$.

\sloppy
\begin{proposition} \label{Lm_Alignment} \textbf{\emph{(Relationship between ISD and ESD updates)}} Fix $t>0$ and let Assumptions~\ref{A1_RL_Existence}, \ref{A2_RL_StrictConcave} and \ref{A4_Differentiability}a hold. Consider a prediction $\theta_{t|t-1} \in \Theta$ and positive-definite penalty $P_t\in \mathbb{R}^{K \times K}$. Compute $\theta_{t|t}$ using the ISD update~\eqref{ProParUpdate}. Then, with probability one,
\begin{equation}
\label{Eq: same sign}
    \langle \theta_{t|t} - \theta_{t|t-1}, \nabla( y_t | \theta_{t|t-1}) \rangle \; \geq\;  0. 
\end{equation} 
If Assumptions~\ref{A3_Boundary} and \ref{A4_Differentiability}b also hold, we may write:
\begin{equation}
    \theta_{t|t} = \theta_{t|t-1} + (P_t + \mathcal{I}_{t|t})^{-1}\nabla(y_t|\theta_{t|t-1}),
\end{equation}
where $\mathcal{I}_{t|t}$ denotes the negative average $K \times K$ Hessian between $\theta_{t|t-1}$ and $\theta_{t|t}$,
\begin{equation}
    \mathcal{I}_{t|t} :=  -\int_{0}^{1} \left.\frac{\partial^2  \log p(y_t|\theta)}{\partial\theta \partial\theta'}\right|_{\substack{\theta \,= \, u \,\theta_{t|t-1} \,+\, (1-u)\,\theta_{t|t}}} \mathrm{d}u.
\end{equation}
\end{proposition} 

For a scalar time-varying parameter ($K=1$), equation~\eqref{Eq: same sign} implies that the implicit and explicit adjustments ($\theta_{t|t}-\theta_{t|t-1}$ and $\theta^\textnormal{ex}_{t|t}-\theta_{t|t-1}$) have the same sign; naturally, so do the gradients $\nabla(y_t|\theta_{t|t})$ and $\nabla(y_t|\theta_{t|t-1})$. For the ISD update, the derivative of $\log p(y_t|\theta)$ evaluated at the update thus has the same sign as the derivative at the prediction. Because the derivative cannot switch signs, the ISD update increases the value of $\log p(y_t|\theta)$ without surpassing the peak. For the ESD update, the derivative at the update and prediction can have opposite signs; hence, the explicit update can overshoot the peak.
For this reason, ESD updates generally fail to ensure $ p(y_t|\theta^\text{ex}_{t|t})>p(y_t|\theta_{t|t-1})$.

The second result of Proposition~\ref{Lm_Alignment} shows that the ISD update can be written as a ``curvature-corrected'' version of the ESD update. Here, $\mathcal{I}_{t|t}$ is the average negative $K\times K$ Hessian between $\theta_{t|t-1}$ and $\theta_{t|t}$; i.e., the average curvature of $\log p(y_t|\theta)$ between these points. The ISD update adjusts the step size by accounting for second-order effects. If the log-likelihood is linear in $\theta$, then $\mathcal{I}_{t|t}=O_K$ such that both updates are identical. If the log-likelihood is (multivariate) quadratic (e.g., as with a Gaussian distribution in terms of the mean), $\mathcal{I}_{t|t}$ is constant and both updates are equivalent, albeit for different penalty matrices. 

To further characterize the ISD update~\eqref{ProParUpdate}, we focus on log-likelihoods $\log p(y_t|\theta)$ that are concave in $\theta$, yielding particularly strong stability and tracking results.

\begin{assumption} \textbf{\emph{(Log-concave observation density)}} \label{A5_Concavity}$\log p(y_t| \theta) + \frac{\alpha_t}{2}\,\|\theta\|^2$ is concave in $\theta$ for some $\alpha_t \geq 0$, $\forall \theta \in \Theta$, with probability one.
\end{assumption} 

Assumption~\ref{A5_Concavity} strengthens Assumption~\ref{A2_RL_StrictConcave} by imposing concavity on the log-likelihood function itself, rather than on its regularized version~\eqref{RegLike}. Concavity strength is indexed by $\alpha_t\ge0$, where $\alpha_t=0$ implies concavity, while $\alpha_t>0$ implies $\alpha_t$-strong concavity. Assumption~\ref{A5_Concavity} allows us to show that, under a common penalty matrix $P_t$, the implicit gradient update is a ``shrunken'' version of the explicit gradient update.

\begin{proposition} \label{Lm_Shrinkage} \textbf{\emph{(Step-size shrinkage)}} Fix $t>0$ and let Assumptions~\ref{A1_RL_Existence} to \ref{A5_Concavity} hold. Take a prediction $\theta_{t|t-1} \in \Theta$ and positive-definite penalty $P_t\in \mathbb{R}^{K \times K}$ as given. Based on the observation $y_t$, compute $\theta_{t|t}$ using the ISD update~\eqref{ProParUpdate} and $\smash{\theta_{t|t}^\textnormal{ex}}$ using the ESD update~\eqref{ExplicitFOC}. Let $\lambda_{\max}(P_t)$ denote the largest eigenvalue of $P_t$. Then, with probability one,
\begin{equation}
\label{Eq_Shrinkage}
    \big\|\theta_{t|t}-\theta_{t|t-1}\big\|^2_{P_t} \; \leq \; \underbrace{\left(\frac{\lambda_{\max}(P_t)}{\lambda_{\max}(P_t)+ \alpha_t} \right)^2}_\textnormal{$\in [0,1]$, contraction coefficient}\big\|\theta^\textnormal{ex}_{t|t}-\theta_{t|t-1}\big\|^2_{P_t}.
\end{equation}
\end{proposition}
\noindent

The contraction coefficient in~\eqref{Eq_Shrinkage}, which returns multiple times in Sections~\ref{Sec_Stability} and~\ref{Sec_Glob}, depends on the ratio between the measure of concavity and the penalization strength. 
Specifically, larger $\alpha_t$ or smaller $\lambda_{\max}(P_t)$ implies more shrinkage. To understand why the explicit step size is larger, note that for concave log-likelihoods, every tangent lies above the curve. Hence, the linear approximation underlying the explicit update \emph{over}states the achievable likelihood gain, implying that the larger step size of the explicit update is suboptimal.

In practice, the shrinkage property in \eqref{Eq_Shrinkage} lets ISD filters use larger learning rates without sacrificing stability. As we will see, the update map is contractive for any positive-definite learning-rate matrix (Lemma~\ref{Lm_StaUp}). By contrast, ESD filters need additional (Lipschitz) conditions on the gradient and smaller learning rates to avoid overshooting (and possibly divergence), which can limit their responsiveness. Our results thus mirror well-known optimization results: implicit gradient methods typically can be more responsive without compromising stability (e.g., \citealp{toulis2017asymptotic}).

\begin{remark}[Beyond log-concavity]  \label{remark1}
Assumption~\ref{A5_Concavity} is maintained mainly because it yields our strongest theoretical guarantees, in the same spirit that convexity and concavity assumptions are central in optimization theory (e.g., \citealp{boyd2004convex}; \citealp{nesterov2018lectures}). This should not be taken to mean that the proposed method is only useful in such settings: the optimization literature indicates that implicit gradient methods can remain effective for non-concave objectives (e.g., \citealp{hare2009computing}; \citealp{grimmer2023landscape}). To emphasize general applicability, we highlight extensions beyond the concave case throughout the paper. In Sections~\ref{Sec_Stability} and~\ref{Sec_Glob}, we develop additional theoretical results on stability and mean squared error (MSE) improvements in non-concave settings. In Sections~\ref{sec_sim} and~\ref{Sec_GED}, we consider non-concave log densities in simulation studies and show accurate static \mbox{(hyper-)}parameter recovery and filtering performance. Finally, in an empirical illustration in Section~\ref{sec: Empirical illustration}, we show that the implicit filter outperforms competing methods even when both Assumptions~\ref{A2_RL_StrictConcave} and~\ref{A5_Concavity} fail.
\end{remark}

\section{Filter stability} \label{Sec_Stability}

We investigate the stability properties of the proposed ISD framework, providing sufficient conditions for filter invertibility, meaning that filtered paths based on identical data but with different initializations converge exponentially fast over time. We remain agnostic with regard to the DGP and use Assumptions~\ref{A1_RL_Existence}--\ref{A5_Concavity}, which relate to the postulated density only.

Our results in this section are presented in three parts: (a) fixing $t$ and examining the update step (Lemma~\ref{Lm_StaUp}), (b) fixing $t$ and considering both the update and prediction steps (Lemma~\ref{Lm_StaPre}), and (c) proving invertibility by considering the composition of all prediction-to-prediction mappings (Theorem~\ref{Thm_Invertability}). We begin by evaluating update stability for a single, fixed time point $t$. Lemma~\ref{Lm_StaUp} shows that the ISD update~\eqref{ProParUpdate} is stable under Assumptions \ref{A1_RL_Existence} through~\ref{A5_Concavity}, while, absent further conditions, the same does not hold for the ESD update~\eqref{ExplicitFOC}.

\begin{lemma} \label{Lm_StaUp} \textbf{\emph{(Prediction-to-update stability)}} Fix $t>0$ and let Assumptions \ref{A1_RL_Existence} to \ref{A5_Concavity} hold. Let $\theta_{t|t-1}$ and $ \tilde{\theta}_{t|t-1}$ denote two predictions in $\Theta$, which are combined with the observation $y_t$ in the ISD update step~(\ref{ProParUpdate}) to yield corresponding parameter updates, $\theta_{t|t}$ and $\tilde{\theta}_{t|t}$. Then, with probability one,
\begin{equation}
    \big\|\theta_{t|t} - \tilde{\theta}_{t|t}\big\|^2_{P_t} \; \leq\; \underbrace{\left(\frac{\lambda_{\max}(P_t)}{\lambda_{\max}(P_t)+ \alpha_t} \right)^2}_\textnormal{$\in [0,1]$, contraction coefficient}\big\|\theta_{t|t-1} - \tilde{\theta}_{t|t-1}\big\|^2_{P_t},
    \label{Eq_PrUpStaISD}
\end{equation}
where $\lambda_{\max}(P_t)$ is the largest eigenvalue of $P_t$. For the ESD update~(\ref{ExplicitFOC}), under the additional assumptions that $\nabla(y_t|\theta)$ is $L_t$-Lipschitz continuous in $\theta$ with probability one and $\lambda_{\min}(P_t) \geq L_t/2$, where $\lambda_{\min}(P_t)$ is the smallest eigenvalue of $P_t$, 
with probability one,
\begin{equation}
    \big\|\theta^\textnormal{ex}_{t|t} - \tilde{\theta}^\textnormal{ex}_{t|t}\big\|^2_{P_t} \leq \underbrace{\frac{\lambda_{\max}(P_t) - \alpha_t[2 - L_t/\lambda_{\min}(P_t)]}{\lambda_{\max}(P_t)}}_\textnormal{$\in [0,1]$, contraction coefficient} \big\|\theta_{t|t-1} - \tilde{\theta}_{t|t-1}\big\|^2_{P_t}.
    \label{Eq_PrUpStaESD}
\end{equation}
\end{lemma}

The first part of Lemma~\ref{Lm_StaUp} shows that the ISD update is non-expansive in the squared $P_t$-norm: it never increases (and possibly shrinks) the distance between any two paths. The contraction coefficient matches that in Proposition~\ref{Lm_Shrinkage}. With a strongly concave log-likelihood (i.e., $\alpha_t>0$), we obtain a strict contraction in the $P_t$-weighted norm whenever $\theta_{t|t-1}\neq\tilde{\theta}_{t|t-1}$. 

The second part shows an analogous result for the ESD update~(\ref{ExplicitFOC}) under two additional conditions: the score is $L_t$-Lipschitz and $\lambda_{\min}(P_t)\geq L_t/2$. Equivalently, eigenvalues of the learning-rate matrix $H_t$ do not exceed $2/L_t$. This  general requirement for explicit gradient methods is well-known in optimization (see \citealp{boyd2004convex}, Eq.~9.17; \citealp{nesterov2018lectures}, Eqns.~1.2.18--22) and machine learning (e.g.,~\citealp{wu2023implicit}). To prove ESD filter stability under a non-Lipschitz gradient (i.e., $L_t=\infty$), we thus require a zero learning rate. If the DGP is unknown, this seems to be a near-necessary condition:
in simulations (Section~\ref{Sec_Sim}), ESD filters can be divergent when the learning rate is positive but the score non-Lipschitz. Below, we focus on proving ISD filter stability; parallel results for ESD filters under additional (Lipschitz) conditions are possible, but not pursued here.

We now turn to the prediction-to-prediction mapping from time step $t$ to $t+1$. To obtain a strictly contractive prediction-to-prediction mapping for the ISD filter, it is sufficient for the update and prediction steps to be non-expansive in the ${P_t}$-weighted norm, provided at least one of them is strictly contractive. That is, when $\alpha_t = 0$, the prediction mapping from $\theta_{t|t}$ to $\theta_{t+1|t}$ must be strictly contractive. When $\alpha_t > 0$, on the other hand, it is sufficient for the prediction step to be non-expansive (e.g., allowing $\Phi=I_K$ and $\omega=0_K$).

A sufficient condition for non-expansiveness (contractiveness) of the prediction step in the ${P_t}$-weighted norm is that $P_t \succeq \Phi' P_t \Phi$ ($P_t \succ \Phi' P_t \Phi$). Here, the notation $X \succeq Y$ ($X \succ Y$) indicates that $X - Y$ has non-negative (strictly positive) eigenvalues for two symmetric real-valued matrices $X$ and $Y$ of the same size. This requirement is equivalent to $\|\Phi\|_{P_t} \leq 1$ ($\|\Phi\|_{P_t} < 1$), where $\|X\|_{P_t}$ is the induced operator norm of a matrix $X \in \mathbb{R}^{K\times K}$, which is also closely related to the discrete Lyapunov equation (e.g., \citealp{anderson2012optimal}). 

\begin{lemma} \label{Lm_StaPre} \textbf{\emph{(Prediction-to-prediction stability)}} Fix $t>0$ and let Assumptions \ref{A1_RL_Existence} to \ref{A5_Concavity} hold. Let $P_t$ be given with $P_t \succeq \Phi' P_t \Phi$. Let $\theta_{t|t-1}$ and $ \tilde{\theta}_{t|t-1}$ denote two predictions in $\Theta$ that are  used in the ISD update step~(\ref{ProParUpdate}) to yield the corresponding parameter updates $\theta_{t|t}$ and $\tilde{\theta}_{t|t}$, which are subsequently passed on to the prediction step~(\ref{GASPred}) to yield predictions $\theta_{t+1|t}$ and $\tilde{\theta}_{t+1|t}$. With probability one,
\begin{equation}
   \big \|\theta_{t+1|t} - \tilde{\theta}_{t+1|t}\big\|_{P_t}^2 \;\leq\; \kappa_t\, \big\|\theta_{t|t-1} - \tilde{\theta}_{t|t-1}\big\|_{P_t}^2,
\end{equation}
where the contraction coefficient $\kappa_t$ is
\begin{equation}
   \kappa_t \;=\; \frac{\lambda_{\max}(P_t)[\lambda_{\max}(P_t)  -\lambda_{\min}(P_t -\Phi'P_t\Phi)]}{(\lambda_{\max}(P_t) + \alpha_t)^2}\; \in\; [0,1].
   \label{Eq_eta}
\end{equation}
Hence the prediction-to-prediction mapping is non-expansive. If either $\alpha_t > 0$ or $P_t \succ \Phi' P_t \Phi$, then, with probability one, $\kappa_t \in [0,1)$; as such, the mapping is contractive. 
\end{lemma}

The contraction coefficient $\kappa_t\in[0,1]$ in the prediction-to-prediction mapping at time~$t$ depends  on the strength of concavity $\alpha_t$, the penalty matrix $P_t$, and the autoregressive matrix $\Phi$. It equals the contraction coefficients in Proposition~\ref{Lm_Shrinkage} and Lemma~\ref{Lm_StaUp} whenever $\lambda_{\min}(P_t-\Phi' P_t\Phi)=0$, as would be the case if $\Phi$ was the identity matrix. Whenever $\lambda_{\min}(P_t-\Phi' P_t\Phi)>0$, the contraction is further strengthened by the autoregressive matrix $\Phi$ in the prediction step, yielding $\kappa_t <1$ even if $\alpha_t=0$. That is, a strict contraction in the prediction-to-prediction mapping can be obtained through either $\alpha_t>0$ or $\lambda_{\min}(P_t-\Phi' P_t\Phi)>0$.

If $\alpha_t=0$, it is sufficient that $\lambda_{\min}(P_t-\Phi'P_t\Phi)>0$. For a scalar time-varying parameter, this holds if $|\Phi|<1$. In the multi-parameter case, it holds under $\Phi'\Phi\prec I_K$ if (a) $\Phi$ and $P_t$ are both diagonal or (b) either $\Phi$ or $P_t$ is a scalar multiple of the identity matrix. If $\Phi$ is symmetric, the condition $\Phi'\Phi\prec I_K$ is in turn equivalent to $\varrho(\Phi)<1$, where $\varrho(\cdot)$ is the spectral radius. If $\alpha_t>0$, the above strict inequalities can be made weak. We can also allow for more general matrices $\Phi$ and $P_t$ by imposing the latter to solve the discrete Lyapunov equation $P_t-\Phi'P_t\Phi=\Delta_t\succ0$, which for $\varrho(\Phi)<1$ has a unique solution $P_t\succ0$ parameterized by $\Delta_t\succ0$ (e.g., \citealp{bof2018lyapunov}, Thm. 3.2). 

As we show in Appendix~\ref{App_StabStudt}, ISD filter stability can be established even without log-concavity of the observation density (i.e., without $\alpha_t\geq 0$ as in Assumption~\ref{A5_Concavity}), provided the prediction mapping is sufficiently contractive to offset any expansion in the update step. We illustrate this for a Student's~$t$ distribution with time-varying location, which is not log-concave; a more general treatment is developed in \citet{donkervanheel2025gradientbased}.

Finally, we study the composition of all prediction-to-prediction mappings, which must be contractive for the initialization effects to vanish exponentially fast.  
A sufficient condition is that all mappings are contractive in a single shared norm over time. Theorem~\ref{Thm_Invertability} ensures the existence of such a norm and provides an invertibility result that is crucial for maximum-likelihood (ML) estimation of static parameters (e.g., \citealp{straumann2006quasi}).

\begin{theorem} \label{Thm_Invertability} \textbf{\emph{(Invertibility)}} For all $t>0$, let Assumptions \ref{A1_RL_Existence} to \ref{A5_Concavity} hold, with either (a) $P_t \succ \Phi' P_t \Phi$ or (b) $P_t \succeq \Phi' P_t \Phi$ and $\alpha_t > 0$. In addition, let there be some $\bar{P}, Q \in \mathbb{R}^{K\times K}$ with $\bar{P} \succ Q \succ O_{K}$ and a sequence $\{\rho_t > 0\}$ such that for all $t>0$, with probability one,
\begin{equation}
 \kappa_t P_t \,+ \, \rho_t Q \;\preceq\; \rho_t \,\bar{P}\; \preceq \; P_t, 
 \label{Eq_SuffInv}
\end{equation}
where $\kappa_t$ is defined in (\ref{Eq_eta}). Take two initial values $\theta_{0|0} \in \Theta$ and $\tilde \theta_{0|0} \in \Theta$, yielding two sequences $\{\theta_{t|t-1}\}$ and $\{\tilde \theta_{t|t-1}\}$, respectively. Then the ISD filter composed of (\ref{ProParUpdate}) and (\ref{GASPred}) is \emph{invertible}, i.e., there exists a constant $c>1$ such that with probability one, 
\begin{equation}
    \lim_{t \to \infty} c^t\big\|\theta_{t|t-1} - \tilde \theta_{t|t-1}\big\|^2\; \rightarrow \; 0.
    \label{Eq_EASInv}
\end{equation}
\end{theorem}

Theorem \ref{Thm_Invertability} gives a sufficient condition for contraction of all prediction-to-prediction mappings in the common ${\bar P}$-weighted norm, where the (time-invariant) matrix $\bar P$ satisfies condition~\eqref{Eq_SuffInv}. For a scalar time-varying parameter, \eqref{Eq_SuffInv} holds for any sequence $\{P_t\}$ whenever $|\Phi|<1$. For the unit-root case $|\Phi|=1$, it suffices that $\{P_t\}$ is uniformly upper bounded, while ${\alpha_t}$ is uniformly bounded away from zero, thereby preventing $\kappa_t$ from approaching unity. 

In the multi-parameter case, (\ref{Eq_SuffInv}) prevents elements of ${P_t}$ from having drastically different dynamics. The condition holds automatically when their time variation is identical; for example, when $P_t=\zeta_t P$ with $\zeta_t>0$ and $P\succ O_{K}$ satisfying $P\succ \Phi'P\Phi$. In general, the presence of $Q$ in (\ref{Eq_SuffInv}) ensures the contraction coefficient remains strictly below unity.

Interestingly, Theorem~\ref{Thm_Invertability} relies on the researcher-postulated density $p(\cdot|\theta_t)$, but not the true observation density $p^0_t(\cdot)$. Invertibility of the ISD filter can thus be guaranteed without restrictions on the DGP.  Because the ISD-filtered path is asymptotically insensitive to initialization, it is likewise asymptotically insensitive to any individual data point. Hence~(\ref{Eq_EASInv}) implies that any two parameter paths $\{\theta_{t|t-1}\}$ and $\{\tilde\theta_{t|t-1}\}$ based on different initializations but (almost) the same data are exponentially almost surely (e.a.s.) convergent.

As we show in Appendix~\ref{App_NumErr}, Theorem~\ref{Thm_Invertability} also prevents numerical errors during implementation in practice from accumulating.\footnote{We thank two anonymous reviewers for suggesting the additional analyses in Appendices \ref{App_NumErr} and \ref{App Lyapunov}.} A longstanding concern among researchers is that numerical errors can induce instability, even for closed-form recursions like Kalman's filter (e.g., \citealp{anderson2012optimal}). For the ISD filter, Theorem~\ref{Thm_Invertability} allows us to demonstrate that differences stemming from (a)~numerical (rounding) errors or (b)~inexact ISD update steps remain uniformly bounded over time; i.e., paths cannot diverge.

As we show in Appendix~\ref{App Lyapunov}, Theorem~\ref{Thm_Invertability} further implies the exponential \emph{Lyapunov} stability of the ISD filter: the limiting path $\{\theta_{t|t-1}\}$ equals the Lyapunov equilibrium. Moreover, if $\{y_t\}$ is stationary and ergodic and each $\theta_{t|t}$ is jointly measurable in $(y_t,\theta_{t\mid t-1})$, Theorem~\ref{Thm_Invertability} implies that the limiting path is also stationary and ergodic (\citealp[Prop.~4.3]{krengel2011ergodic}; \citealp[Thm.~1]{brandt1986stochastic}). This is critical to ensure the consistency of the ML estimator (Section~\ref{Sec_Est}).

Finally, we can derive an invertibility result for ESD filters analogous to that in Theorem~\ref{Thm_Invertability} for ISD filters, but only under the additional Lipschitz and learning-rate conditions given in Lemma~\ref{Lm_StaUp}. Doing so would yield a new multivariate, DGP-agnostic invertibility result for ESD filters, which contrasts with other approaches (e.g., \citealp{blasques2022maximum}) that rely on DGP-dependent expectations. Our simulations (Section~\ref{Sec_Sim}) confirm the necessity of these additional conditions for ESD filters: if they fail, no positive learning rate may exist that guarantees the stability of the ESD-filtered path.

In optimization, \citet{toulis2017asymptotic} similarly show that explicit gradient methods can be divergent. Stability is even harder to achieve in a dynamic setting, as our filters must remain responsive rather than converge. Any filter with a positive probability of divergence will eventually diverge. Theorem~\ref{Thm_Invertability} guarantees that ISD filters based on log-concave densities are stable under DGP-agnostic conditions 
depending only on  $\Phi$, $\{\alpha_t\}$, and $\{P_t\}$.

\section{Filter accuracy}
\label{Sec_Glob}

Ensuring filter stability, as in Theorem~\ref{Thm_Invertability}, is necessary but insufficient for high accuracy. After all, even a trivial filter that yields the same output for all inputs is stable. Therefore, accuracy cannot be assessed solely from the internal characteristics of the filter; instead, the analysis must also take account of the true process.
We consider a misspecified setting (e.g., the true observation density is unknown) and ask whether our updates still improve the quality of the filtered parameter path. In this case, we can only hope to track, as accurately as possible, the \emph{pseudo}-true parameter (e.g., \citealp{beutner2023consistency}).

Assumptions~\ref{A7_Uniqueness}--\ref{A8_BoundedInformation} below thus link the postulated density $p(\cdot|\theta_t)$ to the true density $p^0_t(\cdot)$. Under these assumptions, we will show that the ISD update is contractive in mean squared error (MSE) toward a small, ``noise-dominated'' region of the pseudo-true parameter; moreover, under strong concavity, this contraction is geometric. This result distinguishes filters that are merely stable from those that are also accurate; for example, for the trivially stable filter mentioned above, such improvements are impossible. For the ESD update, an analogous MSE improvement guarantee can be shown to hold, but only if the gradient is Lipschitz continuous and the learning rate sufficiently small. If concavity of the logarithmic density fails, both updates may yield only a \emph{constant} (rather than proportional) MSE reduction.

\begin{assumption} \label{A7_Uniqueness} \textbf{\emph{(Existence and uniqueness of pseudo-true parameter)}}
There exists a $\theta^\star_t \in \Theta$ such that $\smash{\underset{y_t}{\mathbb{E}}[\log p(y_t|\theta^\star_t)] > }$ $\smash{ \underset{y_t}{\mathbb{E}}[\log p(y_t|\theta)]}$ for all $\theta \in \Theta \setminus \{\theta^\star_t\}$.
\end{assumption}
\begin{assumption} \label{A8_BoundedInformation} \textbf{\emph{(Score moments at pseudo-true parameter)}} For each $t$, 
$\underset{y_t}{\mathbb{E}}[\nabla(y_t|\theta^\star_t)] = 0$ and $\underset{y_t}{\mathbb{E}}[\|\nabla(y_t|\theta^\star_t)\|^2] < \infty$.
\end{assumption}

Assumption~\ref{A7_Uniqueness} posits the existence of a unique pseudo-true parameter $\theta_t^\star$ that maximizes the expected (postulated) log likelihood, $\mathbb{E}_{y_t}[\log p(y_t|\theta)]$. If the postulated density is differentiable and strongly concave with probability one (i.e., Assumptions~\ref{A4_Differentiability}a and~\ref{A5_Concavity} hold with $\alpha_t>0$), Assumption~\ref{A7_Uniqueness} is automatically satisfied. If $p(\cdot|\theta_t)$ coincides with $p^0_t(\cdot)$, then $\theta_t=\theta_t^\star$. Assumption~\ref{A7_Uniqueness} thus underscores the need to choose a good candidate $p(\cdot|\theta_t)$, since no filter (including the ISD filter) can eliminate the misspecification gap. Assumption~\ref{A8_BoundedInformation} requires the first moment of the score at $\theta_t^\star$ to be zero and its second moment to be finite.

If $\theta_{t|t-1}$ deviates substantially from $\theta_t^\star$, the update $\theta_{t| t}$ based on $y_t$ often represents an improvement. As $\theta_{t| t-1}$ approaches $\theta_t^\star$, however, further improvements become increasingly hard to achieve. If $\theta_{t\mid t-1}$ is already very accurate, the noisy observation $y_t$ may in fact pull the update $\theta_{t| t}$ away from the pseudo-truth. If $\theta_{t| t-1}=\theta_t^\star$, a deterioration is almost inevitable; this is inherent to stochastic optimization methods, not a limitation of our approach. The neighborhood of $\theta_t^\star$ is therefore called the \emph{noise-dominated region} (NDR; e.g.,\ \citealp[p.\ 15]{ryu2016stochastic}; \citealp[p.\ 3]{patrascu2018nonasymptotic}; \citealp[Fig.\ 1]{lange2024bellman}). 

Since improvements are not always guaranteed, Theorem~\ref{Thm_GlobOpt} explicitly characterizes
the tug of war between contractive and expansive forces, which respectively decrease and increase the MSE relative to the pseudo-true parameter. Which of the two forces dominates largely depends on the accuracy of the prediction. While most authors (e.g., \citealp{asi2019stochastic}) establish upper bounds on the MSE after updating, our equations~\eqref{Eq_Opt_ISD}--\eqref{Eq_Opt_ESD} below provide exact equalities (rather than inequalities). This approach enables us to identify the precise conditions under which updates lead to improvement (see further discussion below).

\begin{theorem} \label{Thm_GlobOpt} \textbf{\emph{(Contraction to the NDR)}} Fix $t>0$ and let Assumptions \ref{A1_RL_Existence} to \ref{A8_BoundedInformation} hold. Then, for the ISD update \eqref{ProParUpdate}, we have
\begin{equation}
\underbrace{\underset{y_t}{\mathbb{E}}\left[\big\|\theta_{t|t} - \theta^\star_t\big\|_{P_t}^2 \right]}_\textnormal{MSE after update} = \underbrace{\big\|\theta_{t|t-1} - \theta^\star_t\big\|_{P_t}^2}_\textnormal{SE before update} 
-\underbrace{\underset{y_t}{\mathbb{E}}\left[\big\|\theta_{t|t} - \theta^\star_t\big\|_{2\mathcal{I}^{\star}_{t|t} + \mathcal{I}^{\star}_{t|t}P_t^{-1}\mathcal{I}^{\star}_{t|t}}^2 \right]}_\textnormal{$\geq 0$, contractive force}
+\underbrace{\underset{y_t}{\mathbb{E}}\left[\big\|\nabla(y_t|\theta^\star_{t})\big\|_{P^{-1}_t}^2\right]}_\textnormal{$\geq 0$, expansive force}.
\label{Eq_Opt_ISD}
\end{equation}
For the ESD update \eqref{ExplicitFOC}, we have 
\begin{equation}
   \underbrace{\underset{y_t}{\mathbb{E}}\left[\big\|\theta_{t|t}^\textnormal{ex} - \theta^\star_t\big\|_{P_t}^2\right]}_\textnormal{MSE after update} = \underbrace{\big\|\theta_{t|t-1} - \theta^\star_t\big\|_{P_t}^2}_\textnormal{SE before update} -  \underbrace{\underset{y_t}{\mathbb{E}}\left[\big\|\theta_{t|t-1} - \theta^{\star}_{t}\big\|^2_{2 \mathcal{I}^{\star}_{t|t-1} }\right]}_\textnormal{$\geq 0$, contractive force} +  \underbrace{\underset{y_t}{\mathbb{E}}\left[\big\|\nabla(y_t|\theta_{t|t-1})\big\|^2_{P_t^{-1}}\right]}_\textnormal{$\geq 0$, expansive force}.
\label{Eq_Opt_ESD}
\end{equation}
Here $\mathcal{I}^{\star}_{t|t}$, $\mathcal{I}^{\star}_{t|t-1} \succeq \alpha_t I_K\succeq O_K$ denote the negative average $K \times K$ Hessians between $\theta_{t|t}$ or $\theta_{t|t-1}$ and $\theta_{t}^{\star}$, that is,
\begin{equation}
    \mathcal{I}^{\star}_{t|t} :=  -\int_{0}^{1} \left.\frac{\partial^2  \log p(y_t|\theta)}{\partial\theta \partial\theta'}\right|_{\substack{\theta \,= \, u \,\theta_{t|t} \,+\, (1-u)\,\theta_t^{\star}}} \mathrm{d}u,
\end{equation}
\begin{equation}
    \mathcal{I}^{\star}_{t|t-1} :=  -\int_{0}^{1} \left.\frac{\partial^2  \log p(y_t|\theta)}{\partial\theta \partial\theta'}\right|_{\substack{\theta \,= \, u \,\theta_{t|t-1} \,+\, (1-u)\,\theta_t^{\star}}} \mathrm{d}u. 
\end{equation}
\end{theorem}

To the best of our knowledge, Theorem~\ref{Thm_GlobOpt} is new in the stochastic optimization literature: we did not find equations \eqref{Eq_Opt_ISD}--\eqref{Eq_Opt_ESD} in \citet{parikh2014proximal}, \citet{polson2015proximal}, \citet{ryu2016stochastic}, \citet{bianchi2016ergodic}, or \citet{asi2019stochastic}. The comparison with this literature is relevant (see Section~\ref{subsec:lit}), because Theorem~\ref{Thm_GlobOpt} treats a single time step. If the second term on the right-hand side of~\eqref{Eq_Opt_ISD} is dropped and the equality replaced by an inequality, we obtain a result similar to Theorem~3.2 of \citet{asi2019stochastic}. Since the expansive force in~\eqref{Eq_Opt_ISD} is bounded by Assumption~\ref{A8_BoundedInformation}, \citet[p.\ 2264]{asi2019stochastic} conclude that implicit updates are ``nondivergent,'' whereas explicit updates lack this guarantee. Our contribution in Theorem~\ref{Thm_GlobOpt} is the inclusion of contractive forces, which are critical in improving updates over predictions; these terms allow us to write equalities rather than inequalities. Because updates should ideally be more accurate than predictions (at least outside the NDR), these contractive forces are key. 

We now analyze equations~\eqref{Eq_Opt_ISD}--\eqref{Eq_Opt_ESD} in detail. On the left-hand side, we have the MSEs of the ISD and ESD updates in the squared $P_t$-weighted norm. Each right-hand side contains three terms. The first term, which is identical for ISD and ESD updates, is the $P_t$-weighted squared error (SE) before updating; i.e., the distance of the prediction from the pseudo-truth. 

The second term is the contractive force, which differs slightly between the two updates. For the ISD update, it is the \emph{post}-update MSE with weight matrix \(2\mathcal{I}^{\star}_{t|t} + \mathcal{I}^{\star}_{t|t}P_t^{-1}\mathcal{I}^{\star}_{t|t}\); for the ESD update, it is the \emph{pre}-update SE with weight matrix \(2\mathcal{I}^{\star}_{t|t-1}\). These forces are proportional to the negative average Hessians between \(\theta_{t|t}\) or \(\theta_{t|t-1}\), respectively, and \(\theta_t^{\star}\). Under strong concavity (\(\alpha_t>0\) in Assumption~\ref{A5_Concavity}), we have \(\mathcal{I}^\star_{t|t},\mathcal{I}^\star_{t|t-1}\succeq \alpha_t I_K\), so the contractive forces grow quadratically with the distance of \(\theta_{t|t}\) (ISD) or \(\theta_{t|t-1}\) (ESD) from \(\theta_t^\star\). While Assumption~\ref{A5_Concavity} guarantees \(\mathcal{I}^\star_{t|t},\mathcal{I}^\star_{t|t-1}\succeq O_{K}\) with probability one, it could be weakened by introducing a notion of ``expected strong concavity,'' as in optimization (e.g., \citealp[Ass.~3]{toulis2021proximal}).

The third term on the right-hand sides of~\eqref{Eq_Opt_ISD}--\eqref{Eq_Opt_ESD}, the expansive force, reveals a key difference between the two updating methods. For the ISD update, it is the expectation of the $P_t^{-1}$-weighted squared norm of the score evaluated at the pseudo-true parameter $\theta^\star_t$, yielding a noise term that is uniformly bounded (Assumption~\ref{A8_BoundedInformation}) and can be made arbitrarily small by increasing $P_t$. For the ESD update, the same term is evaluated at the prediction $\theta_{t|t-1}$, so its magnitude typically increases with the distance of $\theta_{t|t-1}$ from $\theta_t^\star$. Under strong concavity (i.e., $\alpha_t>0$ in Assumption~\ref{A5_Concavity}), the expansive force in the ESD update grows quadratically with this distance and, hence, may dominate the contractive force. Thus, explicit updates ``can be unstable even for relatively simple problems'' \citep[p.~2264]{asi2019stochastic}.

A key takeaway of Theorem~\ref{Thm_GlobOpt} is that, absent further conditions, ESD updates are not universally beneficial. If the prediction is inaccurate and the expansive force dominates the contractive force, ESD filters may diverge (as we often observe in simulations). This is precluded for the ISD update by the fact that the expansive force is uniformly bounded. Moreover, under strong concavity (i.e., $\alpha_t>0$),  the ISD update is geometrically contractive towards the NDR.

\begin{corollary}
\label{Cor_GlobOpt} \textbf{\emph{(Geometric contraction to the NDR)}} Fix $t>0$ and let Assumptions~\ref{A1_RL_Existence} to \ref{A8_BoundedInformation} hold, where Assumption~\ref{A5_Concavity} holds for some $\alpha_t > 0$. Then the ISD update \eqref{ProParUpdate} satisfies
\begin{equation}
\underbrace{\underset{y_t}{\mathbb{E}}\left[\big\|\theta_{t|t} - \theta^\star_t\big\|_{P_t}^2\right]}_{\textnormal{MSE after update}} \leq \underbrace{\left(\frac{\lambda_{\max}(P_t)}{\lambda_{\max}(P_t)+ \alpha_t}\right)^2}_\textnormal{$\in [0,1)$, contraction coefficient} \left(\underbrace{\big\|\theta_{t|t-1} - \theta^\star_t\big\|_{P_t}^2}_{\textnormal{SE before update}} + \underbrace{\underset{y_t}{\mathbb{E}}\left[\big\|\nabla(y_t|\theta^\star_{t})\big\|_{P^{-1}_t}^2\right]}_{\textnormal{(bounded) irreducible noise}}\right).
\label{Eq_CorOpt_ISD}
\end{equation}
\end{corollary} 

The contraction in Corollary~\ref{Cor_GlobOpt} is geometric in the sense that, for large prediction errors, the ratio of the \emph{post}-update MSE to the \emph{pre}-update SE is equal to a constant, which is less than unity. This contraction coefficient, which also appears in Proposition~\ref{Lm_Shrinkage} and Lemma~\ref{Lm_StaUp}, is stronger than comparable results in the literature (e.g.,\ \citealp[Thm.~1]{lange2024bellman}). 

To simplify, let $P_t=P I_K$ with $P\in \mathbb{R}_{>0}$, such that the $P_t$-weighted squared error becomes $P$ times the usual (i.e., Euclidean) squared error. Further, denote the (mean) squared error before and after updating as $\text{SE}_{t|t-1}:=\|\theta_{t|t-1}-\theta_t^\star\|^2$ and $\text{MSE}_{t|t}:=\mathbb{E}_{y_t}\|\theta_{t|t}-\theta_t^\star\|^2$, respectively. If $\alpha_t\geq \alpha>0$ for all $t$ in Assumption~\ref{A5_Concavity}, then inequality~\eqref{Eq_CorOpt_ISD} can be written as
\begin{equation}
\label{eq: nice illustration}
    \text{MSE}_{t|t}\;\leq\;  \left(\frac{P}{P+\alpha} \right)^2 \,\left(\text{SE}_{t|t-1}+\frac{\sigma^2}{P^2}\right), 
\end{equation}
where $\sigma^2\geq \sup_{t}
\mathbb{E}_{y_t}[\|\nabla(y_t|\theta^\star_t)\|^2]$ bounds the gradient noise. 
It is easy to show that $\text{MSE}_{t|t}<\text{SE}_{t|t-1}$ whenever $\text{SE}_{t|t-1}>\sigma^2/(2P\alpha+\alpha^2)$. Thus, updates are helpful in expectation whenever the squared prediction error exceeds $\sigma^2/(2P\alpha+\alpha^2)$. For more accurate predictions, no MSE gain can be guaranteed; hence, $\sigma^2/(2P\alpha+\alpha^2)$ can be interpreted as the squared radius of the NDR, which increases with $\sigma^2$, but decreases with $\alpha$ and $P$. As such, the NDR can be made arbitrarily small by increasing $P$, but only to the detriment of the contraction rate.

\begin{figure}[bth]

\centering
  \hspace{-2cm}
  \begin{tabular}{cc}
  \includegraphics[width=0.45\textwidth]{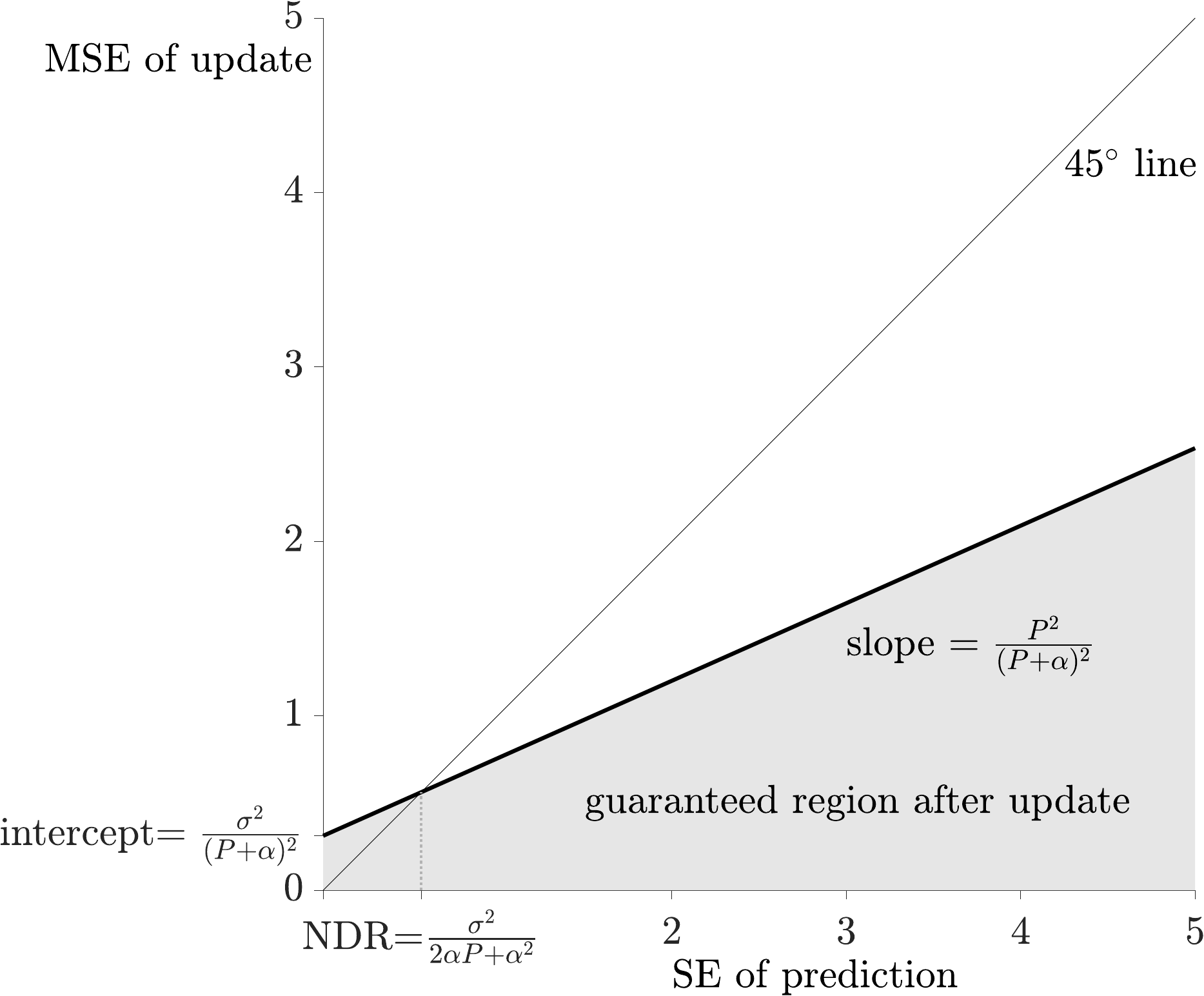}
        &  \includegraphics[width=0.44\textwidth]{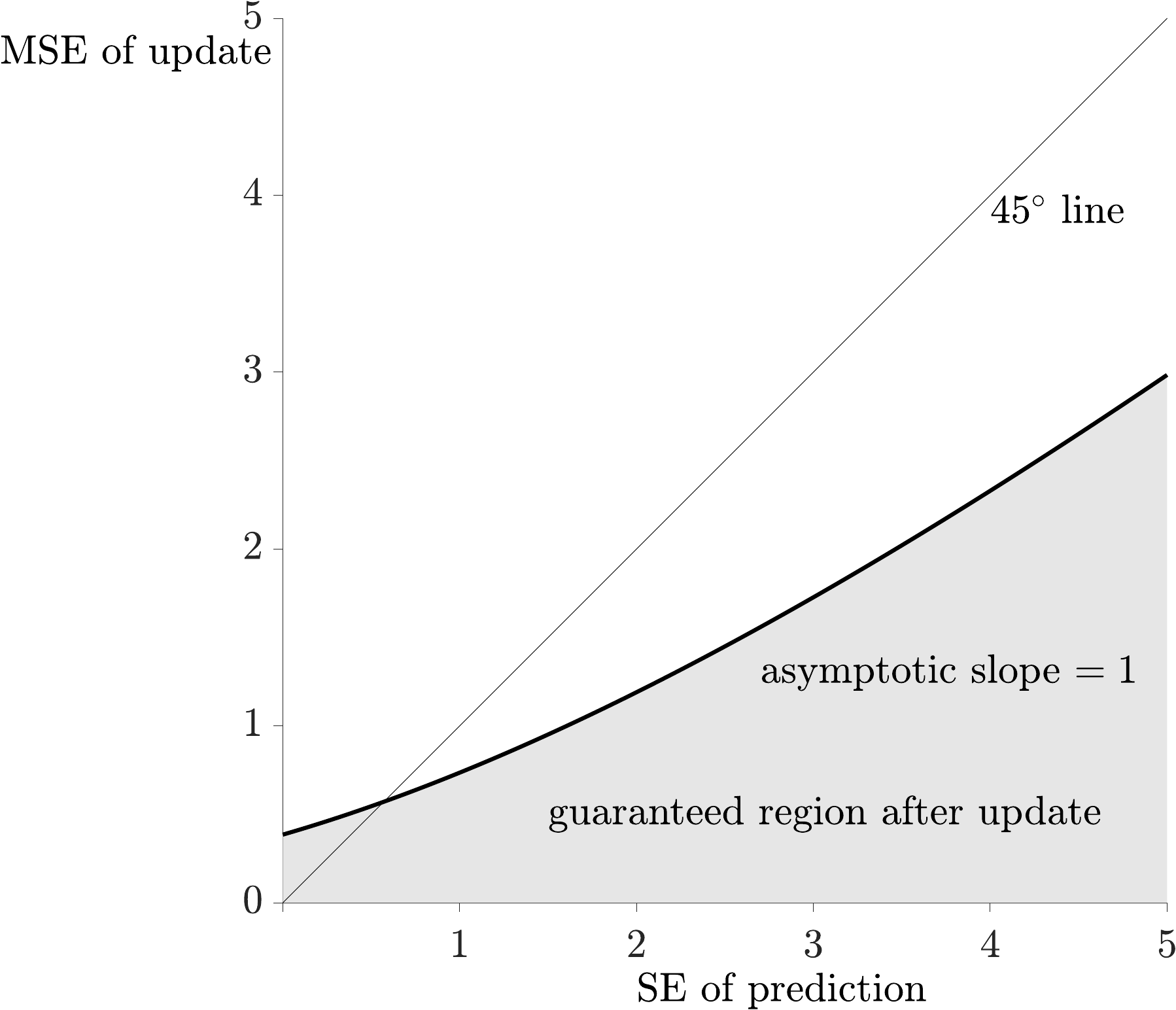}
        \\
       \hspace{1cm} \footnotesize (a) MSE contraction~\eqref{eq: nice illustration} with strong concavity & \hspace{1cm} \footnotesize (b) MSE reduction with non-concavity
  \end{tabular}
  \caption{Illustration of mean squared error (MSE) improvements.}
\label{fig_NDR_new}
\end{figure}

Figure~\ref{fig_NDR_new}(a) illustrates the geometric contraction to the NDR implied by inequality~\eqref{eq: nice illustration}. For strongly concave log densities, the contraction rate $P^2/(P+\alpha)^2<1$ means inaccurate predictions yield large MSE reductions; i.e., the improvement is roughly proportional to the squared prediction error. For non-concave logarithmic densities such as the Student's~$t$ location model, as illustrated in Figure~\ref{fig_NDR_new}(b), updates may only yield constant rather than  proportional MSE improvements (for details, see Appendix~\ref{App MSE nonconcave}).

Returning to the case of log-concave densities as in Corollary~\ref{Cor_GlobOpt}, the penalty $P_t$ affects both the contraction speed and the size of the NDR, thereby creating a trade-off. A larger penalty shrinks the NDR, but worsens the contraction rate. With a static penalty matrix $P_t=P \succ O_{K}$ for all $t$, we can have either a small NDR (large penalty) or rapid contraction (small penalty), but not both. In practice, 
the penalty matrix $P$ that optimizes this trade-off over longer time periods can be estimated via maximum likelihood (see next section).

As we show in Appendix~\ref{App_NDRESD}, ESD updates are also contractive toward the NDR, but only under the additional conditions that (a) the score is Lipschitz in $\theta$ and (b) the learning rate is capped. These conditions are necessary to prevent the expansive force from dominating the contractive force if  $\|\theta_{t|t-1}-\theta_t^\star\|$ is large.

Finally, as in \citet[Prop.\ 2]{lange2024bellman}, the geometric contraction of Corollary~\ref{Cor_GlobOpt} can be used to bound the long-run MSE of $\{\theta_{t|t}\}$ relative to $\{\theta_t^\star\}$, but this requires additional assumptions on the dynamics of the pseudo-true parameter. We leave this to future research; see \citet{donkervanheel2025gradientbased} for related developments.

\section{Estimation of static (hyper-)parameters}
\label{Sec_Est}

The ISD filter's static (hyper-)parameters are unknown and must be estimated. For simplicity, here we consider $P_t= P\succ O_{K}$ for all $t$.\footnote{Alternatively, $H_t=P_t^{-1}$ could be based on (powers of) the time-varying conditional information matrix as is standard in the ESD literature (e.g., \citealp{creal2013generalized}).} We must estimate the penalty matrix $P$ in the update step, $\omega$ and $\Phi$ in the prediction step \eqref{GASPred}, and the static parameter $\psi\in\Psi\subseteq\mathbb{R}^M$ in the observation density. All static parameters can be jointly estimated by maximum likelihood (ML) using the standard prediction-error decomposition (e.g., \citealp{creal2013generalized}):
\begin{equation}
    \hat{\xi} :=\; \underset{\xi \, \in \, \Xi}{\mathrm{argmax}} \; \sum_{t=1}^{T} \log p(y_t|\theta_{t|t-1}, \psi),
    \label{Eq_MLE}
\end{equation}
where $\xi := [\mathrm{vech}(P)^\prime,\ \omega^\prime,\ \mathrm{vec}(\Phi)^\prime,\ \psi^\prime]^\prime$ stacks all static parameters, and $\mathrm{vec}(\cdot)$ and $\mathrm{vech}(\cdot)$ denote the (half-)vectorization operations. The optimization domain is $\Xi\subset\mathbb{R}^{\frac{3}{2}K(K+1)+M}$ with $P\succ O_K$ and $\psi\in\Psi$. The initialization $\theta_{0\mid 0}$ is treated as fixed but could be included in $\xi$. To ensure ISD filter stability by Theorem~\ref{Thm_Invertability}, it is convenient to reparameterize $P$ in terms of $\Phi$ and $\Delta$ using the discrete Lyapunov equation $P-\Phi^\prime P\Phi=\Delta\succ O_K$ (as in Section~\ref{Sec_Stability}). This has the advantage that the standard restrictions $\Delta\succ O_K$ and $\varrho(\Phi)<1$ ensure stability.

Building on the existing theory for ESD filters (e.g., \citealp{blasques2022maximum}), we conjecture that, under standard regularity conditions, the ML estimator $\hat{\xi}$ is consistent for the pseudo-true parameter $\xi^\star$ and asymptotically normal. This requires that  $\xi^\star$ is identified, $\{y_t\}$ is stationary and ergodic with finite moments and near-epoch dependence, while $p(y|\theta,\psi)$ has uniformly bounded derivatives (of a sufficiently low order) such that the law of large numbers and central limit theorem can be applied (\citealp[Thms.~4.6, 4.15]{blasques2022maximum}).

A key ingredient in standard proofs of consistency and asymptotic normality is invertibility in the sense of \citet{bougerol1993kalman} and \citet{straumann2006quasi}. For ESD filters, verifying the required contraction conditions is challenging because the maximal admissible learning rate depends on the (typically unknown) DGP; see the discussion following Theorem~\ref{Thm_Invertability}. For ISD filters, by contrast, Theorem~\ref{Thm_Invertability} provides a strong invertibility result with no restriction on the learning rate and no required knowledge of the DGP. We therefore expect the asymptotic results in the ESD literature to carry over to our setting; in fact, they may be easier to verify given the ISD filter's DGP-agnostic stability properties.

A potential caveat is that Theorem~\ref{Thm_Invertability} requires the postulated logarithmic density to be concave (Assumption~\ref{A5_Concavity}). While sufficient, this condition is likely stronger than necessary. Below, we show that the static (hyper-)parameters can also be accurately recovered when Assumption~\ref{A5_Concavity} fails (Section~\ref{sec_sim}), while filtering accuracy remains high (Section~\ref{Sec_GED}). Although a full asymptotic theory is beyond the scope of the present paper, our simulations below support the conjectured consistency and asymptotic normality of the ML estimates.

\subsection{Simulation study: Static (hyper-)parameter estimation}
\label{sec_sim}

Here we show that the ML estimator~\eqref{Eq_MLE} can accurately  recover the static parameters and present visual support for the conjecture of asymptotic normality around the true values.

We use nine time-varying observation densities from \citet{koopman2016predicting}; for full specifications, see Appendix~\ref{app: DGPs}. We take $y_t\sim p^0(y_t|\theta_t^0)$ with scalar $\theta_t^0\in\mathbb{R}$; all distributions include link functions (e.g., mapping $\theta_t^0$ to $\mathbb{R}_{>0}$ for volatility). As in \cite{koopman2016predicting}, the postulated density $p(\cdot|\theta_t)$ is correctly specified, meaning it matches the functional form of the true density (i.e., including the link function). However, some densities include additional shape parameters (see Appendix~\ref{app: DGPs2}) that are treated as unknown and must be estimated: the degrees of freedom $\nu^0 = 6$ for both Student's $t$ distributions and the shape parameter in the negative binomial, gamma, and Weibull distributions (set to 4, 1.5, and 1.2, respectively). \citet{koopman2016predicting} set $\theta_t^0=\theta_{t|t-1}^{\mathrm{ex}}$, making the ESD filter the DGP; here, we set $\theta_t^0=\theta_{t| t-1}$, making the ISD filter the DGP. In all nine DGPs, we set $\omega^0=0$, $\Phi^0=0.97$, and $H^0=0.10$. 

Assumption~\ref{A5_Concavity} holds for the first seven densities, which are log-concave in the time-varying parameter. The remaining two are not, with Hessians bounded above by \(1/4\) rather than \(0\). To ensure that Assumption~\ref{A2_RL_StrictConcave} holds, we impose a lower bound of \(1/4\) on the penalty parameter, or equivalently an upper bound of \(4\) on the learning rate, although this restriction was never binding in practice. We solve \eqref{ProParUpdate} using Newton steps with a simple line search. 
 In fact, for five of the nine cases, the ISD update has a closed-form expression in terms of the Lambert \(W\) function; see Appendix~\ref{app: DGPs3}. In practice, however, evaluating the Lambert \(W\) function is no faster than our standard Newton-based routine.

\begin{table}[t!]
\centering
\caption{RMSEs of parameter estimates across $1{,}000$ replications.} \label{tab:RMSEs}
\footnotesize
\begin{threeparttable}
\begin{tabular}{@{}llccccccl@{}}

\toprule
\textbf{DGP type}       & \textbf{Distribution} &  \textbf{Log} & \textbf{Sample size} & \multicolumn{4}{c}{\textbf{Static-parameter RMSEs}}
& 
\\
 && \textbf{concavity} & $T (\times 10^3)$ & $\hat{\omega}$ & $\hat{\Phi}$ & $\hat{H}$ & {$\hat{\psi}$}\\ \midrule
Count	& Poisson	& \checkmark & 1 &	.0053&	.0266&	.0249&	
		\\
     &  && 4	&.0020&	.0074&	.0105\\
      & && 16	&.0010&	.0034	&.0051\\
       \midrule
Count&	Negative bin. & \checkmark	&1	&.0053&	.0253	&.0284	&1.725\\
&&&		4	&.0020&	.0081&	.0130&	.533\\
&&&		16	&.0010	&.0039	&.0064	&.238\\
\midrule
Intensity	& Exponential& \checkmark & 1&	.0073&	.0228&	.0273&
\\
&&&		4&	.0025&	.0070	&.0124\\	
	&&&	16&	.0012	&.0033	&.0061\\	
    \midrule
Duration	&Gamma & \checkmark 	&1	&.0072	&.0168	&.0220	&.029
\\
&&&		4	&.0028&	.0062&	.0100	&.016\\
&&&		16	&.0014	&.0030	&.0053	&.007\\
\midrule
Duration 	& Weibull	& \checkmark &1&	.0064&	.0161&	.0193	&.064\\
&&&		4	&.0028	&.0063	&.0103&	.030\\
	&&&	16	&.0013	&.0029	&.0046&	.015\\
    \midrule
Volatility &	Gaussian & \checkmark	&1&	.0114	&.0666	&.0704	\\
&&&		4	&.0024	& .0109	&.0211\\	
&&&		16	& .0010	&.0046	&.0101\\
\midrule
Volatility & Student's $t$ & \checkmark &	1	& .0132	 &.0801&	.0840	&10.233\\
&&&	4	&.0023&	.0149	&.0250&	.605\\
&&&		16	& .0010 &	.0059 &	.0117&	.275
\\
\midrule
Dependence &	Gaussian & \xmark & 	1 &	.0143 &	.1234 &	.1412\\	
&&&		4	&.0023&	.0223	&.0367\\	
&&&		16	&.0008 &	.0074	 &.0156\\	
\midrule
Dependence	 &Student's $t$	& \xmark & 1 &	.0191 &	.1497	&.1740 &	.630\\
&&	&	4 &	.0031 &	.0370 &	.0463 &	.318\\
	&&&	16 &	.0009 &	.0096 &	.0181	& .146\\
\bottomrule
\end{tabular}
\end{threeparttable}
\end{table}

Table~\ref{tab:RMSEs} reports RMSEs of static (hyper-)parameter estimates for all DGPs and sample sizes $T\in\{10^3,4{\cdot}10^3,16{\cdot}10^3\}$ with $1{,}000$ replications. Static parameters are generally well-recovered; RMSEs roughly halve when the sample size $T$ quadruples from $4{,}000$ to $16{,}000$. Recovery remains adequate for the final two densities, but the asymptotics appear to set in later: RMSEs still fall by more than half even for the largest sample size. This may be due to the non-concavity and/or general difficulty of estimating dynamic-dependence models.

\begin{figure}[tb]
\centering
\includegraphics[scale=0.3]{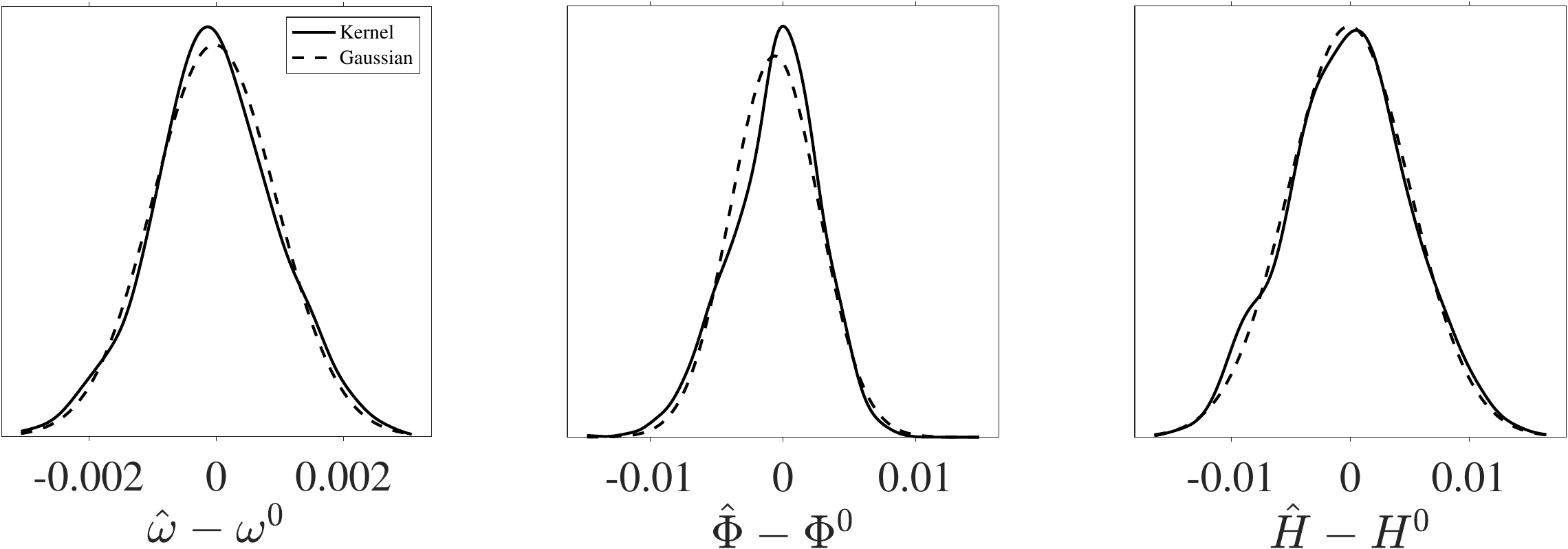}
\caption{Kernel estimate (solid) of estimation errors of the ISD filter's static parameters for the dynamic Poisson distribution with $T=16{,}000$.  \label{Fig:Conjecture}}
\end{figure}

Figure~\ref{Fig:Conjecture} shows the distribution of static-parameter estimation errors to be approximately Gaussian for the dynamic Poisson model based on the largest sample size. The same is true for the other DGPs (see Appendix~\ref{app: DGPs4}), providing visual support for our conjecture of asymptotic normality around the true values.

\section{Simulation studies: Filtering performance}\label{Sec_Sim}

We perform simulations comparing the proposed ISD filter with its ESD counterpart. We consider misspecified settings in which the filters are score driven (ISD or ESD), while the DGP is a state-space model. The various settings below are chosen to highlight specific advantages of ISD over ESD filters. Unless stated otherwise, we use $1{,}000$ replications of length $T$; the first $R$ observations are used to estimate the static parameters. Performance is evaluated through MSEs of the predicted parameters $\{\theta_{t\mid t-1}\}$ and $\{\theta_{t\mid t-1}^{\textnormal{ex}}\}$ relative to the true parameters $\{\theta_t^0\}$, computed in-sample for $t=1,\ldots,R$ and out-of-sample for $t=R+1,\ldots,T$.

\subsection{Dynamic Poisson distribution: Non-Lipschitz gradient}

For each time $t$, $y_t\in \mathbb{N}$ is drawn from a Poisson distribution with a time-varying intensity $\lambda^0_t:=\exp(\theta^0_t)$, i.e.,\ $p(y_t|\theta_t^0)=(\lambda^0_t)^{y_t}\exp(-\lambda^0_t)/y_t!$. The score with respect to the log-intensity parameter $\theta\in \mathbb{R}$ is $y_t-\exp(\theta)$. The negative Hessian and Fisher information are both $\exp(\theta)>0$; hence, the density is log-concave in $\theta$. We specify the state dynamics as $\theta_t^0=0.98\theta_{t-1}^0+\eta_t$, where $\eta_t \distas{\text{i.i.d.}}\mathrm{N}(0,\sigma_\eta^2)$, and we vary the value of $\sigma_\eta$. 

We consider ISD and ESD filters based on the (correctly specified) Poisson distribution with exponential link, with $R=2{,}000$ and $T=10{,}000$. We follow \citet{koopman2016predicting}, who used a time-varying learning rate (for the ESD filter) that scales with the inverse square root of the (model-based) predicted Fisher information quantity; similarly, we set $H^{j}_t=H^{j} \exp(-\theta^{j}_{t|t-1}/2)$ with a static learning-rate coefficient $H^{j}>0$ for $j\in \{\text{im},\text{ex}\}$. For the ESD filter, the scaled score $\exp(-\theta/2)(y_t-\exp(\theta))$ is non-Lipschitz in $\theta\in \mathbb{R}$.

\begin{table}[h!]
  \caption{MSEs of predicted states with dynamic Poisson distribution.  \label{tab:MSEPoisson}}
  \centering
\begin{threeparttable}
  \footnotesize
    \begin{tabular}{llrrrrr}
          &       & $\sigma_\eta=0.10$   & 0.15  & 0.20   &   0.25 & 0.30
          \\
          \hline
 \multicolumn{1}{l}{ISD filter} & In-sample
          
&	0.088	& 0.149 &	0.221&	0.304	& 0.398
\\
 & Out-of-sample &	0.091	& 0.154	& 0.227	& 0.311 &	0.407\\
 \hline
 \multicolumn{1}{l}{ESD filter} & In-sample &	0.088 &	0.149 &	0.220 &	0.311&	0.439
 \\
& Out-of-sample & 	0.091&	0.153	& 0.227 & $\infty$ & $\infty$ 
    \end{tabular}%
\end{threeparttable}
\end{table}%

Table~\ref{tab:MSEPoisson} shows the in-sample MSEs to be comparable. For both filters, MSEs increase with the state variability $\sigma_{\eta}$. The out-of-sample performance is similar only for $\sigma_{\eta}\lesssim 0.20$.  For $\sigma_\eta=0.25$ and $0.30$, the ISD filter's out-of-sample performance remains aligned with the in-sample results, but the ESD filter diverged in 2\% and 26\% of replications, respectively. Larger state innovations induce larger prediction errors, such that the (non-Lipschitz) ESD update may lead to divergence. In their simulation study, \citet{koopman2016predicting} used $\sigma_\eta=0.15$, such that the potential instability of the ESD filter went unnoticed.

\subsection{Dynamic GED: Non-concave and non-Lipschitz gradient}
\label{Sec_GED}

At each time $t$, $y_t\in\mathbb{R}$ follows a generalized error distribution (GED) with time-varying mean $\theta_t^0$. Hence $p^0(y_t|\theta^0_t)=\upsilon \exp(-|(y_t-\theta^0_t)/\sigma|^\upsilon)/(2\sigma\Gamma(\upsilon^{-1}))$, where $\Gamma(\cdot)$ is the Gamma function and $\sigma,\upsilon>0$ are static shape parameters. Here $\sigma$ is a scale parameter, but not the standard deviation. We set $\sigma^2=\Gamma(\upsilon^{-1})\Gamma(3\upsilon^{-1})$, such that, conditional on $\theta_t^0$, the variance of $y_t$ is unity. We vary $\upsilon$, with $\upsilon=1$ corresponding to Laplace and $\upsilon=2$ to Gaussian. For $\upsilon>1$, the log-density is $C^1$ and concave in $\theta\in\mathbb{R}$; for $\upsilon<1$ it is neither. At $\upsilon=1$, it is concave but not $C^1$. The gradient is Lipschitz only for $\upsilon\in(1,2]$. A comparative advantage of the ISD update $\theta_{t|t}$ is that it must lie between $\theta_{t|t-1}$ and $y_t$
at every step. In contrast, the ESD update may overshoot $y_t$, meaning
$
|\theta^{\textnormal{ex}}_{t\mid t}-\theta^{\textnormal{ex}}_{t\mid t-1}|>|y_t-\theta^{\textnormal{ex}}_{t\mid t-1}|
$
may regularly occur.

The true parameter evolves as $\theta_t^0=0.98\,\theta_{t-1}^0+\eta_t$, with $\eta_t\distas{\text{i.i.d.}}\mathcal{N}(0,1)$. The signal-to-noise ratio is roughly one, since the state-innovation and observation-noise variances are both unity. We simulate series with $T=10{,}000$ and use the first $R=1{,}000$ observations to estimate the autoregressive parameter $\Phi\in(-1,1)$ in \eqref{GASPred} and the learning rate $H>0$. For simplicity, we set $\omega=\omega^0=0$ and $\upsilon=\upsilon^0$ (i.e., using the true values).

\begin{table}[h!]
    \caption{ MSEs of predicted states with dynamic GED.}
  \label{tab:MSEGED}%
  \centering  
  \footnotesize
    \begin{tabular}{rlrrrrrrrr}
          &    & ${\upsilon}=0.5$   & \multicolumn{1}{c}{1}     & \multicolumn{1}{c}{1.5}    & \multicolumn{1}{c}{2}      & \multicolumn{1}{c}{2.5}    & \multicolumn{1}{c}{3}      & \multicolumn{1}{c}{3.5}   & \multicolumn{1}{c}{4}  \\
          \hline
\multicolumn{1}{l}{ISD filter}
& In-sample &	1.55 &	1.64 &	1.58&	1.57	& 1.57 &	1.58 &	1.58	& 1.59
\\
	& Out-of-sample &	1.64	& 1.67	&1.59	&1.58	& 1.58	& 1.59&	1.59&	1.60
    \\
\hline
 \multicolumn{1}{l}{ESD filter} & In-sample &	4.60	& 1.90 &	1.64 &	1.57 &	1.61 &	1.72	& 1.86 &	2.08 \\
 &	Out-of-sample &	22.68 &	2.11	& 1.65 &	1.58 &	1.62  & $\infty$ & $\infty$ & $\infty$
 \\
    \end{tabular}%
\end{table}%

Table~\ref{tab:MSEGED} reports in- and out-of-sample MSEs. For the Gaussian case (i.e., $\upsilon=2$), the performance is identical. Generally, the ISD filter attains lower MSEs (in the range ${\sim}1.55$–${\sim}1.67$ for all $\upsilon$), with out-of-sample MSEs closely tracking in-sample MSEs. For the ESD filter, in-sample MSEs are always higher, and the ESD filter diverged in $2\%$, $58\%$, and $80\%$ of  out-of-sample replications for $\upsilon=3$, $3.5$, and $4$, respectively, yielding infinite MSEs. This occurs because the gradient is polynomial of degree $\upsilon-1$ in the prediction error---i.e., excessively large for inaccurate predictions---so poor predictions induce poor updates and, possibly, divergence. For $\upsilon<1$, the MSE of the ESD filter remains finite (as the gradient is unbounded only for near-perfect predictions), but
MSEs can still be very large (e.g., ${\sim}23$ for $\upsilon=0.5$).

\subsection{Dynamic Gamma distribution: Two time-varying parameters}
\label{sec: dynamic gamma}

For each time $t$, an observation $y_t\in \mathbb{R}_{>0}$ is drawn from a Gamma distribution with two dynamic parameters $a_t\in \mathbb{R}_{>0}$ and $b_t\in \mathbb{R}_{>0}$, which are collected in the (true) state vector $\theta^0_t=(a_t,b_t)'\in \mathbb{R}_{>0}^2$, such that $p(y_t|\theta^0_t)=(b_t)^{a_t} y_t^{a_t-1}\exp(-b_t y_t)/\Gamma(a_t)$. The same parametrization is used in \citet[eq.\ 3]{fearnhead2004exact}, albeit in a static context. Conditional on $\theta_t^0$, the mean and variance of $y_t$ are $a_t/b_t$ and $a_t/b_t^2$, respectively. The ISD filter can be applied directly to $\theta_t=(a_t,b_t)'$, because optimization~\eqref{ProParUpdate} guarantees that both elements remain positive: the optimization domain $\Theta$ is the positive quadrant in $K=2$ dimensions. For each $y_t$, the Gamma log density is (jointly) concave in $(a_t,b_t)\in \mathbb{R}_{>0}^2$; hence, Theorem~\ref{Thm_Invertability} guarantees ISD filter stability. For the ESD filter, positivity of the time-varying parameters must be enforced through exponential link functions. The resulting log density is neither concave nor $L$-smooth in the transformed (i.e., logarithmic) parameters, meaning no theoretical guarantees can be made for the ESD filter.

\begin{figure}[h!]
\center
\begin{tabular}{cc}
    \includegraphics[scale=0.22]{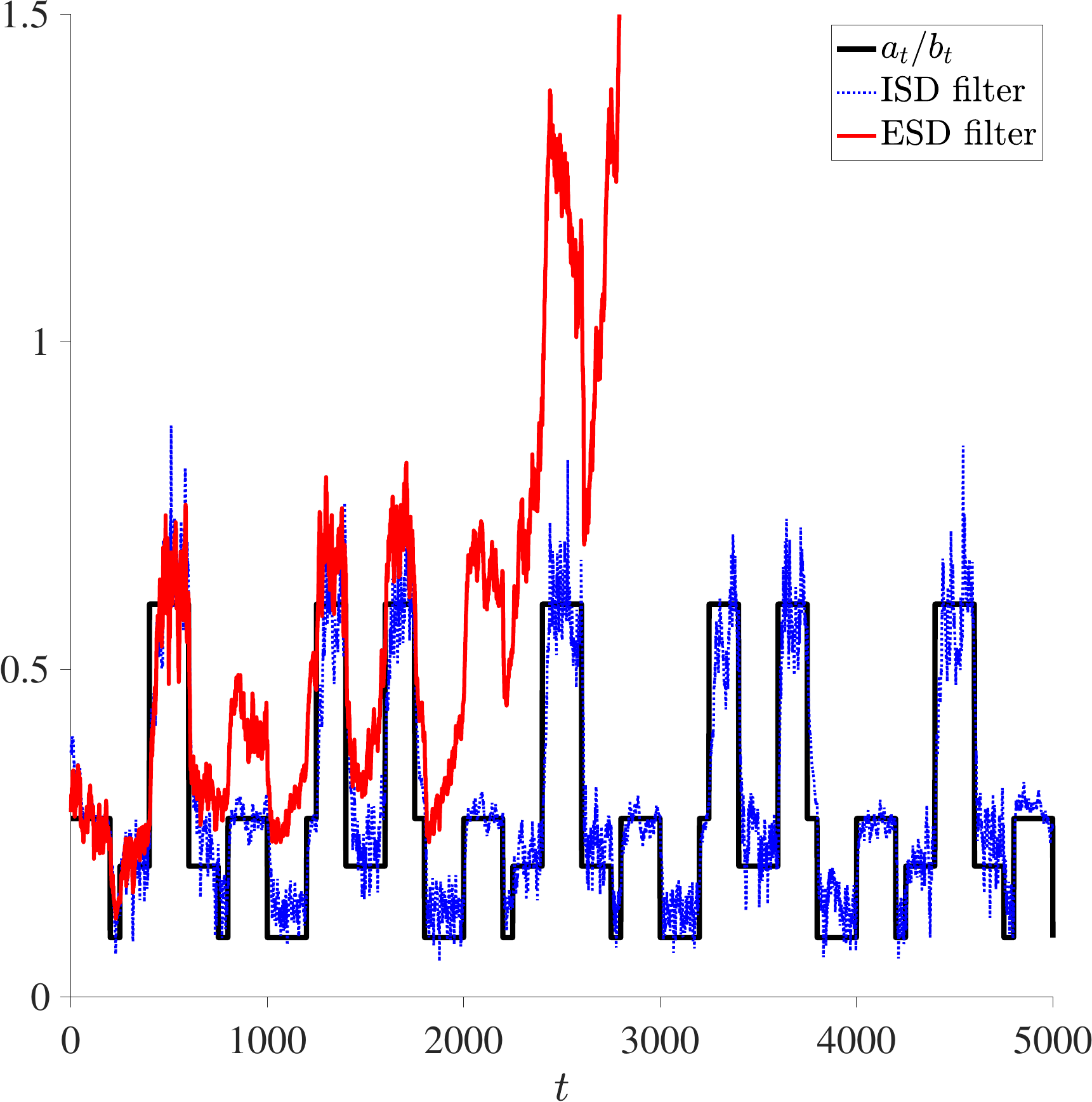}  &    \includegraphics[scale=0.22]{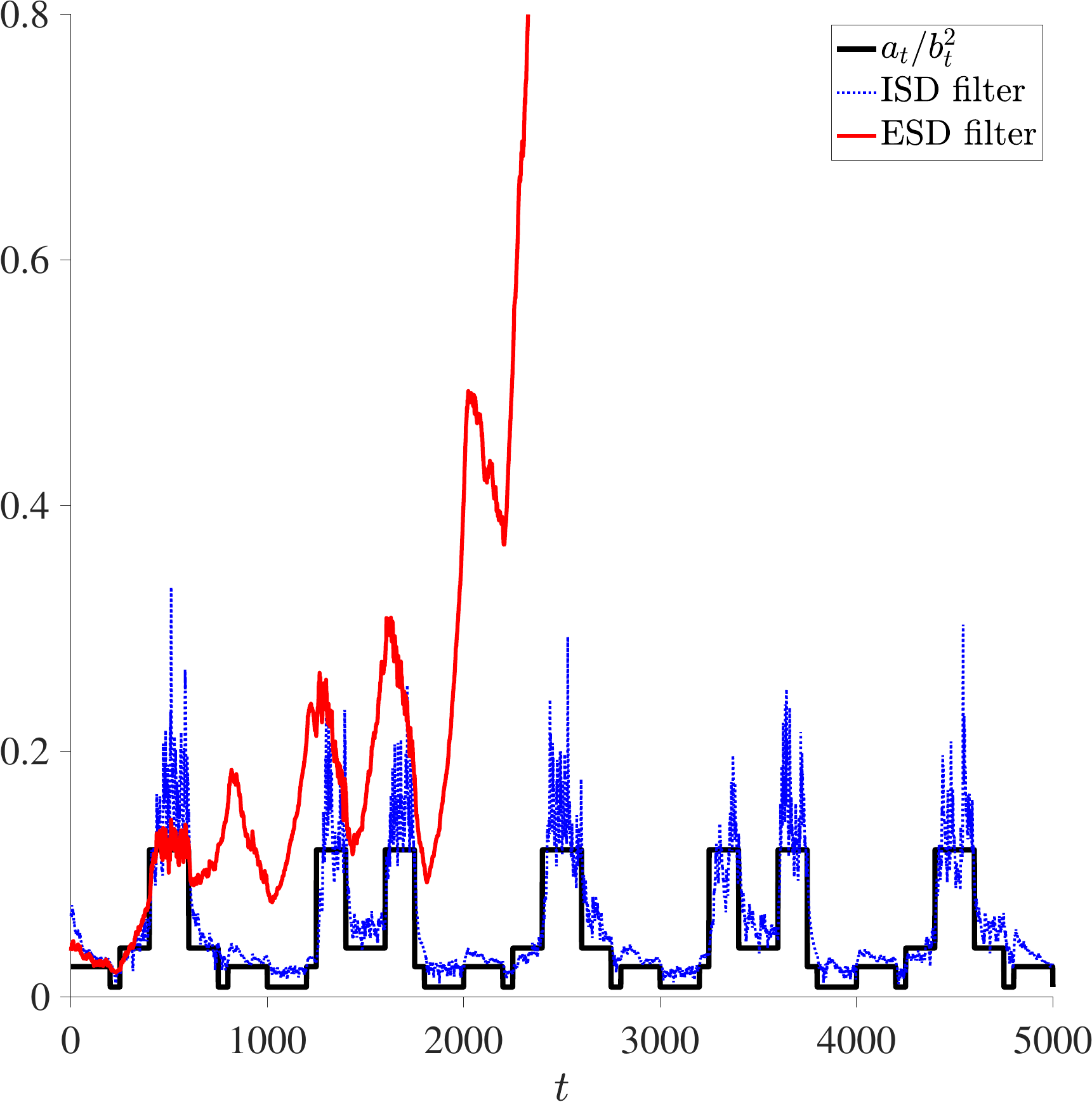}
    \\
   \footnotesize (a) True and predicted mean of $y_t$ & \footnotesize (b) True and predicted variance of $y_t$
\end{tabular}
\caption{Illustration of filtering performance with dynamic Gamma distribution.}
  \label{Fig_Gamma}
\end{figure}

Figure~\ref{Fig_Gamma} shows filtering results for a single time series \(\{y_t\}\) of length \(T=5{,}000\) generated by the non-smooth DGP \(a_t=2+\text{sign}\{\sin(2\pi t/400)\}\) and \(b_t=8+3\,\text{sign}\{\cos(2\pi t/1000)\}\). Only the first \(R=500\) observations are used to estimate the static parameters \(\omega\), \(\Phi\) (diagonal), and \(P\) (positive definite). The figure plots the theoretical mean and variance, \(a_t/b_t\) and \(a_t/b_t^2\), together with their ISD and ESD predictions. The two filters perform similarly in sample, but out of sample ESD diverges, whereas ISD continues to track both moments relatively well.

\subsection{Dynamic Dirichlet distribution: Fat-tailed state innovations}

At each time $t$, the observation $y_t\in[0,1]^N$ is drawn from a homogeneous Dirichlet distribution with dynamic concentration parameter $\lambda_t^0=\exp(\theta_t^0)>0$ and density
$p^0(y_t|\lambda^0_t)=\Gamma(\lambda_t^0 N) / \Gamma(\lambda_t^0)^N \prod_{i=1}^N y_{it}^{\lambda^0_t-1}$, where $\Gamma(\cdot)$ denotes the Gamma function. The $N$ elements of $y_t$ are nonnegative and sum to one. The log density is concave in $\theta_t^0$, but the gradient is not Lipschitz, meaning no stability guarantees exist for the ESD filter. The true process is $\theta_t^0=\omega^0+\Phi^0 \,\theta_{t-1}^0+\sigma_\eta\eta_t$, with i.i.d.\ increments $\{\eta_t\}$ of unit variance, distributed normally or as a Student's \emph{t} with $\nu^0=5$. We set $\omega^0=0.1$, $\Phi^0=0.95$, and $\sigma_\eta^2=0.195$; hence, $\mathbb{E}(\theta_t^0)=\mathbb{V}(\theta_t^0)=2$. We vary $N$ from $2$ to $100$. For each $N$, we simulate $100$ series of length $T=10{,}000$, using the first $R=1{,}000$ observations to estimate $\omega$, $\Phi$ and $H$.

\begin{figure}[h!]\center
\includegraphics[width=0.7\textwidth]{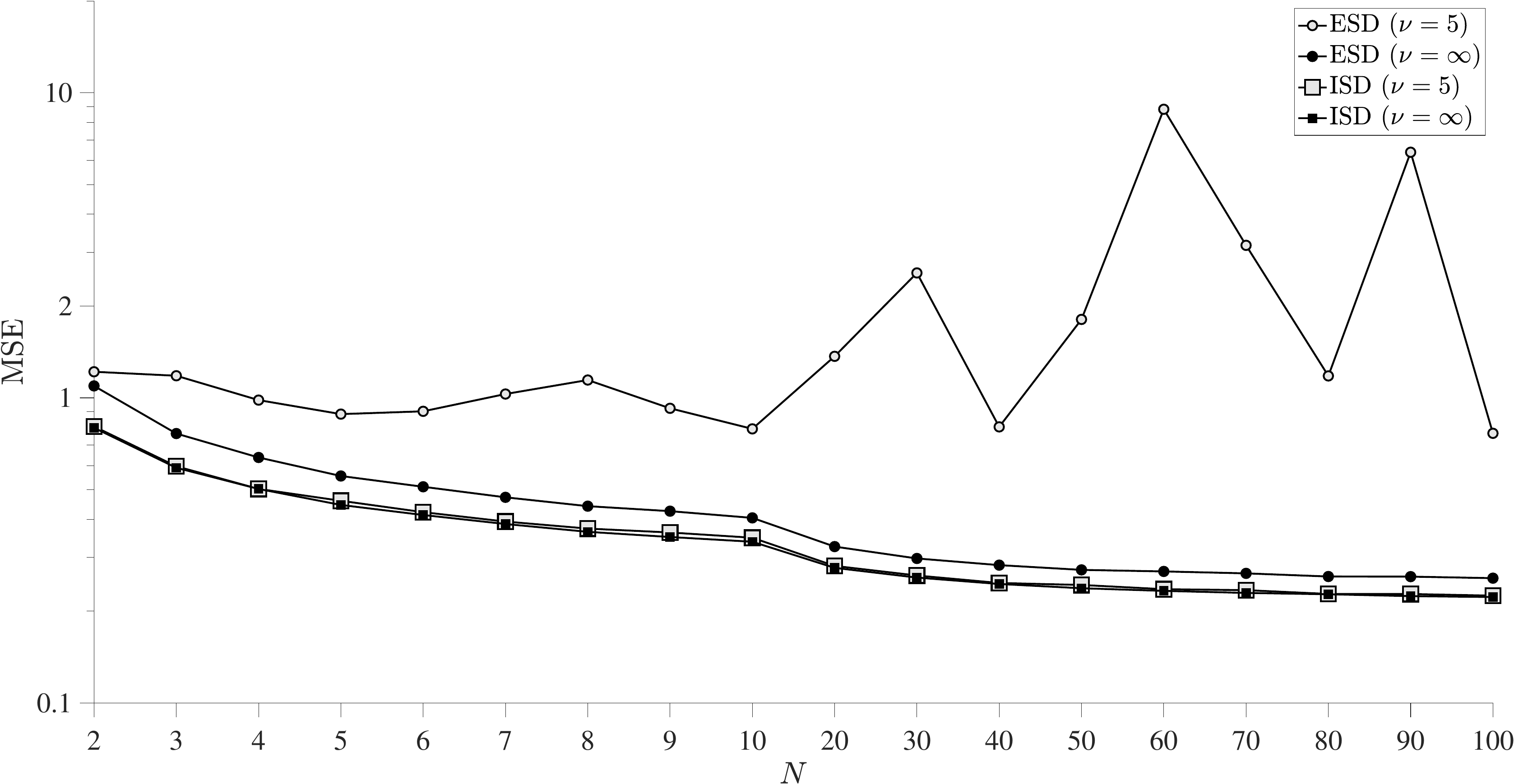}  
\caption{MSE of out-of-sample predicted states for $N$-dimensional observations from dynamic Dirichlet distribution with Gaussian or Student's $t$ state innovations.}
  \label{Fig_Dirichlet}
\end{figure}

Figure~\ref{Fig_Dirichlet} shows the out-of-sample MSEs of predicted states for ISD and ESD filters under Gaussian and fat-tailed state innovations. Across both settings and for all $N$, the ISD filter outperforms its ESD counterpart. Under fat-tailed state innovations, the ISD filtering performance remains accurate, whereas the ESD filter exhibits unreliable behavior. Moreover, ESD filtering performance does not consistently improve as $N$ increases. In particular, the ESD filtering MSEs sometimes exceed the unconditional variance $\mathbb{V}(\theta_t^0)=2$. Abrupt state changes combined with a non-Lipschitz gradient appear to prevent the ESD filter from exploiting the greater information content of higher-dimensional draws.

\subsection{Dynamic network flows: High-dimensional state space}
\label{sec: Network}

At each time \(t\), we observe an \(N\times N\) matrix of bilateral trade or information flows, \(Y_t=(Y_{ij,t})\), on a directed network with \(N\in\{3,5,10,20,30,40,50\}\) nodes; self-flows are excluded. Node-specific sender and receiver effects are denoted by \(\theta_{1,t}^0,\theta_{2,t}^0\in\mathbb{R}^N\). The full state vector is \(\theta_t^0=\big((\theta_{1,t}^0)',(\theta_{2,t}^0)'\big)'\in\mathbb{R}^{2N}\), so the state dimension is \(K=2N\). Conditional on \(\theta_t^0\), the flows are independent Gamma random variables with common shape parameter \(\psi_0=5\) and mean matrix \(M_t^0=\exp\!\big(\bar{m}+\theta_{1,t}^0\iota' + \iota(\theta_{2,t}^0)'\big)\), where \(\bar m\in \mathbb{R}^{N\times N}\) is a baseline matrix of logarithmic means, \(\iota\) is a vector of ones, and the exponential is applied elementwise. This specification of \(M_t^0\) is closely related to standard gravity models of bilateral trade with exporter/importer effects; e.g., \citet{anderson2003gravitas}, \citet{head2014gravity}, and \citet{fally2015structural}.

The latent state evolves as \(\theta_t^0=\Phi_0\theta_{t-1}^0+\eta_t\), with \(\eta_t\distas{\text{i.i.d.}}\mathcal N(0,\Sigma_{\eta})\), where \(\Phi_0\) and \(\Sigma_{\eta}\) are diagonal. For each replication, the diagonal entries of \(\Phi_0\) are drawn independently from a uniform random variate \(\mathcal U[0.80,0.995]\), the standard deviations are drawn from \(\mathcal U[0.2,0.5]\), and the initial state is drawn from its unconditional distribution. The off-diagonal entries of \(\bar m\) are drawn from \(\bar m_{ij}\distas{\text{i.i.d.}} \mathcal N(0,0.40^2)\).

As our focus here is on filtering performance, we treat \(\bar m\), \(\Phi_0\), and \(\psi_0\) as known, while the particle filter (PF) also uses \(\Sigma_{\eta}\). We compare the PF with \(10^3\) and \(10^4\) particles (denoted PF3 and PF4) to ISD and ESD filters  with estimated scalar learning-rate parameters. The ISD update is implemented by Newton's method with backtracking (see Appendix~\ref{appendix:gamma_flow_details}). Since the Gamma log density is concave in \(\theta_t^0\), our theoretical guarantees apply to the ISD filter. However, since the gradient is not Lipschitz, no guarantees for the ESD filter are available.

For each \(N\), we generate \(100\) replications of length \(T=1{,}000\).  For the ISD and ESD filters, the first \(R=50\) observations are used to tune the scalar learning-rate parameters, and the remaining \(T-R=950\) observations are used for out-of-sample evaluation. Accuracy is measured by computing the MSE of the predicted log-intensity surface, \(\theta_{1,t|t-1}\iota' + \iota\theta_{2,t|t-1}'\), excluding diagonal elements, relative to its true simulated counterpart.

\begin{figure}[t!]
\center
\begin{tabular}{cc}
    \includegraphics[scale=0.45]{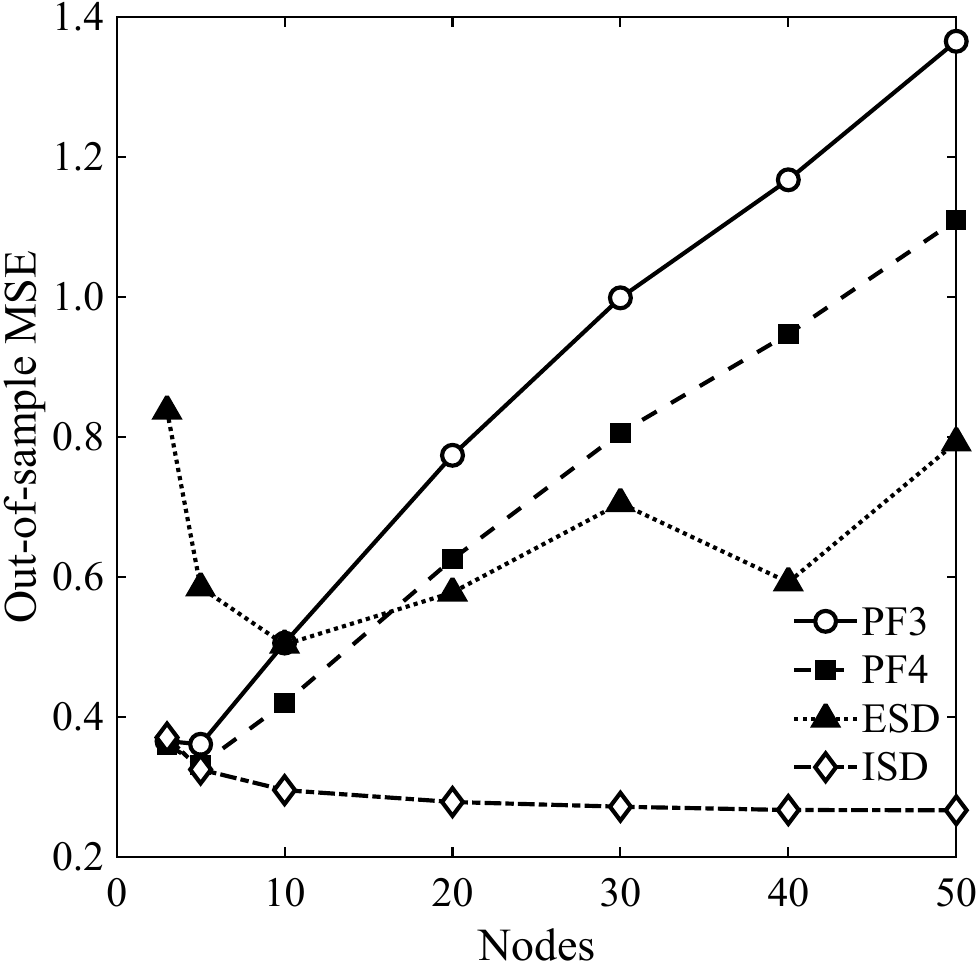}  &    \includegraphics[scale=0.45]{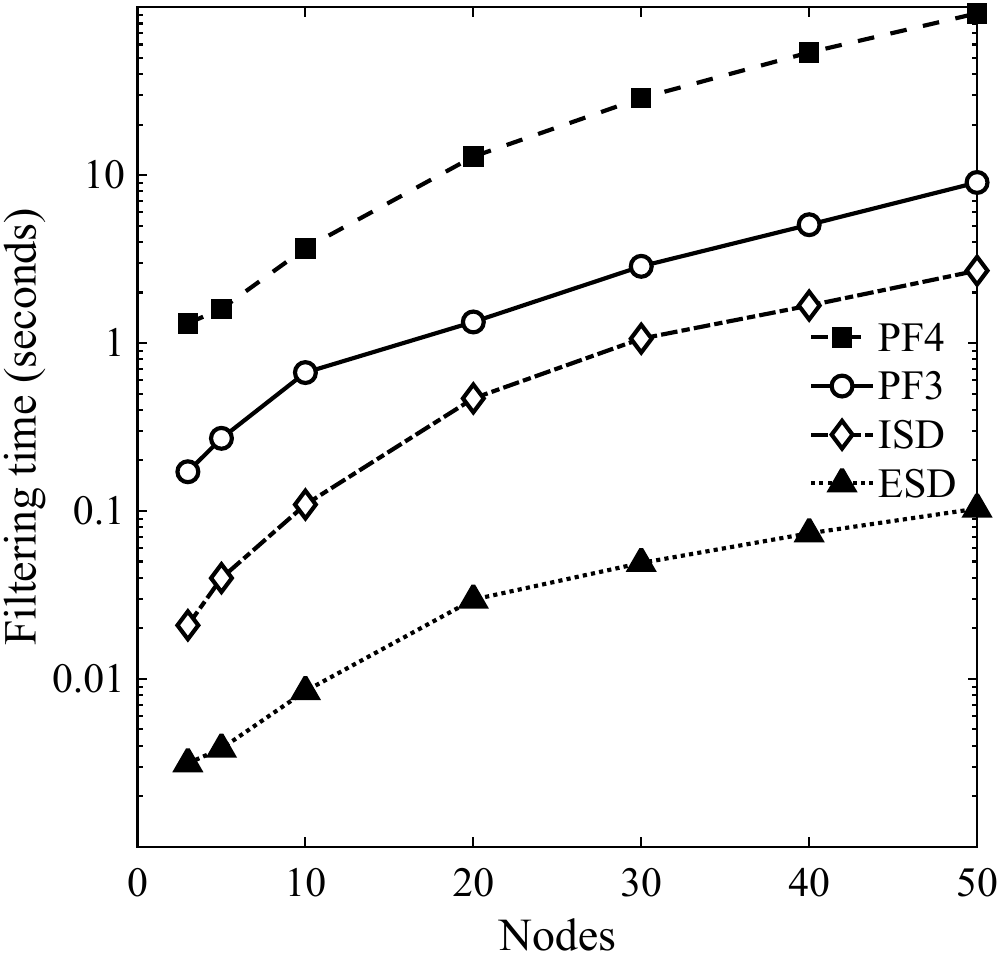}
    \\
   \footnotesize (a) MSE & \footnotesize (b) Time (seconds)
\end{tabular}
\caption{Out-of-sample prediction MSE and filtering time for the dynamic network model with $N$ nodes, \(N\times N\) observations and \(K=2N\) states.}
  \label{Fig_Network}
\end{figure}

Figure~\ref{Fig_Network}(a) shows that the ISD filter is competitive for all \(N\) while clearly outperforming the alternatives for \(N\geq 10\). For \(N=50\), its MSE is about three times lower than that of the ESD filter and about four to five times lower than that of PF3 and PF4. It is also the only method whose performance improves with \(N\). For \(N=10,20,30,40,\) and \(50\), the ESD filter exhibits instability: among the 100 replications, between 1 and 8 runs diverge, while no divergence occurs for \(N=3\) or \(5\). The reported MSE for the ESD filter is based only on the non-divergent runs. Figure~\ref{Fig_Network}(b) shows that the ESD filter is fastest, as expected, followed by the ISD filter. For \(N=50\), despite solving a 100-dimensional optimization problem over \(1{,}000\) time steps, the ISD filter takes about \(2.5\) seconds per replication, making it three times faster than PF3 and thirty times faster than PF4.

To interpret these results, we distinguish statistical accuracy from computational cost. In terms of accuracy, the particle filter suffers from the curse of dimensionality: although the conditional mean is optimal under MSE loss, approximating it becomes prohibitively difficult as \(N\) grows (e.g., \citealp{bengtsson2008curse}). By contrast, the ISD update exploits the richer cross-sectional information contained in each observation, so its performance improves with \(N\). In terms of computation, the ISD filter solves at each time step a smooth, strongly concave optimization problem, initialized at the one-step-ahead prediction, which is typically close to the optimum. As a result, only a few Newton steps are usually needed, even in high dimensions. Hence, in this design, the ISD filter is both more accurate and faster than the PF. These findings are consistent with the arguments and simulation evidence in \citet[Sec.~1.2 and Sec.~9.3]{lange2024bellman}, which suggest that, in high-dimensional settings, optimization-based methods such as the ISD filter can outperform integration-based methods such as the PF, even when the object of interest is the conditional mean.

\section{Empirical illustrations}
\label{Sec_Emp}

We consider three applications for which the ISD update~\eqref{ProParUpdate} allows a closed-form solution.

\subsection{Linear regression with time-varying slope}
\label{Sec_Emp:linreg}

The capital asset pricing model (CAPM), a standard benchmark in finance, links expected excess returns of individual assets linearly to those of the market. However, empirical evidence (e.g., \citealp{jagannathan1996conditional}) shows that assuming a constant market coefficient $\beta$ can be unrealistic, especially in equity markets. We study possible time variation in the CAPM market $\beta$ using the ISD filter. We model the excess asset return $y_t$ as
\begin{equation}
    \label{Eq_CAPM}
    y_t \;= \;\alpha \,+\, \beta_{t}\,x_t \,+\,  \varepsilon_t, \qquad \varepsilon_t \distas{\text{i.i.d.}}\textnormal{N}(0,\sigma^2),
\end{equation}
where $\alpha$ is a static intercept, $x_t$ is the excess market return, and $\varepsilon_t$ is an i.i.d.\ Gaussian disturbance with mean zero and variance $\sigma^2>0$. The ISD update \eqref{ProParUpdate} applied to a prediction $\beta_{t|t-1}$ for $\beta_t$ in \eqref{Eq_CAPM} can be solved analytically (see Appendix~\ref{AppC} for details), yielding
\begin{equation}\label{Eq_Beta} 
\beta_{t|t} \;= \;\beta_{t|t-1} \,+\, \frac{\sigma^2}{\sigma^2 + H x_t^2}\, H\, \nabla(y_t | \beta_{t|t-1}, x_t),   
\end{equation} 
where $H = P^{-1}>0$ is a scalar learning-rate parameter and $\nabla(y_t | \beta_{t|t-1}, x_t):=x_t( y_t - \beta_{t|t-1}\,x_t)/\sigma^2 $ is the explicit score (i.e., evaluated at $\beta_{t|t-1}$). For the prediction step, we use the linear first-order specification~\eqref{GASPred}. Equation~\eqref{Eq_Beta} illustrates Proposition~\ref{Lm_Shrinkage} for linear regression: its right-hand side includes the shrinkage factor $\sigma^2/(\sigma^2 + H x_t^2)\in(0,1]$, which is absent (i.e., equal to one) in the ESD update. In the ISD update, larger values of $x_t^2$ induce greater shrinkage. Indeed, the specific functional forms of the explicit score $\nabla(y_t | \beta_{t|t-1}, x_t)$ and the shrinkage factor as a function of $x_t$ imply that $\beta_{t|t}\to 0$ as $|x_t|\to \infty$.

In fact, the ISD update \eqref{Eq_Beta} is bounded in both $x_t^2$ and the learning rate $H>0$. The latter follows from $H$ appearing both in the shrinkage factor and in front of the score. This boundedness in $H$ is not limited to the present regression example: ISD filters are robust under a wide range of learning rates, whereas ESD filters require more careful fine-tuning. In stochastic-gradient methods, similar arguments arise when comparing the least-mean-square (LMS) filter to its normalized version (NLMS; see~\citealp{nagumo1967learning} and~\citealp{diniz1997adaptive}).

\begin{figure}[t!]
\center
\begin{tabular}{cc}
    \hspace*{-2.25cm}
    \includegraphics[scale=0.2]{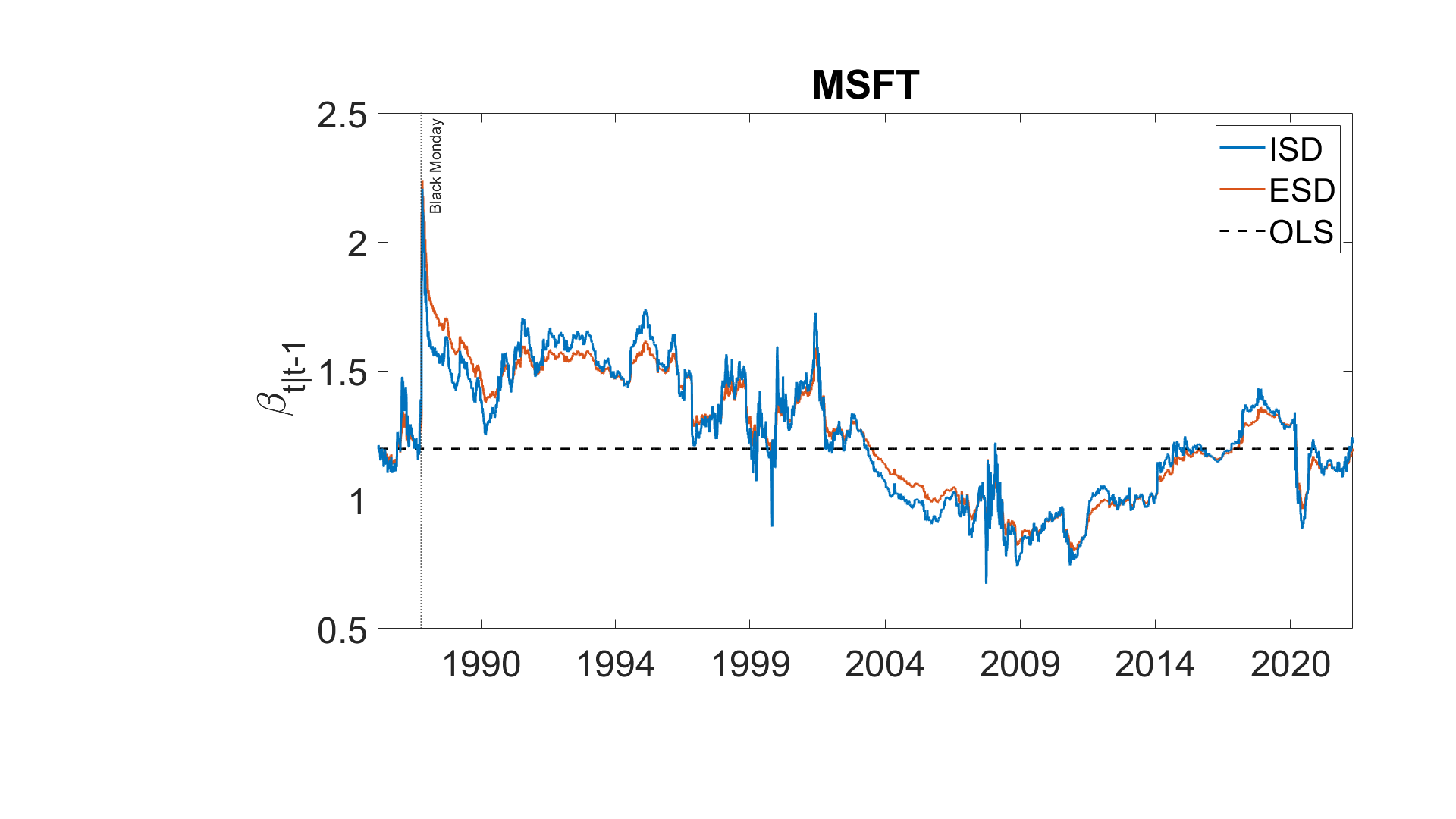}  &  \hspace*{-0.5cm}\includegraphics[scale=0.2]{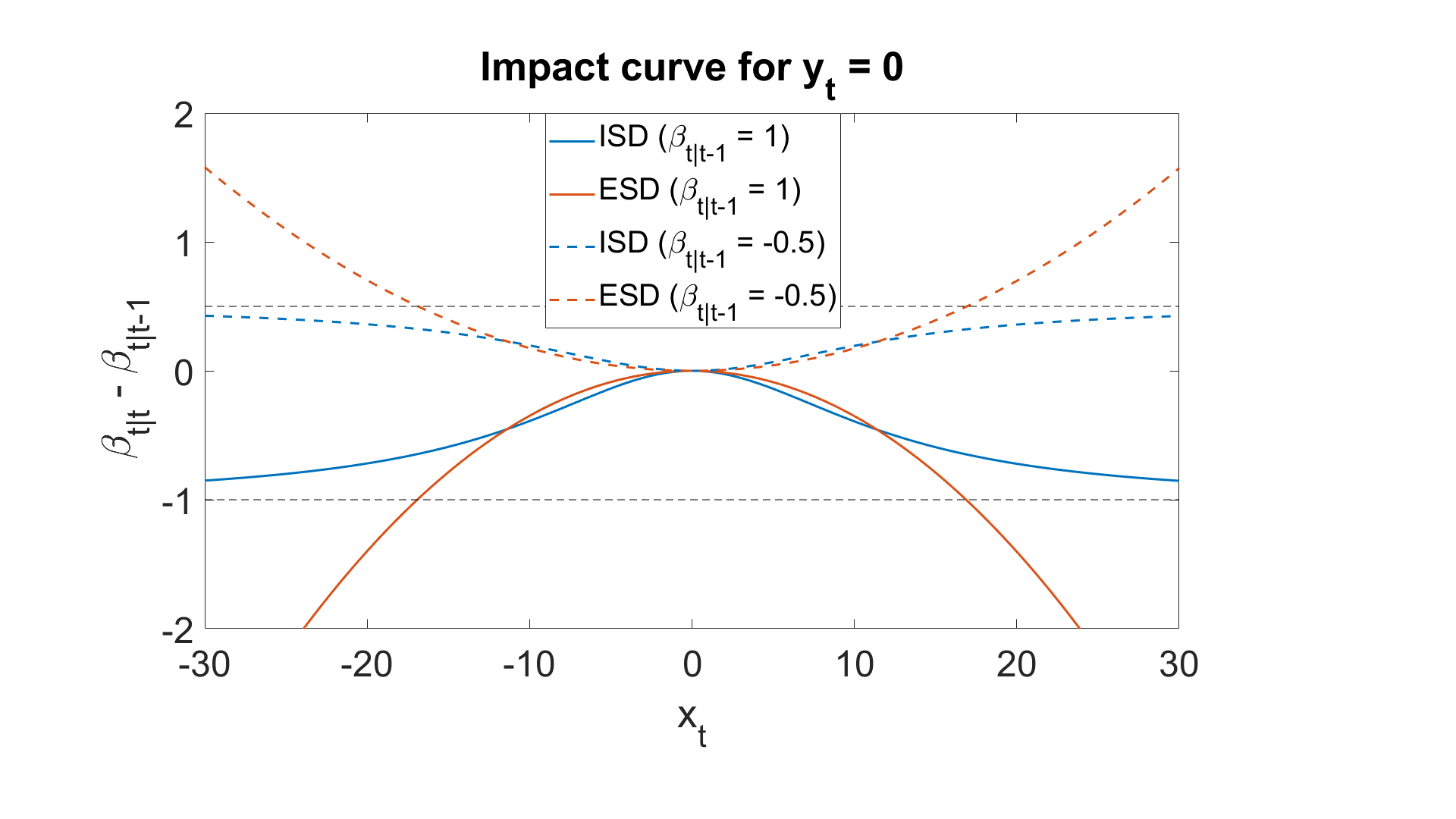}
    \vspace*{-0.5cm}
    \\    
   \footnotesize (a) Time evolution of $\beta_{t|t-1}$ & \footnotesize \hspace*{-1.25cm} (b) Estimated impact curve
\end{tabular}
\caption{Dynamic CAPM using the ISD and ESD filters for MSFT from March 1986 until April 2022. Vertical dotted lines in panel (a) mark Black Monday on October 19, 1987.}
  \label{Fig_Beta} 
\end{figure}

We apply the ISD filter to the dynamic regression model~\eqref{Eq_CAPM} with daily excess returns of Microsoft (MSFT) from 14 March 1986 to 29 April 2022, obtained from Yahoo Finance.\footnote{\url{https://finance.yahoo.com/quote/MSFT/history?p=MSFT}} For the market return and risk-free rate, we use the series from Kenneth French's database.\footnote{\url{https://mba.tuck.dartmouth.edu/pages/faculty/ken.french/data\_library.html}} Figure~\ref{Fig_Beta}(a) shows the evolution of $\beta_{t|t-1}$ for the ISD and ESD filters. In Figure~\ref{Fig_Beta}(b), we plot the slope adjustments $\beta_{t|t}-\beta_{t|t-1}$ as a function of $x_t$ for a fixed $y_t = 0$ and two predictions ($\beta_{t|t-1} = 1$ and $\beta_{t|t-1} = -0.5$); we refer to these as ``impact curves.''

In panel~(a), the ISD and ESD filters display similar paths $\{\beta_{t\mid t-1}\}$, although the ESD filter recovers more slowly from large shocks such as Black Monday (1987). Panel~(b) explains why: the ESD impact curve is quadratic (and hence unbounded) in the exogenous variable $x_t$, which necessitates a small learning rate to curb outliers. Accordingly, its estimated rate is only $\hat{H}=0.0092$, suppressing responsiveness in the rest of the sample (e.g., around 1994 and 2004, when the ESD filter lags behind its ISD counterpart).

By contrast, the ISD filter is not overly sensitive to outliers like Black Monday and is more responsive to persistent movements (e.g., the downward trend around 2004). Because the ISD impact curve redescends to zero as $|x_t|\to\infty$, it admits a larger estimated learning rate, $\hat{H}=0.0169$; roughly twice that for the ESD filter. This allows the ISD filter to pick up gradual changes more quickly, without being thrown off by large values of $x_t^2$.

\subsection{Time-varying growth-at-risk}
\label{sec: GAR}

For policymakers, monitoring macroeconomic downside risk is crucial. A popular approach is growth-at-risk (GaR), which focuses on the conditional lower quantiles of GDP growth. GaR and its link to financial/economic conditions are typically estimated via quantile regressions (QRs; \citealp{koenker2001quantile}) with exogenous predictors (e.g., \citealp{adrian2019vulnerable}).

We update a time-varying conditional quantile by postulating an asymmetric Laplace distribution with a time-varying location. Maximizing this density is equivalent to minimizing \citeauthor{koenker1978regression}'s (\citeyear{koenker1978regression}) QR check function (see \citealp{koenker1999goodness}). The ESD update for the $\tau$-level quantile with $0<\tau<1$, denoted by $q^\textnormal{ex}_{t|t}(\tau)$, is
\begin{equation}
\label{QR_caviar}
    q^\textnormal{ex}_{t|t}(\tau) = q_{t|t-1}(\tau) - 1[y_t < q_{t|t-1}(\tau)] \frac{H(1-\tau)}{\sigma} + 1[y_t > q_{t|t-1}(\tau)] \frac{H\tau}{\sigma},
\end{equation}
where $y_t$ is GDP growth, $1[\cdot]$ is the indicator, and $H>0$ and $\sigma>0$ are constant learning-rate and dispersion parameters. 
The ESD update \eqref{QR_caviar} adjusts the quantile by $-H(1-\tau)/\sigma$ when $y_t<q_{t|t-1}(\tau)$ and by $+H\tau/\sigma$ when $y_t>q_{t|t-1}(\tau)$. If $y_t=q_{t|t-1}(\tau)$, no adjustment is made, such that $q^\textnormal{ex}_{t|t}(\tau)=q_{t|t-1}(\tau)$; in this case, the ESD update allows a subgradient interpretation. Aside from this probability-zero event, the ESD quantile update $q^\textnormal{ex}_{t|t}(\tau)$ coincides with (the limiting version of) \citeauthor{engle2004caviar}'s (\citeyear{engle2004caviar}) adaptive CAViaR update.

Due to their design, ESD/CAViaR updates can overshoot $y_t$. While the ISD update $q_{t|t}(\tau)$ is similar to the ESD/CAViaR update, it cannot surpass the observation $y_t$:
\begin{equation}
\label{QR_Propar} 
    q_{t|t}(\tau) =
    \left\{ \begin{array}{l@{\hspace{1cm}}l}  
       \min\{y_t,q^\textnormal{ex}_{t|t}(\tau)\}, & y_t > q_{t|t-1}(\tau),\\
       \max\{y_t,q^\textnormal{ex}_{t|t}(\tau) \}, & y_t\leq q_{t|t-1}(\tau).
    \end{array} \right.
\end{equation}
As before, the ISD update is a shrunken version of its  ESD version: if $y_t$ is above (below) $q_{t|t-1}(\tau)$, ISD follows the ESD update upward (downward), but never beyond $y_t$. This capping of ISD updates is advantageous, as updating beyond $y_t$ would decrease model fit.

When tracking multiple quantiles with QRs, quantile crossing poses a serious problem.  For ESD/CAViaR updates, crossings cannot generally be avoided. For the ISD update~\eqref{QR_Propar}, as we show in Appendix~\ref{App_QR}, quantiles remain properly ordered if all updates use the same $H$ and $\sigma$. To see why, note that for an observation lying between any two predicted quantiles, neither update can surpass the observation, thus preventing the quantiles from crossing. To ensure that quantiles also remain ordered  in the prediction step, we specify
\begin{equation}
\label{QR_Prediction}
    q_{t+1|t}(\tau) \;=\; c(\tau)\,(1-\Phi) \,+\, \Phi\, q_{t|t}(\tau)\, +\, \gamma\, x_t,
\end{equation}
with a common autoregressive parameter $\Phi\in[0,1)$, long-run quantile levels $c(\tau)$ that are strictly ordered in $\tau$, and a common slope $\gamma$ relating to the exogenous predictor $x_t$.\footnote{While it may be useful to allow different quantiles to have different sensitivities to the exogenous input $x_t$, this has the potential to introduce quantile crossings. Moreover, for our application, the likelihood improvement of quantile-specific slopes $\gamma(\tau)$ is too small to justify the additional model complexity.}

We estimate the 5, 10, 25, and 50 percent GaR using ISD and ESD filters. For $\{y_t\}$, we use quarterly U.S.\ GDP growth from 1971Q1 to 2021Q4, while for $\{x_t\}$ we take the National Financial Conditions Index (NFCI), with quarterly values obtained by averaging weekly data (cf. 
\citealp{adrian2019vulnerable}). Both series are available in the FRED database.\footnote{See \url{https://fred.stlouisfed.org/series/GDP} and \url{https://fred.stlouisfed.org/series/NFCI}.} To reduce the number of parameters, we use targeting (e.g., \citealp{engle2002dynamic}) and set $c(\tau)$ to its full-sample empirical counterpart. The remaining static parameters are estimated using composite likelihood, comparable to \citet{zou2008composite}. We fix $\sigma=1$, which can be treated as a nuisance parameter (\citealp{geraci2007quantile}).
The postulated log-likelihood equals the sum of four Laplace log-densities, of which three are asymmetric and one is symmetric.
\begin{figure}[t!]
\vspace*{-0.5cm}
    \hbox{
\hspace*{-2.25cm}
    \includegraphics[scale=0.19]{figs/fig7.png}}
      \vspace*{-0.8cm}
    \caption{Growth-at-risk estimates for the ISD and ESD models for $\tau = 0.05$, $\tau = 0.10$, $\tau = 0.25$, and $\tau = 0.50$, 1971Q1 to 2021Q4.} 
    \label{Fig_QR}
\end{figure}

Figure \ref{Fig_QR} displays the 5, 10, 25, and 50 percent GaR estimates from the ESD/CAViaR and ISD filters, involving \eqref{QR_caviar} and~\eqref{QR_Propar}, respectively. The ISD-filtered quantiles are more responsive, showing larger downward adjustments at the onset of COVID-19 (2020Q2) as well as faster post-crisis recovery. The enhanced stability of the ISD update~\eqref{QR_Propar} allows its estimated learning rate to exceed that of its explicit counterpart by a factor of five ($\hat{H}=4.002$ vs.\ $\hat{H}=0.804$). While the ESD-filtered quantiles exhibit regular crossings, the ISD-filtered versions remain properly ordered. For instance, the ESD median ($\tau=0.50$) often overshoots $y_t$ and occasionally crosses the first quartile ($\tau=0.25$). In line with \citeauthor{adrian2019vulnerable} (\citeyear{adrian2019vulnerable}), we find a negative NFCI effect ($\hat{\gamma}=-0.052$ for ISD; $\hat{\gamma}=-0.019$ for ESD), implying that tighter financial conditions exert downward pressure on the GDP quantiles.

\subsection{T-bill rate spreads}
\label{sec: Empirical illustration}

Monitoring term spreads---leading indicators of business-cycle recessions---is essential for companies and policymakers alike. We use \citeauthor{harvey2014filtering}'s (\citeyear{harvey2014filtering}) Student's~$t$ location model, which has been widely used (e.g., \citealp[sec.\ 3.1]{caivano2016robust}; \citealp[sec.\ 6.3]{blasques2018feasible}; \citealp[sec.\ 5]{blasques2022maximum}), including for T-bill rate spreads (\citealp{artemova2022score1}). However, the Student's $t$ density is not log-concave in the location parameter (i.e., Assumption~\ref{A5_Concavity} fails). Moreover, if the penalty parameter $P$ is sufficiently small, optimization~\eqref{RegLike} is not strongly concave (i.e., Assumption~\ref{A2_RL_StrictConcave} may also fail). Here we show that the ISD filter can perform well even if Assumptions~\ref{A2_RL_StrictConcave} and \ref{A5_Concavity} fail.

The ISD update for the Student's~$t$ location model admits a closed-form solution: it reduces to solving a cubic (for details, see Appendix~\ref{App T Bill}). Standard software provides closed-form (albeit lengthy) expressions for up to three roots, of which at least one is real valued. The lack of concavity means that up to three stationary points (two of them local maxima) may exist; in this case, we select the global maximum by comparing objective values in~\eqref{RegLike}. Multiplicity of stationary points is precluded if the penalty term is strong enough for Assumption~\ref{A2_RL_StrictConcave} to hold, i.e., if $P>1/8$ or, equivalently, $H<8$ (for details, see Appendix~\ref{App T Bill}). 
As the update is well-defined for all $P>0$, 
we do not impose this condition.

The ISD update $\theta_{t|t}$ is a weighted average of the prediction $\theta_{t|t-1}$ and the observation $y_t$. The weight of $y_t$ lies in the interval $(0,H/(1+H)]\subseteq (0,1)$ for any $H>0$ (see Appendix~\ref{App T Bill}), such that $\theta_{t|t}\in [\min\{y_t,\theta_{t|t-1}\},\max\{y_t,\theta_{t|t-1}\}]$. In contrast, the ESD update with $H>1$ can overshoot the data and lead to ``zig-zagging,'' as we will see below.

Following \citet{artemova2022score1}, we analyze the spread between three- and six-month T-bill rates from the FRED-QD database (series ``TB6M3M'') scaled by a factor of ten. The sample comprises 249 quarterly observations from 1959:Q1 to 2021:Q1. For these data, \citet{artemova2022score1} found that the ESD filter with a Student's $t$ density performed best.

\begin{table}[H]
\centering
\caption{Parameter estimates for Student's $t$ location model with T-bill spread data.}
\footnotesize
    \begin{threeparttable}
\begin{tabular}{cc@{\hspace{1cm}}ccccc@{\hspace{1cm}}c}
Distribution & Filter  & $\hat{\omega}$ & $\hat{\Phi}$ & $\hat{H}$ & $\hat{\sigma}^2$ & $\hat{\nu}$ & LL 
\\
\hline
Student's $t$ 
& ISD &  0.235  &  0.751 &   23.717 &   0.387 &  2.060&	$-364.4$ 
\\
& ESD & 0.353  &  0.714  &  2.194   & 0.516 &   2.632	& $-370.6$
\\
\hline
\end{tabular}
\end{threeparttable}
\label{Tab:parameter estimates}
\end{table}

Table~\ref{Tab:parameter estimates} reports ML estimates and log-likelihood values for ESD and ISD filters based on the Student's $t$ with scale $\sigma>0$ and $\nu>0$ degrees of freedom. The ISD filter achieves a higher likelihood and a larger learning rate ($\hat{H}\approx 24$ vs.\ $\hat{H}\approx 2.2$). Interestingly, it fits the data best when the regularized objective~\eqref{RegLike} is non-concave (as $\hat{H}\approx 24>8$).

\begin{figure}[H]
\centering
  \begin{tabular}{c@{\hspace{1mm}}c}
     \hspace{-0.5cm}   \includegraphics[width=0.47\textwidth]{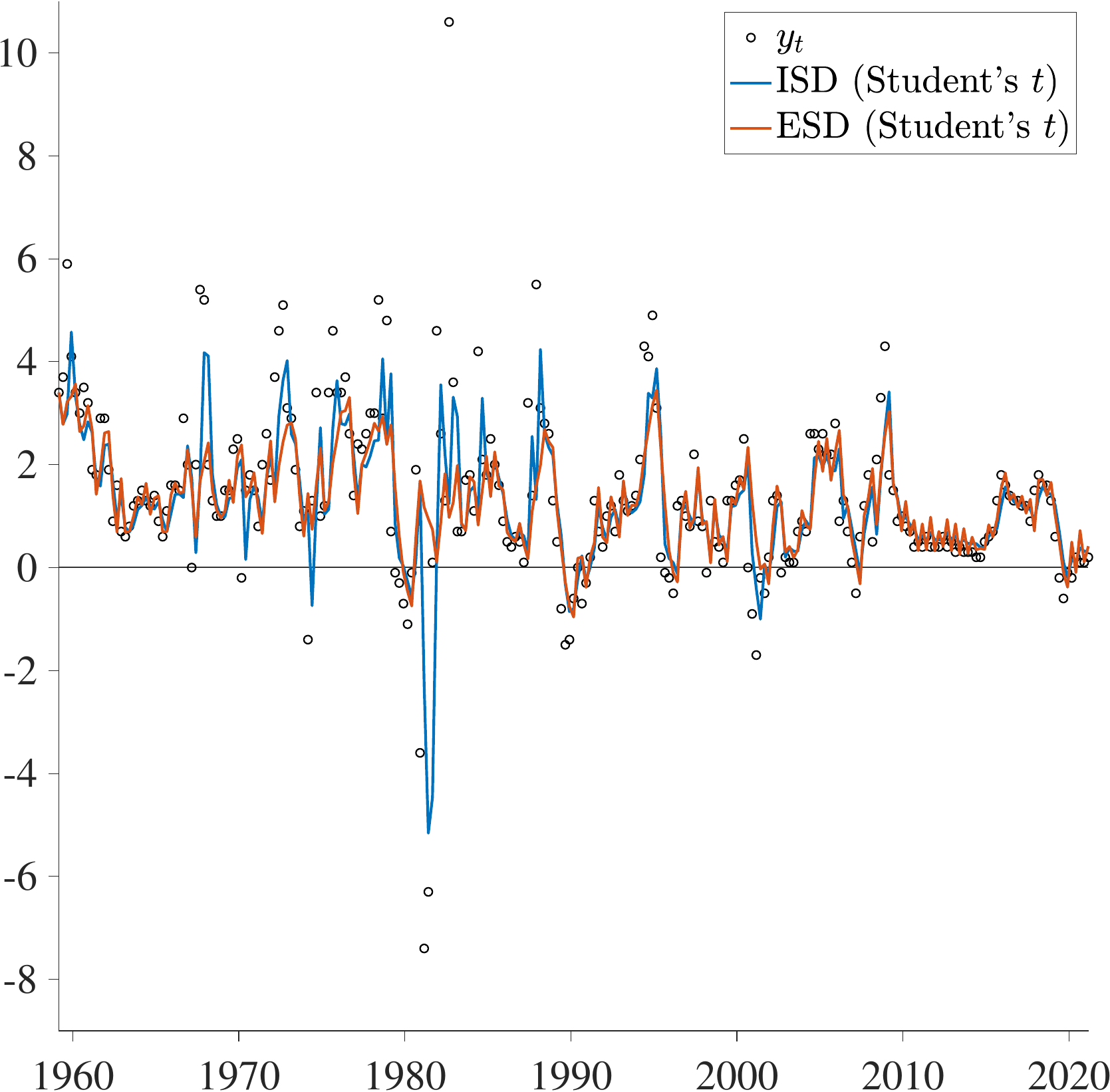}
        &
        \includegraphics[width=0.47\textwidth]{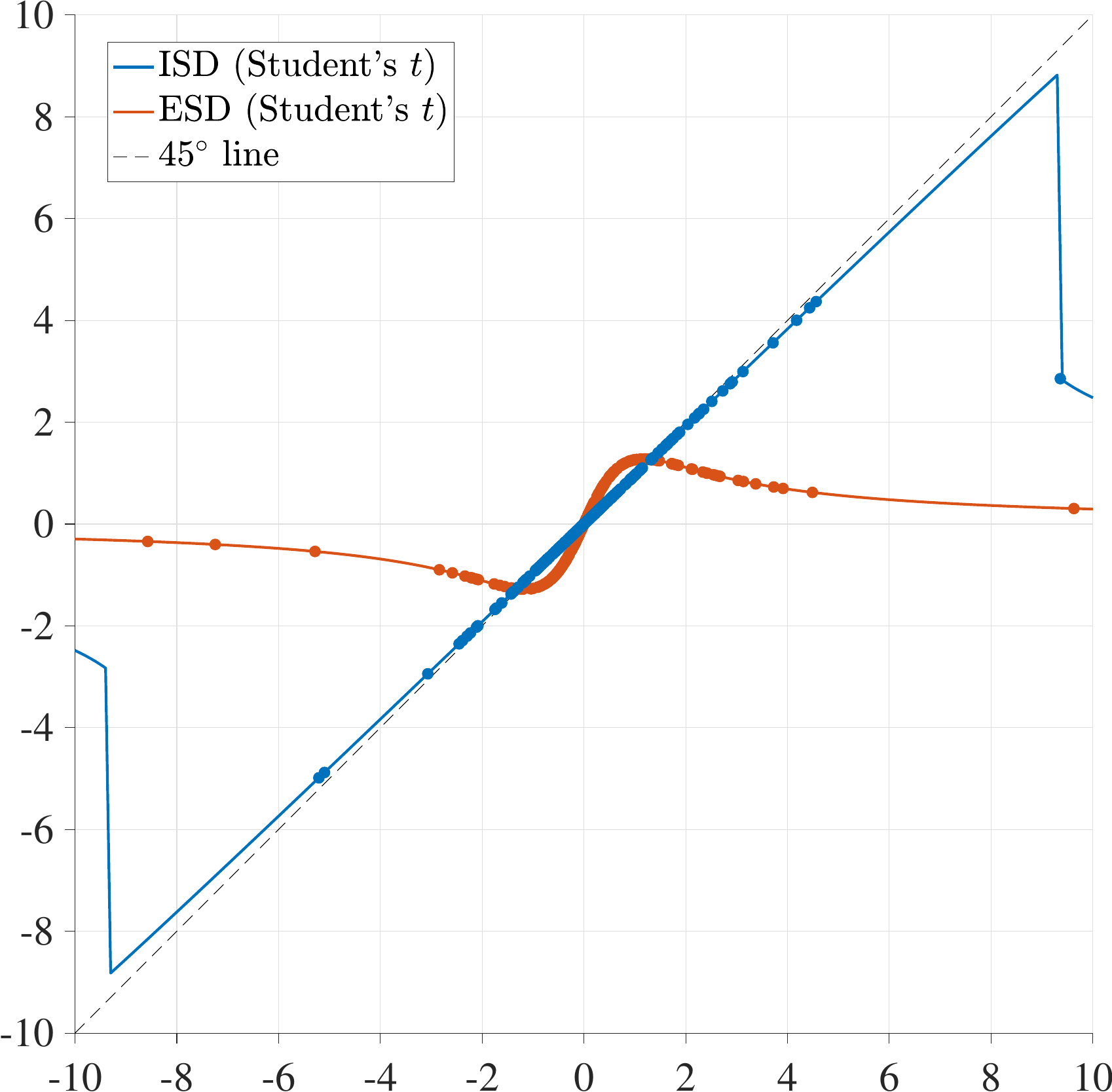}
        \\
        \footnotesize (a) T-bill rate spread and predicted paths
         &
         \footnotesize (b) Impact curves ($y_t-\theta_{t|t-1}$ vs.\ $\theta_{t|t}-\theta_{t|t-1}$)
  \end{tabular}
  \caption{T-bill spread data with filtered paths and impact curves.}
  \label{fig}
\end{figure}

Figure~\ref{fig} shows the associated predicted-parameter paths and impact curves. The ESD path is relatively unresponsive in the 1960s–1990s, but overly responsive after 2010, producing a zig-zagging pattern around the data, consistent with $\hat{H}\approx 2.2>1$. By contrast, the ISD filter's impact curve is more responsive to large shocks without overshooting the data, allowing it to track peaks and troughs in the volatile decades, without zig-zagging after 2010. Both filters largely ignore the 1982:Q3 outlier (10.6). For the ISD impact curve, the discontinuity around this observation reflects a bifurcation point, where the second local maximum starts to dominate. The non-concavity of the objective function~\eqref{RegLike} due to $\hat{H}>8$ and the associated multiplicity of stationary points allow the ISD filter to largely disregard this outlier. Relative to the ESD filter, the ISD filter lowers the in-sample MSE by roughly 18\%, which may be economically significant.

\section{Conclusion}
\label{Sec_Con}

We have introduced an implicit score-driven (ISD) framework for updating time-varying parameters in observation-driven models. At each time step, the ISD filter maximizes the logarithmic observation density subject to a quadratic penalty centered at the one-step-ahead prediction; the associated first-order condition is an implicit stochastic-gradient update. For the class of (possibly misspecified) log-concave densities, we showed that the ISD filter is invertible, while its updates are contractive in mean squared error toward the \mbox{(pseudo-)}true parameter at every time step. The popular class of explicit score-driven (ESD) filters---also known as dynamic conditional score (DCS; \citealp{harvey2013dynamic}) models or generalized autoregressive score (GAS; \citealp{creal2013generalized}) models---arises within our framework by locally linearizing the logarithmic density around the prediction. 
However, to obtain guarantees comparable to those of the ISD filter, especially under model misspecification, ESD filters require the score  to be Lipschitz continuous and the learning rate to be sufficiently small.
We illustrated the theoretical benefits of our approach in simulations and its practical relevance in empirical applications to asset prices, growth-at-risk, and T-bill rates.

{\small
\putbib[ref]
}
\end{bibunit}

\clearpage
\appendix

\begin{center}
\Large \textbf{Online supplement to:  \\
``Implicit score-driven filters
\\ for time-varying parameter models''}\\[3pt]
\normalsize
Rutger-Jan Lange, Bram van Os, and Dick van Dijk
\\
\today
\end{center}

\pagenumbering{arabic}
\renewcommand*{\thepage}{S\arabic{page}}
\setcounter{page}{1}

\setcounter{equation}{0}
\renewcommand{\theequation}{\Alph{section}.\arabic{equation}}
\setcounter{table}{0}
\renewcommand{\thetable}{\Alph{section}.\arabic{table}}
\setcounter{figure}{0}
\renewcommand{\thefigure}{\Alph{section}.\arabic{figure}}
\renewcommand{\thesubsection}{\Alph{section}.\arabic{subsection}}


\phantomsection
\addcontentsline{toc}{chapter}{Online Supplement}

\startcontents[sections]
\printcontents[sections]{l}{1}{\setcounter{tocdepth}{2}}

\clearpage

\begin{bibunit}[chicago]

\section{Proofs of main results}
\label{appendix A}

\subsection{Proposition \ref{Lm_Alignment}:  Relationship between ISD and ESD updates}

By Assumption \ref{A2_RL_StrictConcave} we have that the regularized log likelihood $f(\theta|y_t,\theta_{t|t-1})$ is concave in $\theta$ with probability one in $y_t$. As a result, we have for almost every $y_t$ that
\begin{equation}
    f(\theta_{t|t}|y_t,\theta_{t|t-1}) \leq f(\theta_{t|t-1}|y_t,\theta_{t|t-1}) + \langle \theta_{t|t} - \theta_{t|t-1},\nabla(y_t |\theta_{t|t-1}) \rangle,
\end{equation}
reordering and using the fact that $\theta_{t|t}$ maximizes $f(\theta|y_t,\theta_{t|t-1})$ we obtain the desired result:
\begin{equation}
    \langle \theta_{t|t} - \theta_{t|t-1},\nabla(y_t |\theta_{t|t-1}) \rangle \geq f(\theta_{t|t}|y_t,\theta_{t|t-1}) - f(\theta_{t|t-1}|y_t,\theta_{t|t-1}) \geq 0.
\end{equation}

In the scalar case, the first-order condition (FOC) and strict positivity of the learning rate imply that $\nabla(y_t |\theta_{t|t-1}) \nabla(y_t |\theta_{t|t}) \geq 0$ . Furthermore, $\nabla(y_t |\theta_{t|t}) = 0$ produces $\theta_{t|t} = \theta_{t|t-1}$, in turn implying that $\nabla(y_t |\theta_{t|t-1}) = \nabla(y_t |\theta_{t|t})= 0$. Conversely, if $\nabla(y_t |\theta_{t|t-1}) = 0$, we have that $\theta_{t|t} = \theta_{t|t-1}$, as filling in $\theta_{t|t-1}$ solves the FOC (and Assumption \ref{A2_RL_StrictConcave} implies uniqueness of $\theta_{t|t}$). Therefore, $\nabla(y_t |\theta_{t|t-1}) = 0$ if and only if $\nabla(y_t |\theta_{t|t}) = 0$. Combining this with $\nabla(y_t |\theta_{t|t-1}) \nabla(y_t |\theta_{t|t}) \geq 0$, we obtain $\mathrm{sgn}(\nabla( y_t | \theta_{t|t})) = \mathrm{sgn}(\nabla( y_t | \theta_{t|t-1}))$, which is the score-equivalence as defined in \citet{blasques2015information}.

Next, we use second-order differentiability (Assumption \ref{A4_Differentiability}b) to write the ISD update as a curvature-corrected ESD update. Specifically, we have that
\begin{equation}
\label{Eq_MMVT}
    \nabla(y_t |\theta_{t|t}) - \nabla(y_t |\theta_{t|t-1}) = -\mathcal{I}_{t|t}(\theta_{t|t} - \theta_{t|t-1}),    
\end{equation}
where $\mathcal{I}_{t|t}$ is the negative average Hessian between $\theta_{t|t-1}$ and $\theta_{t|t}$:
\begin{equation}
\label{Eq_NegHes}
    \mathcal{I}_{t|t} :=  -\int_{0}^{1} \left.\frac{\partial^2  \log p(y_t|\theta)}{\partial\theta \partial\theta'}\right|_{\substack{\theta \,= \, u \,\theta_{t|t-1} \,+\, (1-u)\,\theta_{t|t}}} \mathrm{d}u.
\end{equation}
Roughly put, relationship \eqref{Eq_MMVT} can be viewed as a multivariate analog of the mean-value theorem. The integral form is necessary because, unlike the scalar case, there need not be any single point $\theta$ for which the Hessian is exactly equal to $-\mathcal{I}_{t|t}$. Using the ISD FOC for $\theta_{t|t} - \theta_{t|t-1}$ on the right-hand side of \eqref{Eq_MMVT} and rearranging yields:
\begin{equation}
    (I_K + \mathcal{I}_{t|t}P_t^{-1})\nabla(y_t |\theta_{t|t}) = \nabla(y_t |\theta_{t|t-1}).
    \label{Eq_MMVT2}
\end{equation}
Next, we note that $P_t + \mathcal{I}_{t|t}$ is positive definite, which follows from strict concavity of the regularized log-likelihood (Assumption \ref{A2_RL_StrictConcave}). That is, this assumption implies that the penalty matrix $P_t$ strictly exceeds (in an eigenvalue sense) the Hessian of $\log p(y_t|\theta)$ for any $\theta$ with probability one in $y_t$. From the definition of $\mathcal{I}_{t|t}$, it follows that $P_t \succ -\mathcal{I}_{t|t} \Rightarrow P_t + \mathcal{I}_{t|t} \succ O_K$. Therefore $P_t + \mathcal{I}_{t|t}$ is invertible; premultiplying with $(P_t + \mathcal{I}_{t|t})^{-1}$ gives:
\begin{equation}
    P_t^{-1}\nabla(y_t |\theta_{t|t}) = (P_t + \mathcal{I}_{t|t})^{-1}\nabla(y_t |\theta_{t|t-1}).
    \label{Eq_MMVT3}
\end{equation}
Using \eqref{Eq_MMVT3} together with the ISD FOC produces the final result,
\begin{equation}
    \theta_{t|t} = \theta_{t|t-1} + (P_t + \mathcal{I}_{t|t})^{-1}\nabla(y_t|\theta_{t|t-1}).
\end{equation}

\subsection{Proposition \ref{Lm_Shrinkage}: Step-size shrinkage}
Using the FOCs, the difference between the implicit and explicit updates is given by:
\begin{equation}
    \theta^\textnormal{ex}_{t|t} - \theta_{t|t} = \theta_{t|t-1} + P_t^{-1} \nabla(y_t |\theta_{t|t-1}) - \theta_{t|t-1} -P_t^{-1} \nabla(y_t |\theta_{t|t}),
\end{equation}
whereby rearranging and using the definition of $\mathcal{I}_{t|t}$ from \eqref{Eq_NegHes} yields 
\allowdisplaybreaks
\begin{align}
    \theta^\textnormal{ex}_{t|t} - \theta_{t|t-1} &= \theta_{t|t} - \theta_{t|t-1} - P_t^{-1} [\nabla(y_t |\theta_{t|t}) - \nabla(y_t |\theta_{t|t-1})] \\
    &= (I_K + P_t^{-1}\mathcal{I}_{t|t})(\theta_{t|t} - \theta_{t|t-1}).
\end{align}

Pre-multiplying with $P_t^{1/2}$, which denotes the symmetric square root of $P_t$, and taking the quadratic norm yields
\allowdisplaybreaks
\begin{align}
\label{Eq_StepSizeShrinkageFirst}
      \|&\theta^\textnormal{ex}_{t|t} - \theta_{t|t-1}\|_{P_t}^2   = \|\theta_{t|t} - \theta_{t|t-1}\|^2_{(I_K + P_t^{-1}\mathcal{I}_{t|t})'P_t (I_K + P_t^{-1}\mathcal{I}_{t|t})} \\
      &= \|\theta_{t|t} - \theta_{t|t-1}\|^2_{P_t + 2\mathcal{I}_{t|t} + \mathcal{I}_{t|t}P_t^{-1}\mathcal{I}_{t|t}}\\
    &=\|\theta_{t|t} - \theta_{t|t-1}\|^2_{P_t} + 2\|\theta_{t|t} - \theta_{t|t-1}\|^2_{\mathcal{I}_{t|t}} + \|\mathcal{I}_{t|t}(\theta_{t|t} - \theta_{t|t-1})\|^2_{P_t^{-1}}\\ 
      &\geq \|\theta_{t|t} - \theta_{t|t-1}\|^2_{P_t} + 2\lambda_{\min}(\mathcal{I}_{t|t})\|\theta_{t|t} - \theta_{t|t-1}\|^2 + 
      \lambda_{\min}(P_t^{-1})\| \theta_{t|t} - \theta_{t|t-1}\|_{\mathcal{I}_{t|t}^2}^2\\
      &\geq \|\theta_{t|t} - \theta_{t|t-1}\|^2_{P_t} + 2\alpha_t\|\theta_{t|t} - \theta_{t|t-1}\|^2 + \frac{\lambda_{\min}(\mathcal{I}_{t|t}^2)}{\lambda_{\max}(P_t)}\| \theta_{t|t} - \theta_{t|t-1}\|^2\\
     &\geq \|\theta_{t|t} - \theta_{t|t-1}\|^2_{P_t} + \left(2\alpha_t + \frac{\alpha_t^2}{\lambda_{\max}(P_t)}\right)\|\theta_{t|t} - \theta_{t|t-1}\|^2\\    
      &\geq \left(1 + \frac{2\alpha_t}{\lambda_{\max}(P_t)} + \frac{\alpha_t^2}{\lambda_{\max}(P_t)^2}\right) \|\theta_{t|t} - \theta_{t|t-1}\|^2_{P_t}
      \\
      &= \left( \frac{\lambda_{\max}(P_t) + \alpha_t}{\lambda_{\max}(P_t)}\right)^2 \|\theta_{t|t} - \theta_{t|t-1}\|^2_{P_t}.    
      \label{Eq_StepSizeShrinkageEnd}
\end{align}
Here the fifth line uses $\lambda_{\min}(\mathcal{I}_{t|t})\geq \alpha_t$ which follows from concavity (Assumption \ref{A5_Concavity}) and the definition of $\mathcal{I}_{t|t}$ in \eqref{Eq_NegHes} as the average negative Hessian. Because $\alpha_t \geq 0$ it also follows that $\lambda_{\min}(\mathcal{I}_{t|t}^2) \geq \alpha_t^2$, which is used in the sixth line. The final line then uses $\|x\|^2_{P_t} \leq \lambda_{\max}(P_t)\|x\|^2$, which implies that $\|x\|^2 \geq \frac{1}{\lambda_{\max}(P_t)}\|x\|^2_{P_t}$ for arbitrary $K\times 1$ vector $x$ and using that $\lambda_{\max}(P_t) > 0$. Note that $2\alpha_t + \frac{\alpha_t^2}{\lambda_{\max}(P_t)} \geq 0$, such that the correct sign is maintained. Finally, rearranging the last expression above gives the desired result:
\begin{equation}
    \big\|\theta_{t|t}-\theta_{t|t-1}\big\|^2_{P_t} \; \leq \; \left(\frac{\lambda_{\max}(P_t)}{\lambda_{\max}(P_t)+ \alpha_t}\right)^2\big\|\theta^\textnormal{ex}_{t|t}-\theta_{t|t-1}\big\|^2_{P_t}.
    \label{Eq_FinStepShrink}
\end{equation}

\subsection{Lemma \ref{Lm_StaUp}: Prediction-to-update stability}
Consider two predictions $\theta_{t|t-1}$ and $\tilde{\theta}_{t|t-1}$ that are updated based on the observation $y_t$ to $\theta_{t|t}$ and $\tilde{\theta}_{t|t}$, respectively. For the ISD update, we may write
\allowdisplaybreaks
\begin{align}
    \theta_{t|t} - \tilde{\theta}_{t|t} &= \theta_{t|t-1} - \tilde{\theta}_{t|t-1} + P_t^{-1}[
    \nabla(y_t|\theta_{t|t}) - \nabla(y_t|\tilde{\theta}_{t|t})] \\
    & = \theta_{t|t-1} - \tilde{\theta}_{t|t-1} - P_t^{-1}\tilde{\mathcal{I}}_{t|t}(\theta_{t|t} - \tilde{\theta}_{t|t}),
    \label{Eq_LemPreUp_ISD}
\end{align}
where $\tilde{\mathcal{I}}_{t|t}$ is the negative average Hessian between $\theta_{t|t}$ and $\tilde{\theta}_{t|t}$:
\begin{equation}
\label{Eq_NegHes_StaISD}
    \tilde{\mathcal{I}}_{t|t} :=  -\int_{0}^{1} \left.\frac{\partial^2  \log p(y_t|\theta)}{\partial\theta \partial\theta'}\right|_{\substack{\theta \,= \, u \,\theta_{t|t} \,+\, (1-u)\,\tilde{\theta}_{t|t}}} \mathrm{d}u.
\end{equation}
Reordering of \eqref{Eq_LemPreUp_ISD} yields
\begin{equation}
    \theta_{t|t-1} - \tilde{\theta}_{t|t-1} = (I_K + P_t^{-1}\tilde{\mathcal{I}}_{t|t})(\theta_{t|t} - \tilde{\theta}_{t|t}).
\end{equation}
Next, pre-multiplying with $P_t^{1/2}$ and taking the quadratic norm gives
\allowdisplaybreaks
\begin{align}
    \|\theta_{t|t-1} - \tilde{\theta}_{t|t-1}\|^2_{P_t} &= \|\theta_{t|t} - \tilde{\theta}_{t|t}\|^2_{P_t + 2\tilde{\mathcal{I}}_{t|t} + \tilde{\mathcal{I}}_{t|t}P_t^{-1}\tilde{\mathcal{I}}_{t|t}}. 
\end{align}
Using the same steps as in \eqref{Eq_StepSizeShrinkageFirst}-\eqref{Eq_FinStepShrink}, we obtain
\begin{equation}
    \big\|\theta_{t|t}-\tilde{\theta}_{t|t}\big\|^2_{P_t} \; \leq \; \left(\frac{\lambda_{\max}(P_t)}{\lambda_{\max}(P_t)+ \alpha_t}\right)^2\big\| \theta_{t|t-1}-\tilde{\theta}_{t|t-1}\big\|^2_{P_t}.
\end{equation}

For the ESD update, we may write
\allowdisplaybreaks
\begin{align}
    \theta^\textnormal{ex}_{t|t} - \tilde{\theta}^\textnormal{ex}_{t|t} &= \theta_{t|t-1} - \tilde{\theta}_{t|t-1} + P_t^{-1}[
    \nabla(y_t|\theta_{t|t-1}) - \nabla(y_t|\tilde{\theta}_{t|t-1})] \\
    & = (I_K - P_t^{-1}\tilde{\mathcal{I}}_{t|t-1})(\theta_{t|t-1} - \tilde{\theta}_{t|t-1}),  
\end{align}
where $\tilde{\mathcal{I}}_{t|t-1}$ is the negative average Hessian between $\theta_{t|t-1}$ and $\tilde{\theta}_{t|t-1}$:
\begin{equation}
\label{Eq_NegHes_StaESD}
    \tilde{\mathcal{I}}_{t|t-1} :=  -\int_{0}^{1} \left.\frac{\partial^2  \log p(y_t|\theta)}{\partial\theta \partial\theta'}\right|_{\substack{\theta \,= \, u \,\theta_{t|t-1} \,+\, (1-u)\,\tilde{\theta}_{t|t-1}}} \mathrm{d}u.
\end{equation}
Next, pre-multiplying with $P_t^{1/2}$ and taking the quadratic norm gives
\allowdisplaybreaks
\begin{align}
    \|\theta^\textnormal{ex}_{t|t} &- \tilde{\theta}^\textnormal{ex}_{t|t}\|^2_{P_t} = \|\theta_{t|t-1} - \tilde{\theta}_{t|t-1}\|^2_{(I_K - P_t^{-1}\tilde{\mathcal{I}}_{t|t-1})'P_t (I_K - P_t^{-1}\tilde{\mathcal{I}}_{t|t-1})}\\
    &=  \|\theta_{t|t-1} - \tilde{\theta}_{t|t-1}\|^2_{P_t - 2\tilde{\mathcal{I}}_{t|t-1} + \tilde{\mathcal{I}}_{t|t-1}P_t^{-1}\tilde{\mathcal{I}}_{t|t-1}}\\
    &= \|\theta_{t|t-1} - \tilde{\theta}_{t|t-1}\|^2_{P_t} + \|\tilde{\mathcal{I}}^{1/2}_{t|t-1}(\theta_{t|t-1} - \tilde{\theta}_{t|t-1})\|^2_{\tilde{\mathcal{I}}^{1/2}_{t|t-1}P_t^{-1}\tilde{\mathcal{I}}^{1/2}_{t|t-1} - 2I_K}\\
    &\leq \|\theta_{t|t-1} - \tilde{\theta}_{t|t-1}\|^2_{P_t} + \lambda_{\max}(\tilde{\mathcal{I}}^{1/2}_{t|t-1}P_t^{-1}\tilde{\mathcal{I}}^{1/2}_{t|t-1} - 2I_K)\|\theta_{t|t-1} - \tilde{\theta}_{t|t-1}\|_{\tilde{\mathcal{I}}_{t|t-1}}^2,
    \label{Eq_LemStaPreESD}
\end{align}
where $\lambda_{\max}(\tilde{\mathcal{I}}^{1/2}_{t|t-1}P_t^{-1}\tilde{\mathcal{I}}^{1/2}_{t|t-1} - 2I_K) = \lambda_{\max}(\tilde{\mathcal{I}}_{t|t-1}P_t^{-1}) - 2 \leq \lambda_{\max}(\tilde{\mathcal{I}}_{t|t-1})\lambda_{\max}(P_t^{-1}) - 2 \leq L_t/\lambda_{\min}(P_t) - 2 \leq 0$ and the final argument follows from the assumption $\lambda_{\min}(P_t) \geq L_t/2$. This means that the final term in \eqref{Eq_LemStaPreESD} is negative. Continuing, we obtain
\allowdisplaybreaks
\begin{align}
   \|\theta^\textnormal{ex}_{t|t} - \tilde{\theta}^\textnormal{ex}_{t|t}\|^2_{P_t} &\leq \|\theta_{t|t-1} - \tilde{\theta}_{t|t-1}\|^2_{P_t} - [2 - L_t/\lambda_{\min}(P_t)]\|\theta_{t|t-1} - \tilde{\theta}_{t|t-1}\|_{\tilde{\mathcal{I}}_{t|t-1}}^2\\
    &\leq \|\theta_{t|t-1} - \tilde{\theta}_{t|t-1}\|^2_{P_t} - \alpha_t[2 - L_t/\lambda_{\min}(P_t)]\|\theta_{t|t-1} - \tilde{\theta}_{t|t-1}\|^2\\
    &\leq \left(1 - \frac{\alpha_t[2 - L_t/\lambda_{\min}(P_t)]}{\lambda_{\max}(P_t)} \right) \|\theta_{t|t-1} - \tilde{\theta}_{t|t-1}\|^2_{P_t}\\
    &= \frac{\lambda_{\max}(P_t) - \alpha_t[2 - L_t/\lambda_{\min}(P_t)]}{\lambda_{\max}(P_t)} \|\theta_{t|t-1} - \tilde{\theta}_{t|t-1}\|^2_{P_t},
    \label{Eq_LemStaPreESD2}
\end{align}
where the second line uses $\lambda_{\min}(\tilde{\mathcal{I}}_{t|t-1})\geq \alpha_t \geq 0$ by concavity (Assumption \ref{A5_Concavity}) and the third line uses that $-\|x\|^2 \leq -\frac{1}{\lambda_{\max}(P_t)}\|x\|^2_{P_t}$ for arbitrary $K\times 1$ vector $x$. Finally, we note that $\lambda_{\max}(P_t) -\alpha_t\left[2 - L_t/\lambda_{\min}(P_t)\right] \geq \lambda_{\max}(P_t) -\alpha_t\left[2 - \alpha_t/\lambda_{\max}(P_t)\right] = \frac{1}{\lambda_{\max}(P_t)}(\lambda_{\max}(P_t)^2 - 2\alpha_t\lambda_{\max}(P_t) + \alpha_t^2) = \frac{1}{\lambda_{\max}(P_t)}(\lambda_{\max}(P_t) - \alpha_t)^2 \geq 0$, such that contraction coefficient is indeed contained in the unit interval:
\begin{equation}
    \|\theta^\textnormal{ex}_{t|t} - \tilde{\theta}^\textnormal{ex}_{t|t}\|^2_{P_t} \leq \underbrace{\frac{\lambda_{\max}(P_t) - \alpha_t[2 - L_t/\lambda_{\min}(P_t)]}{\lambda_{\max}(P_t)}}_\text{$\in [0,1]$, contraction coefficient} \|\theta_{t|t-1} - \tilde{\theta}_{t|t-1}\|^2_{P_t}.
\end{equation}

\subsection{Lemma \ref{Lm_StaPre}: Prediction-to-prediction stability}
The update-to-prediction mapping from time $t$ to $t+1$ can be written as
\allowdisplaybreaks
\begin{align}
         \|\theta_{t+1|t} - \tilde{\theta}_{t+1|t}\|^2_{P_t} &= \|\Phi(\theta_{t|t} - \tilde{\theta}_{t|t})\|^2_{P_t} = -\|\theta_{t|t} - \tilde{\theta}_{t|t}\|^2_{P_t - \Phi'P_t\Phi} + \|\theta_{t|t} - \tilde{\theta}_{t|t}\|^2_{P_t} \\& \leq -\lambda_{\min}(P_t - \Phi'P_t\Phi)\|\theta_{t|t} - \tilde{\theta}_{t|t}\|^2 + \|\theta_{t|t} - \tilde{\theta}_{t|t}\|^2_{P_t}\\&\leq \varepsilon_{1,t} \|\theta_{t|t} - \tilde{\theta}_{t|t}\|^2_{P_t}, 
         \label{Eq_App_scP}
\end{align}
where the second line uses that $\lambda_{\min}(P_t - \Phi'P_t\Phi) \geq 0$ by positive semi-definiteness of $P_t - \Phi'P_t\Phi$, while the last line uses $-\|\cdot\|^2 \leq -\lambda_{\max}(P_t)^{-1} \|\cdot\|_{P_t}^2$. Here $\varepsilon_{1,t}$ is given by
\begin{equation}
    \varepsilon_{1,t} = \frac{\lambda_{\max}(P_t) - \lambda_{\min}(P_t -\Phi'P_t\Phi)}{\lambda_{\max}(P_t)}.
\end{equation}
By positive definiteness of $P_t$ it follows that $\Phi'P_t\Phi$ is positive semi-definite due to its quadratic form. Therefore, we have that $0 \leq \lambda_{\max}(\Phi'P_t\Phi) = \lambda_{\max}(P_t -(P_t-\Phi'P_t\Phi)) \leq  \lambda_{\max}(P_t) +  \lambda_{\max}( -(P_t-\Phi'P_t\Phi)) = \lambda_{\max}(P_t) - \lambda_{\min}(P_t-\Phi'P_t\Phi) \leq \lambda_{\max}(P_t)$, such that $\varepsilon_{1,t} \in [0,1]$. If $P_t - \Phi'P_t\Phi$ is positive definite, we have that $\varepsilon_{1,t} \in [0,1)$.

In addition, from Lemma \ref{Lm_StaUp}, we have
\begin{equation}
  \|\tilde{\theta}_{t|t}-\theta_{t|t}\|^2_{P_t} \leq \varepsilon_{2,t}\|\tilde{\theta}_{t|t-1}-\theta_{t|t-1}\|^2_{P_t}, \quad  \varepsilon_{2,t} = \left(\frac{\lambda_{\max}(P_t)}{\lambda_{\max}(P_t) + \alpha_t}\right)^2,
  \label{Eq_App_scU}
\end{equation}
where $\varepsilon_{2,t} \in [0,1]$ if $\alpha_t\geq 0$ and $\varepsilon_{2,t} \in [0,1)$ if $\alpha_t > 0$. Combining (\ref{Eq_App_scP}) and (\ref{Eq_App_scU}), we obtain \begin{equation}
    \|\theta_{t+1|t} - \tilde{\theta}_{t+1|t}\|^2_{P_t} \leq \kappa_t \| \theta_{t|t-1} - \tilde \theta_{t|t-1} \|^2_{P_t},
\end{equation}
where $\kappa_t$ is given by
\allowdisplaybreaks
\begin{align}
\kappa_t &= \varepsilon_{1,t} \varepsilon_{2,t} =  \frac{\lambda_{\max}(P_t)  - \lambda_{\min}(P_t -\Phi'P_t\Phi)}{\lambda_{\max}(P_t)} \frac{\lambda_{\max}(P_t)^2}{(\lambda_{\max}(P_t) + \alpha_t)^2} 
\\
&= \frac{\lambda_{\max}(P_t)[\lambda_{\max}(P_t)  -\lambda_{\min}(P_t -\Phi'P_t\Phi)]}{(\lambda_{\max}(P_t) + \alpha_t)^2}.
\end{align}
If either $\alpha_t > 0$ or $P_t - \Phi'P_t\Phi \succ O_K$ we obtain $\kappa_t \in [0,1)$, which concludes the proof.

\subsection{Theorem \ref{Thm_Invertability}: Invertibility}
\label{App_ProofInv}
By assumption, there exists a $\bar{P}$ such that we have for all $P_t$ that $\kappa_t P_t \prec \rho_t \bar{P} \preceq P_t$ for some $\rho_t > 0$. This condition implies that the prediction-to-prediction mapping from time $t$ to $t+1$ is strictly contracting in the norm $\|\cdot\|_{\rho_t \bar{P}}$. To see this, we may write 
\allowdisplaybreaks
\begin{align}
    \|\theta_{t+1|t} - \tilde{\theta}_{t+1|t}\|^2_{\rho_t \bar{P}} &\leq \|\theta_{t+1|t} - \tilde{\theta}_{t+1|t}\|^2_{P_t} \leq \kappa_t \| \theta_{t|t-1} - \tilde \theta_{t|t-1} \|^2_{P_t} \\&= - \| \theta_{t|t-1} - \tilde \theta_{t|t-1} \|^2_{\rho_t \bar{P} - \kappa_t P_t} + \|\theta_{t|t-1} - \tilde \theta_{t|t-1} \|^2_{\rho_t \bar{P}} \\
    &\leq \delta_t \|\theta_{t|t-1} - \tilde \theta_{t|t-1} \|^2_{\rho_t \bar{P}},
\end{align}
where $\delta_t$ is given by
\begin{equation}
    \delta_t = \frac{\lambda_{\max}(\rho_t \bar{P}) -\lambda_{\min}(\rho_t \bar{P} - \kappa_t P_t)}{\lambda_{\max}(\rho_t \bar{P})}.
\end{equation}
Due to the condition $\rho_t \bar{P} - \kappa_t P_t \succeq \rho_t Q \succ 0$, we obtain that $\delta_t \in [0,\delta]$, where $\delta$ is given by
\begin{equation}
    \delta = \frac{\lambda_{\max}(\rho_t \bar{P}) -\lambda_{\min}(\rho_t Q)}{\lambda_{\max}(\rho_t \bar{P})} = \frac{\lambda_{\max}(\bar{P}) -\lambda_{\min}(Q)}{\lambda_{\max}(\bar{P})},
    \label{Eq_delta}
\end{equation}
where due to positive definiteness of $\bar{P}$ and $Q$ we have that $\delta \in (0,1)$. It now follows that
\begin{equation}
    \|\theta_{t+1|t} - \tilde{\theta}_{t+1|t}\|^2_{\bar{P}} \leq \delta \| \theta_{t|t-1} - \tilde \theta_{t|t-1} \|^2_{\bar{P}},
    \label{Eq_PbarContraction}
\end{equation}
such that every prediction-to-prediction mapping is strictly contracting in a common norm $\|\cdot\|^2_{\bar{P}}$ with at least strength of contraction $\delta \in (0,1)$. Therefore, for any $c \in (1, \frac{1}{\delta})$, we have
\begin{equation}
    \lim_{t \to \infty} c^t\|\theta_{t|t-1} - \tilde \theta_{t|t-1}\|_{\bar{P}}^2 \rightarrow 0,
\end{equation}
By norm equivalence it follows that this difference converges to 0 in any norm.

\subsection{Theorem \ref{Thm_GlobOpt}: Contraction to the NDR}
\label{App_ProofGlobOpt}
We write the first-order condition of the ISD update as follows
\begin{equation}
    P_t^{1/2}(\theta_{t|t}-\theta_{t|t-1}) = P_t^{-1/2}\nabla(y_t |\theta_{t|t}),
\end{equation}
where adding $P^{-1/2}_t \nabla(y_t|\theta^\star_{t}) - P_t^{1/2}\theta^\star_{t}$ to both sides and rearranging gives
\begin{equation}
    P_t^{1/2}(\theta_{t|t}-\theta^\star_t) +  P_t^{-1/2}(\nabla(y_t|\theta^\star_{t}) -\nabla(y_t |\theta_{t|t})) = P_t^{1/2}(\theta_{t|t-1}-\theta^\star_t) + P^{-1/2}_t \nabla(y_t|\theta^\star_{t}).
\end{equation}
Next, we write the difference in gradients, $\nabla(y_t|\theta^\star_{t}) -\nabla(y_t |\theta_{t|t})$, as the product of the negative average Hessian and the difference in two points. That is,
\begin{equation}
    (P_t^{1/2} + P_t^{-1/2}\mathcal{I}^{\star}_{t|t})(\theta_{t|t}-\theta^\star_t) = P_t^{1/2}(\theta_{t|t-1}-\theta^\star_t) + P^{-1/2}_t \nabla(y_t|\theta^\star_{t}),
    \label{Eq_CNDR_ISD}
\end{equation}
where $\mathcal{I}^{\star}_{t|t}\succeq \alpha_t I_K\succeq O_K$ is the negative average $K \times K$ Hessian between $\theta_{t|t}$ and $\theta_{t}^{\star}$:
\begin{equation}
    \mathcal{I}^{\star}_{t|t} :=  -\int_{0}^{1} \left.\frac{\partial^2  \log p(y_t|\theta)}{\partial\theta \partial\theta'}\right|_{\substack{\theta \,= \, u \,\theta_{t|t} \,+\, (1-u)\,\theta_t^{\star}}} \mathrm{d}u.
\end{equation}
Next, we consider the quadratic norm of \eqref{Eq_CNDR_ISD} and take the expectation over $y_t$ with respect to the DGP. Reordering and using that $\underset{y_t}{\mathbb{E}}[\nabla(y_t|\theta^\star_t)] = 0$ by Assumption \ref{A8_BoundedInformation} gives:
\begin{equation}
\underbrace{\underset{y_t}{\mathbb{E}}\left[\big\|\theta_{t|t} - \theta^\star_t\big\|_{P_t}^2 \right]}_\text{MSE after update} = \underbrace{\big\|\theta_{t|t-1} - \theta^\star_t\big\|_{P_t}^2}_\text{SE before update} 
-\underbrace{\underset{y_t}{\mathbb{E}}\left[\big\|\theta_{t|t} - \theta^\star_t\big\|_{2\mathcal{I}^{\star}_{t|t} + \mathcal{I}^{\star}_{t|t}P_t^{-1}\mathcal{I}^{\star}_{t|t}}^2 \right]}_\text{$\geq 0$, contractive force}
+\underbrace{\underset{y_t}{\mathbb{E}}\left[\big\|\nabla(y_t|\theta^\star_{t})\big\|_{P^{-1}_t}^2\right]}_\text{$\geq 0$, expansive force}.
\label{Eq_ConExpISD}
\end{equation}
The positivity of the contractive force is apparent from the positive semi-definiteness of $\mathcal{I}^{\star}_{t|t}$, which implies that also $2\mathcal{I}^{\star}_{t|t} + \mathcal{I}^{\star}_{t|t}P_t^{-1}\mathcal{I}^{\star}_{t|t}$ is positive semi-definite.

For the second result, we write the ESD update as follows
\begin{equation}
    P_t^{1/2}(\theta_{t|t}^\textnormal{ex}-\theta_{t|t-1}) = P_t^{-1/2}\nabla(y_t |\theta_{t|t-1}),
\end{equation}
where subtracting $P_t^{1/2}\theta^\star_{t}$ on both sides and taking the quadratic norm yields:
\begin{equation}
    \|\theta_{t|t}^\textnormal{ex} - \theta^{\star}_{t}\|^2_{P_t} = \|\theta_{t|t-1} - \theta^{\star}_{t}\|^2_{P_t} + 2\langle \theta_{t|t-1}-\theta^\star_t,  \nabla(y_t|\theta_{t|t-1}) \rangle + \|\nabla(y_t|\theta_{t|t-1})\|^2_{P_t^{-1}}.
\end{equation}
We now take the expectation over $y_t$ with respect to the DGP. Using that $\underset{y_t}{\mathbb{E}}[\nabla(y_t|\theta^\star_t)] = 0$ by Assumption~\ref{A8_BoundedInformation}, we also subtract $2\underset{y_t}{\mathbb{E}}[\langle \theta_{t|t-1}-\theta^\star_t,  \nabla(y_t|\theta^{\star}_{t}) \rangle] = 0$ from the right-hand side and write the difference in gradients, $\nabla(y_t |\theta_{t|t-1}) - \nabla(y_t|\theta^\star_{t})$, as the product of the negative average Hessian and the difference in two points. This gives
\begin{equation}
    \underbrace{\underset{y_t}{\mathbb{E}}\left[\big\|\theta_{t|t}^\textnormal{ex} - \theta^\star_t\big\|_{P_t}^2\right]}_\text{MSE after update} = \underbrace{\big\|\theta_{t|t-1} - \theta^\star_t\big\|_{P_t}^2}_\text{SE before update} -  \underbrace{\underset{y_t}{\mathbb{E}}\left[\big\|\theta_{t|t-1} - \theta^{\star}_{t}\big\|^2_{2 \mathcal{I}^{\star}_{t|t-1} }\right]}_\text{$\geq 0$, contractive force} +  \underbrace{\underset{y_t}{\mathbb{E}}\left[\big\|\nabla(y_t|\theta_{t|t-1})\big\|^2_{P_t^{-1}}\right]}_\text{$\geq 0$, expansive force}.
    \label{Eq_ConExpESD}
\end{equation}
where the positivity of the contractive force follows from the positive semi-definiteness of $\mathcal{I}^{\star}_{t|t-1}\succeq \alpha_t I_K\succeq O_K$, which is the negative average $K \times K$ Hessian between $\theta_{t|t-1}$ and $\theta_{t}^{\star}$:
\begin{equation}
    \mathcal{I}^{\star}_{t|t-1} :=  -\int_{0}^{1} \left.\frac{\partial^2  \log p(y_t|\theta)}{\partial\theta \partial\theta'}\right|_{\substack{\theta \,= \, u \,\theta_{t|t-1} \,+\, (1-u)\,\theta_t^{\star}}} \mathrm{d}u.
\end{equation}

\subsection{Corollary \ref{Cor_GlobOpt}: Geometric contraction to the NDR}

Equation~\eqref{Eq_ConExpISD} in the proof of Theorem~\ref{Thm_GlobOpt} says
\begin{equation}
\underbrace{\underset{y_t}{\mathbb{E}}\left[\big\|\theta_{t|t} - \theta^\star_t\big\|_{P_t}^2 \right]}_\text{MSE after update} = \underbrace{\big\|\theta_{t|t-1} - \theta^\star_t\big\|_{P_t}^2}_\text{SE before update} 
-\underbrace{\underset{y_t}{\mathbb{E}}\left[\big\|\theta_{t|t} - \theta^\star_t\big\|_{2\mathcal{I}^{\star}_{t|t} + \mathcal{I}^{\star}_{t|t}P_t^{-1}\mathcal{I}^{\star}_{t|t}}^2 \right]}_\text{$\geq 0$, contractive force}
+\underbrace{\underset{y_t}{\mathbb{E}}\left[\big\|\nabla(y_t|\theta^\star_{t})\big\|_{P^{-1}_t}^2\right]}_\text{$\geq 0$, expansive force},
\end{equation}
where, as in Theorem~\ref{Thm_GlobOpt}, the negative average Hessian between  $\theta_t^\star$ and $\theta_{t|t}$ is defined as
\begin{equation}
   \mathcal{I}^{\star}_{t|t} :=  -\int_{0}^{1} \left.\frac{\partial^2  \log p(y_t|\theta)}{\partial\theta \partial\theta'}\right|_{\substack{\theta \,= \, u \,\theta_{t|t} \,+\, (1-u)\,\theta_t^{\star}}} \mathrm{d}u. 
\end{equation}
The equality above can equivalently be written as
\begin{equation}
\underset{y_t}{\mathbb{E}} \left[ \big\| \theta_{t|t}-\theta^\star_{t}\big\|_{P_t+2\mathcal{I}^\star_{t|t}+\mathcal{I}^\star_{t|t}P_t^{-1}\mathcal{I}^\star_{t|t}}^2\right] =\big\| \theta_{t|t-1}-\theta^\star_{t}\big\|_{P_t}^2+\underset{y_t}{\mathbb{E}}  \left[\big \| \nabla(y_t|\theta_t^\star)\big\|^2_{P_t^{-1}}\right].\label{to use again 2}
\end{equation}

Next, using $\mathcal{I}^\star_{t|t}\succeq \alpha_t I_K$ with $\alpha_t\geq 0$ and $P_t\succ O_{K}$, the term on the left-hand side can be bounded below by writing
\begin{align}
     & \| \theta_{t|t}-\theta^\star_{t}\|_{P_t+2\mathcal{I}^\star_{t|t}+\mathcal{I}^\star_{t|t}P_{t}^{-1}\mathcal{I}^\star_{t|t}}^2
      \\
   & =\|\theta_{t|t} - \theta^\star_{t}\|^2_{P_t} + 2\|\theta_{t|t} - \theta^\star_{t}\|^2_{\mathcal{I}^\star_{t|t}} + \|\mathcal{I}^\star_{t|t}(\theta_{t|t} - \theta^\star_{t})\|^2_{P_t^{-1}}\\ 
      &\geq \|\theta_{t|t} - \theta^\star_{t}\|^2_{P_t} + 2\lambda_{\min}(\mathcal{I}^\star_{t|t})\|\theta_{t|t} - \theta^\star_{t}\|^2 + 
      \lambda_{\min}(P_t^{-1})\| \theta_{t|t} - \theta^\star_{t}\|_{(\mathcal{I}^\star_{t|t})^2}^2
      \\
      &\geq \|\theta_{t|t} - \theta^\star_{t}\|^2_{P_t} + 2\alpha_t\|\theta_{t|t} - \theta^\star_{t}\|^2 + \frac{\lambda_{\min}((\mathcal{I}^\star_{t|t})^2)}{\lambda_{\max}(P_t)}\| \theta_{t|t} - \theta^\star_{t}\|^2\\
     &\geq \|\theta_{t|t} - \theta^\star_{t}\|^2_{P_t} + \left(2\alpha_t + \frac{\alpha_t^2}{\lambda_{\max}(P_t)}\right)\|\theta_{t|t} - \theta^\star_{t}\|^2
     \\   
      &\geq  \left(1 + \frac{2\alpha_t}{\lambda_{\max}(P_t)} + \frac{\alpha_t^2}{\lambda_{\max}(P_t)^2}\right) \|\theta_{t|t} - \theta^\star_{t}\|^2_{P_t}
      \\
      &= \left( \frac{\lambda_{\max}(P_t) + \alpha_t}{\lambda_{\max}(P_t)}\right)^2 \|\theta_{t|t} - \theta^\star_{t}\|^2_{P_t}.   \label{to use again} 
\end{align}
Here, we have used the same arguments as in equations~\eqref{Eq_StepSizeShrinkageFirst}-\eqref{Eq_FinStepShrink}. Combining~\eqref{to use again} with~\eqref{to use again 2} yields
\begin{equation}
 \left( \frac{\lambda_{\max}(P_t) + \alpha_t}{\lambda_{\max}(P_t)}\right)^2 \underset{y_t}{\mathbb{E}}\left[ \big\|\theta_{t|t} - \theta^\star_{t}\big\|^2_{P_t}\right]\;  \leq \; \big\| \theta_{t|t-1}-\theta^\star_{t}\big\|_{P_t}^2+ \underset{y_t}{\mathbb{E}}\left[ \big\| \nabla(y_t|\theta_t^\star)\big\|^2_{P_t^{-1}}\right].
\end{equation}
Multiplying both sides by the square of $\lambda_{\max}(P_t)/(\lambda_{\max}(P_t)+\alpha_t)$ yields the final result:
\begin{equation}
\underset{y_t}{\mathbb{E}}\left[\big\|\theta_{t|t} - \theta^\star_t\big\|_{P_t}^2\right] \leq \underbrace{\left(\frac{\lambda_{\max}(P_t)}{\lambda_{\max}(P_t)+ \alpha_t}\right)^2}_\text{$\in [0,1)$, contraction coefficient} \left(\big\|\theta_{t|t-1} - \theta^\star_t\big\|_{P_t}^2 + \underset{y_t}{\mathbb{E}}\left[\big\|\nabla(y_t|\theta^\star_{t})\big\|_{P^{-1}_t}^2\right]\right),
\end{equation}
where the contraction coefficient falls within the interval $[0,1)$ using $\alpha_t > 0$ and $P_t \succ O_K$.

\setcounter{equation}{0}
\section{Further theoretical results}
\label{appendix B}

\subsection{Stability without log-concavity for Student's $t$}
\label{App_StabStudt}

\textbf{Summary.} Here we give a DGP-agnostic stability result for the ISD filter using a postulated density that fails to be log-concave. Specifically, we consider the Student's $t$ distribution with unknown location parameter $\mu_t$, scale parameter $\sigma>0$, and degrees of freedom $\nu > 0$. To ensure that Assumption~\ref{A2_RL_StrictConcave} holds, we ensure that the penalty parameter $P=H^{-1}>0$ is sufficiently large; specifically, as we show below, $P>(\nu+1)/(8\nu\sigma^2)$ is sufficient. Under this condition, the prediction-to-prediction mapping is contractive if the following DGP-agnostic sufficient condition holds:
\begin{equation}
    |\Phi| < 1 - H \frac{\nu+1}{8\nu \, \sigma^2} = \frac{P - \frac{\nu+1}{8\nu \, \sigma^2}}{P}.
    \label{Eq_tphibound}
\end{equation}
The proof of this claim proceeds in two parts. First, we consider Hessian bounds leading to the condition $P>(\nu+1)/(8\nu\sigma^2)$. Second, we consider the prediction-to-prediction mapping, which leads to~\eqref{Eq_tphibound}.

\textbf{Hessian bounds.} Here we show that, for the Student's $t$ density with level $\mu$, the Hessian of the logarithmic density with respect to $\mu$ satisfies
\begin{equation}
\label{Eq_tHesbounds}
    \frac{-(\nu+1)}{\nu \, \sigma^2} \leq \mathcal{H}(y_t|\mu,\sigma,\nu) \leq \frac{\nu+1}{8\nu \, \sigma^2}.
\end{equation}
The positivity of the upper bound means that the  Student's $t$ distribution fails to be log-concave in terms of the location parameter $\mu$. We note that these bounds are identical to those in \citet[Sec.\ 5]{blasques2022maximum} up to the scaling factor $S_t = \sigma^2/(1 + \nu^{-1}) = \frac{\nu \, \sigma^2}{\nu+1}$. To ensure that the ISD objective function~\eqref{RegLike} is strictly concave (Assumption~\ref{A2_RL_StrictConcave}), it is sufficient that the penalty exceeds the non-concavity of the log density. That is, we impose $P = H^{-1}> \mathcal{H}(y_t|\mu,\sigma,\nu)$, or equivalently, $H \mathcal{H}(y_t|\mu,\sigma,\nu) < 1$.

To prove~\eqref{Eq_tHesbounds}, we note that the score is
\begin{align}
    \nabla(y_t|\mu,\sigma,\nu) &= \frac{\partial}{\partial\mu} \log p(y_t|\mu,\sigma,\nu) \\
    &=  \frac{\partial}{\partial\mu} \left(-\frac{\nu+1}{2}\log\left(1+\frac{(y_t - \mu)^2}{\nu \,\sigma^2}   \right) + c\right) 
    \\
    &= \frac{\nu+1}{\nu \, \sigma^2}\frac{y_t - \mu}{1+\frac{(y_t - \mu)^2}{\nu \,\sigma^2}},\label{students t score}
\end{align}
where $c$ is a constant that does not depend on $\mu$. Taking another derivative with respect to $\mu$ yields the following Hessian:
\begin{align}
    \mathcal{H}(y_t|\mu,\sigma,\nu) &= \frac{\partial}{\partial\mu} \nabla(y_t|\mu,\sigma,\nu) \\
    &=  \frac{\partial}{\partial\mu} \frac{\nu+1}{\nu \, \sigma^2}\frac{y_t - \mu}{1+\frac{(y_t - \mu)^2}{\nu \,\sigma^2}}\\
    &= \frac{\nu+1}{\nu \, \sigma^2} \frac{-(1+\frac{(y_t - \mu)^2}{\nu \, \sigma^2}) - (y_t - \mu)\frac{-2(y_t-\mu)}{\nu \sigma^2}}{(1+\frac{(y_t - \mu)^2}{\nu \, \sigma^2})^2}\\
    &= \frac{\nu+1}{\nu \, \sigma^2} \frac{\frac{(y_t - \mu)^2}{\nu \, \sigma^2} - 1}{(1+\frac{(y_t - \mu)^2}{\nu \, \sigma^2})^2},
\end{align}
where the penultimate line uses the quotient rule for derivatives. To bound the Hessian, we use the substitution $x = \frac{y_t - \mu}{\sqrt{\nu} \,\sigma}$ such that $\mathcal{H}(y_t|\mu,\sigma,\nu)  =  \frac{\nu+1}{\nu \, \sigma^2} \frac{x^2 - 1}{(1+x^2)^2}$. Maximizing the Hessian with respect to $x$ can be done by solving the first-order condition:
\begin{align}
        0 &= \frac{\partial}{\partial x}\frac{\nu+1}{\nu \, \sigma^2} \frac{x^2 - 1}{(1+x^2)^2} \\
        &= \frac{\nu+1}{\nu \, \sigma^2} \frac{2x(1+x^2)^2 - 4x(x^2 -1)(1+x^2)}{(1+x^2)^4}\\
        &= \frac{\nu+1}{\nu \, \sigma^2} \frac{2x(1+x^2) - 4x(x^2 -1)}{(1+x^2)^3} \\
        &= \frac{\nu+1}{\nu \, \sigma^2} \frac{2x + 2x^3 - 4x^3 + 4x}{(1+x^2)^3}\\
        &= \frac{\nu+1}{\nu \, \sigma^2} \frac{6x - 2x^3}{(1+x^2)^3}\\ 
        &= \frac{\nu+1}{\nu \, \sigma^2} \frac{2x(3-x^2)}{(1 + x^2)^3},
\end{align}
where the second line uses again the quotient rule for derivatives and the third line multiplies by $\frac{(1+x^2)^{-1}}{(1+x^2)^{-1}} = 1$. The final line shows that the solutions are $x = 0$ and $x = \pm \sqrt{3}$. Filling in these stationary points, we have that $x=0$ yields $\frac{\nu+1}{\nu \, \sigma^2} \frac{0^2 - 1}{(1+0^2)^2} = \frac{-(\nu+1)}{\nu \, \sigma^2} < 0$ and $x = \pm \sqrt{3}$ gives $\frac{\nu+1}{\nu \, \sigma^2} \frac{3 - 1}{(1+3)^2} = \frac{\nu+1}{8\nu \, \sigma^2} > 0$. In other words, $x=0$ is the minimizer and $x = \pm \sqrt{3}$ the maximizer. In sum, we obtain the Hessian bounds given in~\eqref{Eq_tHesbounds}.

\textbf{Prediction-to-prediction stability.} Here we investigate the stability of the prediction-to-prediction recursion under the assumption that $P>(\nu+1)/(8\nu\sigma^2)$. The difference between two updates $\mu_{t|t} - \tilde{\mu}_{t|t}$, based on predictions $\mu_{t|t-1}$, $\tilde{\mu}_{t|t-1} \in \mathbb{R}$ using the same ISD update step with learning rate $H = P^{-1} > 0$ and the same observation $y_t$, can be written as
\begin{align}
\label{Eq_tStab}
   \mu_{t|t} - \tilde{\mu}_{t|t} &= (\mu_{t|t-1} - \tilde{\mu}_{t|t-1}) + H[\nabla(y_t|\mu_{t|t},\sigma,\nu) -\nabla(y_t|\tilde{\mu}_{t|t},\sigma,\nu)]\\
    &= (\mu_{t|t-1} - \tilde{\mu}_{t|t-1}) + H\,\mathcal{H}(y_t|\mu_t^{\dagger},\sigma,\nu)(\mu_{t|t} - \tilde{\mu}_{t|t}), 
\end{align}
where $\mathcal{H}(y_t|\mu_t^{\dagger},\sigma,\nu):= \left. \frac{\partial^2  \log p(y_t|\mu,\sigma,\nu)}{(\partial\mu)^2}\right|_{\substack{\mu = \mu_t^{\dagger}}} = \left. \frac{\partial  \nabla(y_t|\mu,\sigma,\nu)}{\partial\mu}\right|_{\substack{\mu = \mu_t^{\dagger}}}$ is the Hessian of the logarithmic density evaluated at some midpoint $\mu_t^{\dagger} \in [\min\{\mu_{t|t},\tilde{\mu}_{t|t}  \},\max\{\mu_{t|t},\tilde{\mu}_{t|t}  \}]$ using the mean-value theorem. As $P > \frac{\nu+1}{8\nu \, \sigma^2}$ by assumption, we may write \eqref{Eq_tStab} as
\begin{equation}
   \mu_{t|t} - \tilde{\mu}_{t|t} = [1 - H\,\mathcal{H}(y_t|\mu_t^{\dagger},\sigma, \nu)]^{-1}(\mu_{t|t-1} - \tilde{\mu}_{t|t-1}).
   \label{Eq_tStab2}
\end{equation}
Using the prediction step, $\mu_{t+1|t} = \omega + \Phi \mu_{t|t}$, and squaring both sides yields:
\begin{align}
     (\mu_{t+1|t} - \tilde{\mu}_{t+1|t})^2 &= (\omega + \Phi \mu_{t|t} - \omega - \Phi \tilde{\mu}_{t|t})^2\\
     &=\Phi^2 (\mu_{t|t} - \tilde{\mu}_{t|t})^2\\
     &= \frac{\Phi^2}{[1 - H\,\mathcal{H}(y_t|\mu_t^{\dagger},\sigma, \nu)]^2}(\mu_{t|t-1} - \tilde{\mu}_{t|t-1})^2\\
     &\leq \frac{\Phi^2}{(1 - H \frac{\nu+1}{8\nu \, \sigma^2})^2}(\mu_{t|t-1} - \tilde{\mu}_{t|t-1})^2,
\end{align}
where the third line uses result \eqref{Eq_tStab2}, while the last line uses the  upper bound in \eqref{Eq_tHesbounds}. To obtain a contraction, a sufficient condition  reads:
\begin{equation}
\left| \frac{\Phi^2}{(1 - H \frac{\nu+1}{8\nu \, \sigma^2})^2}\right| < 1.
    \label{Eq_tphibound2}
\end{equation}
Note that $\frac{\nu+1}{8\nu \, \sigma^2}<P$ or, equivalently, $ \frac{\nu+1}{8\nu \, \sigma^2}H< 1$. By taking the square root, we see that condition~\eqref{Eq_tphibound2} is equivalent to condition~\eqref{Eq_tphibound}. 

As the penalty approaches its imposed lower bound to guarantee strict concavity of the regularized objective~\eqref{RegLike}, we have that the right-hand side shrinks to 0, i.e., $\lim_{P \downarrow \frac{\nu+1}{8\nu \, \sigma^2}}\frac{P - \frac{\nu+1}{8\nu \, \sigma^2}}{P} = 0$. In other words, for smaller $P$ associated with more aggressive updating, we find that the possible expansion of the update step grows. To compensate, we then require that the prediction step becomes more contractive, meaning that we need that $|\Phi| \rightarrow 0$. Conversely, as $P$ tends to infinity, we have that the right-hand side of \eqref{Eq_tphibound} tends to 1, i.e., $\lim_{P \uparrow \infty}\frac{P - \frac{\nu+1}{8\nu \, \sigma^2}}{P} = 1$. In this limiting case, we have that the update step approaches the identity mapping, which is neither contractive nor expansive; the classic condition $|\Phi| < 1$ then suffices to obtain a contraction. Equivalently, as $\Phi \rightarrow 1$, we require $P \rightarrow \infty$. This means that tracking more persistent states (requiring $\Phi$ closer to 1) necessitates slower updating ($H = P^{-1}$ closer to 0).

We conclude that concavity (Assumption~\ref{A5_Concavity}) is not a necessary condition for stability of the ISD filter. For the example above, the proof is fairly straightforward because the time-varying parameter is a scalar (thereby circumventing eigenvalue inequalities and allowing for a mid-point version of the mean-value theorem) and because the Hessian of the Student's $t$ distribution with respect to the location parameter admits a natural upper bound. We leave the general case for future research.

\subsection{Bounded cumulative numerical error of the ISD filter}
\label{App_NumErr}

For a given initialization $\theta_{0|0} \in \Theta$ and data set $\{y_t\}$, let $\{\theta_{t|t}\}$ and $\{\theta_{t|t-1}\}$ denote the updated and predicted parameter path, respectively, produced by recursively applying the ISD update~\eqref{ProParUpdate} and prediction step~\eqref{GASPred}. In addition, let $\{\tilde{\theta}_{t|t}\}$ and $\{\tilde{\theta}_{t|t-1}\}$ denote perturbed parameter paths based on the same initialization $\theta_{0|0}$ and data $\{y_t\}$, but exposed to numerical errors at each update step $t$: $\gamma_{t|t} := \tilde{\theta}_{t|t} - \bar{\theta}_{t|t}$, where $\bar{\theta}_{t|t} := \underset{\theta \in \Theta}{\mathrm{argmax}} \; \left\{ \log p(y_t|\theta) \, -\,  \frac{1}{2}\big\|\theta - \tilde{\theta}_{t|t-1}\big\|_{P_t}^2 \right\}$ is the ISD updated parameter based on prediction $\tilde{\theta}_{t|t-1}$ without numerical error. We assume that the numerical errors $\{\gamma_{t|t}\}$ are bounded, that is, $\|\gamma_{t|t}\| \leq \epsilon < \infty , \forall t\geq 1$, and that the conditions of Theorem~\ref{Thm_Invertability} (filter invertibility) are satisfied. The latter implies the existence of a common norm $\|\cdot\|_{\bar{P}}$ for some positive definite $K \times K$ matrix $\bar{P}$ in which the filter is contractive at each point in time with at least contraction strength $\sqrt{\delta} < 1$, see \eqref{Eq_delta}--\eqref{Eq_PbarContraction} for details. To focus on the effect of numerical error in the update step (which may require numerical optimization methods), we assume here, for simplicity, that there are no numerical errors in the linear prediction step \eqref{GASPred}. Also taking into account such numerical errors is nevertheless straightforward and would produce a comparable end result.

Using the definition of $\gamma_{t|t}$, we may write for $t\geq 1$:
\begin{align}
    \| \theta_{t+1|t} - \tilde{\theta}_{t+1|t} \|_{\bar{P}} &= \| \theta_{t+1|t} - \bar{\theta}_{t+1|t} - \Phi \gamma_{t|t} \|_{\bar{P}} \\
    &\leq \| \theta_{t+1|t} - \bar{\theta}_{t+1|t}\|_{\bar{P}} + \| \Phi \gamma_{t|t} \|_{\bar{P}}\\
    &\leq \sqrt{\delta} \| \theta_{t|t-1} - \tilde{\theta}_{t|t-1}\|_{\bar{P}} +  \|\gamma_{t|t} \|_{\Phi'\bar{P}\Phi}\\
    &\leq \sqrt{\delta} \| \theta_{t|t-1} - \tilde{\theta}_{t|t-1}\|_{\bar{P}}+  \sqrt{\lambda_{\max}(\Phi'\bar{P}\Phi)} \epsilon,
    \label{Eq_NumErrRecursion}
\end{align}
where the first line uses the definition of the prediction step and the numerical error $\gamma_{t|t}$, that is, $\tilde{\theta}_{t+1|t} = \omega + \Phi \tilde{\theta}_{t|t} = \omega + \Phi (\bar{\theta}_{t|t} + \gamma_{t|t}) = \bar{\theta}_{t+1|t} + \Phi \gamma_{t|t}$, where $\bar{\theta}_{t+1|t}: = \omega + \Phi \bar{\theta}_{t|t}$ the prediction based on the updated parameter $\bar{\theta}_{t|t}$. The second line uses the triangle inequality. Furthermore, the third line uses the stability result that says that applying the same ISD prediction-to-prediction mapping to any two predictions (here $\theta_{t|t-1}$ and $\tilde{\theta}_{t|t-1}$) yielding two new predictions for the next period (here $\theta_{t+1|t}$ and $\bar{\theta}_{t+1|t}$) produces a contraction with at least strength $\sqrt{\delta} < 1$ in the norm $\|\cdot\|_{\bar{P}}$, see again \eqref{Eq_delta}--\eqref{Eq_PbarContraction} for details. The final line uses that $\|\cdot\|_{A} \leq \sqrt{\lambda_{\max}(A)}\| \cdot \|$ for any positive definite matrix $A$ and that $\|\gamma_{t|t}\| \leq \epsilon$.

Repeated application of the recursive relationship \eqref{Eq_NumErrRecursion} gives
\begin{align}
    \label{Eq_FinGeoSumNumErr1}
    \| \theta_{t+1|t} - \tilde{\theta}_{t+1|t} \|_{\bar{P}} &\leq  \left(\sqrt{\delta}\right)^t \| \theta_{1|0} - \tilde{\theta}_{1|0}\|_{\bar{P}} + \sum_{i=0}^{t-1}  \left(\sqrt{\delta}\right)^{i} \sqrt{\lambda_{\max}(\Phi'\bar{P}\Phi)} \epsilon\\
    &= \frac{1 -  \left(\sqrt{\delta}\right)^{t}}{1 - \sqrt{\delta}}\sqrt{\lambda_{\max}(\Phi'\bar{P}\Phi)} \epsilon,
    \label{Eq_FinGeoSumNumErr2}
\end{align}
where the second line uses that both the path without numerical errors ($\{\theta_{t|t-1}\}$) and the one with numerical errors ($\{\tilde{\theta}_{t|t-1}\}$) share the same initialization $\theta_{0|0}$ and prediction mapping such that $\theta_{1|0} = \tilde{\theta}_{1|0}$; this means that the first term in \eqref{Eq_FinGeoSumNumErr1} disappears. A standard result for the finite sum of a contractive geometric series is applied to the other term.

Converting to the Euclidean norm and taking the limit of $t$ to infinity, we obtain the following bound for the cumulative numerical error:
\begin{align}
    \limsup_{t \to \infty} \| \theta_{t+1|t} - \tilde{\theta}_{t+1|t} \| &\leq \sqrt{\frac{1}{\lambda_{\min}(\bar{P})}} \limsup_{t \to \infty} \| \theta_{t+1|t} - \tilde{\theta}_{t+1|t} \|_{\bar{P}}\\
    &\leq \sqrt{\frac{{\lambda_{\max}(\Phi'\bar{P}\Phi)}}{\lambda_{\min}(\bar{P})}}\frac{\epsilon}{1 - \sqrt{\delta}},
    \label{Eq_InfGeoSumNummErr}
\end{align}
where the first line uses that $\|\cdot\| \leq \sqrt{ (1/\lambda_{\min}(A))}\|\cdot\|_{A}$ for any positive definite matrix $A$ and the second line uses the result from \eqref{Eq_FinGeoSumNumErr2} with $t \to \infty$. The final expression in \eqref{Eq_InfGeoSumNummErr} demonstrates that the maximum cumulative numerical error is bounded and a linear function of the maximum error for a single ISD update step, $\epsilon$. In other words, numerical errors do not accumulate endlessly, and there is no risk of divergence as the effect of past errors is limited. Specifically, the effect of past numerical errors decays geometrically with strength $\sqrt{\delta} < 1$, producing a convergent geometric sum. As a result, the bound in \eqref{Eq_InfGeoSumNummErr} scales hyperbolically with the decay rate $\sqrt{\delta}$. We also observe that the cumulative bound is increasing in $\Phi'\Phi$ (directly and via $\sqrt{\delta}$), reflecting that more persistent dynamics lead to less decay of past errors and thus a larger cumulative error bound. The dependence of the bound on the eigenvalues of $\bar{P}$ simply stems from the conversions between the Euclidean norm $\|\cdot\|$ and the weighted norm $\|\cdot\|_{\bar{P}}$. These insights can be used to guide an appropriate choice for the numerical error tolerance of a single ISD update, $\epsilon$, and underscore the importance of filter invertibility for numerical stability.

\subsection{Theorem~\ref{Thm_Invertability} implies Lyapunov stability}
\label{App Lyapunov}

To connect Theorem~\ref{Thm_Invertability} with the concept of Lyapunov stability, define the Lyapunov function $V(x) = \|x \|^2_{\bar{P}} = x'\bar{P}x$ and the path difference $\varepsilon_{t+1} := \theta_{t+1|t} - \tilde{\theta}_{t+1|t}$. Next, we refer to the proof of Theorem~\ref{Thm_Invertability} and write equation~\eqref{Eq_PbarContraction} in terms of $V(\cdot)$ and $\varepsilon_t$ as
\begin{equation}
\label{Eq_V}
V(\varepsilon_{t+1}) \leq \delta V(\varepsilon_{t}).
\end{equation}
This shows that $\Delta V(\varepsilon_{t+1}) := V(\varepsilon_{t+1}) -  V(\varepsilon_{t}) \leq (\delta - 1) V(\varepsilon_{t}) \leq -q\,\varepsilon_{t}^2 \leq 0$, with $q:= (1-\delta)\lambda_{\min}(\bar{P}) > 0$, which is a nonlinear discrete Lyapunov inequality where the left-hand side is the time derivative. Because the right-hand side is strictly negative, except for $\varepsilon_{t} = 0_K$, the system is asymptotically stable with unique equilibrium $\varepsilon_{t} = 0_K$ (\citealp[Thm.\ 5.5]{bof2018lyapunov}), which occurs at $\theta_{t|t-1} = \tilde \theta_{t|t-1}$. The fact that this convergence is exponential can be directly deduced from \eqref{Eq_V}, or indirectly, by noting that $\Delta V(\varepsilon_{t+1}) \leq -\tilde{\delta} V(\varepsilon_{t})$ with $\tilde{\delta} = (1-\delta) > 0$, is a sufficient condition, which mirrors the classical condition $\dot V \leq - \tilde{\delta} V$ in continuous time (\citealp[Thm.\ 20.3]{poznak2008advanced}).

\subsection{MSE reductions without log-concavity for Student's $t$}
\label{App MSE nonconcave}

\textbf{Summary.} When log-concavity fails, MSE improvements for inaccurate predictions may be merely constant rather than proportional to the squared prediction error. To show this, we consider the case in which the researcher postulates a (possibly misspecified) Student's~$t$ density with unknown location $\mu$. As we show below, the score is $\mathcal{O}(-1/\mu)$ and the resulting MSE improvement approaches a constant as $|\mu|\to\infty$. If the true density is Cauchy with location zero, the MSE after updating can be computed in closed form for any $\mu$. This closed-form expression is used to draw the MSE curve in Figure~\ref{fig_NDR_new}(b). This example shows that, without additional information, better-than-constant MSE reductions cannot generally be obtained without log-concavity. This fact holds for both ISD and ESD updates.

\textbf{Score limits.} To align with the literature (e.g., \citealp[Sec.\ 5]{blasques2022maximum}), we take $\nu\sigma^2/(\nu+1)$ times the logarithmic density of the Student's $t$ distribution, where $\nu\sigma^2/(\nu+1)$  can be viewed as a scaling factor. The score based on the \emph{scaled} log-likelihood contribution then reads
\begin{align}
\nabla(y_t|\mu) &:= \frac{\partial}{\partial\mu}\left[ \frac{\nu \sigma^2}{\nu+1} \log p(y_t|\mu,\sigma,\nu) \right]
\\
&= \frac{\nu \sigma^2}{\nu+1} \frac{\partial}{\partial\mu} \left[-\frac{\nu+1}{2}\log\left(1+\frac{(y_t - \mu)^2}{\nu \,\sigma^2}   \right)\right]
\\  
&= \frac{-\nu \sigma^2}{2} \frac{\partial}{\partial\mu} \log\left(1+\frac{(y_t - \mu)^2}{\nu \,\sigma^2}   \right)
\\  
    &= \frac{y_t - \mu}{1+\frac{(y_t - \mu)^2}{\nu \,\sigma^2}},
\end{align}
where $p(y_t|\mu,\sigma,\nu)$ is the density function of a Student's $t$ distribution with location $\mu$, scale $\sigma>0$, and degrees of freedom $\nu>0$. For $|\mu|$ much larger than $|y_t|$, the scaled score becomes
\begin{align}
\nabla(y_t|\mu)&=\frac{y_t - \mu}{1+\frac{(y_t - \mu)^2}{\nu \,\sigma^2}}
\\
&=  \frac{(y_t-\mu)^{2}}{\nu\sigma^2+(y_t-\mu)^{2}} \,  \frac{\nu\sigma^2}{y_t-\mu} \; =\; \mathcal{O}\left(\frac{-\nu\sigma^2}{\mu}\right) .
\end{align}
Next, we consider the expectation over $y_t$ (using the true density) of (a) $\mu$ times the score and (b) the squared score. As $|\mu|$ becomes large, we claim that
\begin{equation}
\label{limit}
\lim_{|\mu| \to \infty}\; \underset{y_t}{\mathbb{E}}[\,\mu\, \nabla(y_t|\mu)\,] = -\nu\sigma^2, \qquad \lim_{|\mu|\to \infty}\;  \underset{y_t}{\mathbb{E}}[\,(\nabla(y_t|\mu))^2\,]=0,
\end{equation}
provided the limit and the expectation can be exchanged. This is justified by Lebesgue's dominated-convergence theorem, which requires the existence of a dominating function of $y_t$, independent of $\mu$, with finite expectation. For $\mu \nabla(y_t| \mu)$, it is straightforward to verify that a dominating function can be found, which grows at most linearly in $|y_t|$:
\begin{align}
    |\mu\nabla(y_t|\mu)|&=\left|\mu\frac{y_t - \mu}{1+\frac{(y_t - \mu)^2}{\nu \,\sigma^2}}\right|
    \\&=\left|(\mu-y_t)\frac{y_t - \mu}{1+\frac{(y_t - \mu)^2}{\nu \,\sigma^2}}+y_t \frac{y_t - \mu}{1+\frac{(y_t - \mu)^2}{\nu \,\sigma^2}}\right|
    \\
    &\leq \left|\nu\sigma^2\frac{(y_t - \mu)^2}{\nu \,\sigma^2+(y_t - \mu)^2}\right|  + \left|y_t \frac{y_t - \mu}{1+\frac{(y_t - \mu)^2}{\nu \,\sigma^2}}\right| \\
    &\leq \nu\sigma^2 + C\, |y_t|,
\end{align}
where the third line uses the triangle inequality $|a + b| \leq |a| + |b|$ for $a,b \in \mathbb{R}$ and the final line that $|\nabla(y_t|\mu)|\leq C$ for some constant $C<\infty$. In fact, it can be shown that $C=\sqrt{\nu}\sigma/2$ works, because we can use the inequality  $|z|/(1+z^2) \leq 1/2, \forall z\in \mathbb{R}$ with the specific value $z=(y_t-\mu)/(\sqrt{\nu}\sigma)$. Thus, for the first limit in~\eqref{limit} to hold, it is sufficient that the true law satisfies $\mathbb{E}[|y_t|]<\infty$. For the second limit in~\eqref{limit}, it is enough to note that $|\nabla(y_t|\mu)|^2\leq C^2$; in this case, dominated convergence applies immediately.

\textbf{MSE improvement without log-concavity.}  To see why the limits in~\eqref{limit} imply a constant MSE improvement for inaccurate predictions, focus on the ESD update with learning rate $H>0$, for which
\begin{alignat}{3}
\mu^\text{ex}_{t|t}-\mu_t^\star
&= \mu_{t|t-1}-\mu_t^\star+H\nabla(y_t|\mu_{t|t-1}),
\end{alignat}
where $\mu_t^\star$ is the (pseudo-)true location parameter. Squaring both sides yields
\begin{alignat}{2}
    (\mu^\text{ex}_{t|t}-\mu_t^\star)^2 &= (\mu_{t|t-1}-\mu^\star_{t})^2 +2H(\mu_{t|t-1}-\mu^\star_{t})\, \nabla(y_t|\mu_{t|t-1}) + H^2\,\big( \nabla(y_t|\mu_{t|t-1})\big)^2.
\end{alignat}
Without loss of generality, consider the case $\mu_t^\star=0$. Take an expectation over $y_t$, conditional on the true density, to yield
\begin{alignat}{2}
 \underset{y_t}{\mathbb{E}} \left[  \big(\mu^\text{ex}_{t|t}\big)^2  \right] &=  (\mu_{t|t-1} )^2 +2 H\underbrace{\underset{y_t}{\mathbb{E}} \left[ \mu_{t|t-1} \nabla(y_t|\mu_{t|t-1}) \right]}_{\text{approaches }-\nu\sigma^2 \text{ as }|\mu_{t|t-1}|\to \infty}
+ H^2 \underbrace{\underset{y_t}{\mathbb{E}} \left[ \big( \nabla(y_t|\mu_{t|t-1}) \big)^2 \right]}_{\text{approaches }0\text{ as }|\mu_{t|t-1}|\to \infty}.
\label{MSE constant reduction}
\end{alignat}
Letting $|\mu_{t\mid t-1}|$ tend to infinity in~\eqref{MSE constant reduction} and using the limits in~\eqref{limit}, we obtain that for highly inaccurate predictions the MSE decrease is at most $2H\nu\sigma^2$. Hence, the improvement is bounded by a constant rather than growing proportionally to the squared prediction error. Note that this result is the same as for the case $\mu_t^\star\neq0$ because $\underset{y_t}{\mathbb{E}} \left[ \mu_t^\star \nabla(y_t|\mu_{t|t-1}) \right]$ approaches 0 as $|\mu_{t|t-1}|\to \infty$ using dominated convergence with  $|\nabla(y_t|\mu)|\leq C$.

Above arguments rely on the existence of $\mathbb{E}[|y_t|]$ for dominated convergence to apply. However, as we will see below, the same conclusion continues to hold even if $y_t$ is drawn from a Cauchy distribution, which has no moments at all.

\textbf{MSE improvement under Cauchy data.}  
If the true distribution is Cauchy with location zero, the expectations in~\eqref{MSE constant reduction} can be computed in closed form.

\begin{lemma}[Student's $t$ score moments under Cauchy data] Let $\mu\in \mathbb{R}$, $\sigma,\nu,\gamma>0$ and define $\xi:=\sqrt{\nu}\sigma+\gamma$.  Then
\label{lemma integrals} 
\begin{align}
    \int_{-\infty}^\infty \frac{y - \mu}{1+\frac{(y - \mu)^2}{\nu \,\sigma^2}}\; \frac{1}{\pi \gamma}\;\frac{1}{1+y^2/\gamma^2}\; \mathrm{d}y\; &= \; \frac{-\mu\, \nu\,\sigma^2}{\mu^2+\xi^2},
    \\
        \int_{-\infty}^\infty \left(\frac{y - \mu}{1+\frac{(y - \mu)^2}{\nu \,\sigma^2}}\right)^2\; \frac{1}{\pi \gamma}\;\frac{1}{1+y^2/\gamma^2}\; \mathrm{d}y \; &=\frac{\nu^{3/2}\,\sigma^3}{2(\mu^2+\xi^2)^2}
\Bigl(\sqrt{\nu}\sigma \mu^2 +\xi \mu^2+\gamma \xi^2\Bigr).
\end{align}
As $|\mu|\to \infty$, these quantities are of order $\mathcal{O}(-1/\mu)$ and $\mathcal{O}(1/\mu^2)$, respectively. 
\end{lemma}

Lemma~\ref{lemma integrals}, whose proof is available on request, shows that the expected score is $\mathcal{O}(-\nu\sigma^2/\mu)$, in line with the analysis above based on Lebesgue's dominated convergence. 

Hence, if the true distribution is Cauchy with location zero and scale $\gamma>0$, while the researcher postulates a Student's $t$ density with location $\mu_{t\mid t-1}$, scale $\sigma>0$, and degrees of freedom $\nu>0$, then the post-update MSE can be computed in closed form by substituting the expressions in Lemma~\ref{lemma integrals} into equation~\eqref{MSE constant reduction}. The resulting post-update MSE is a function of $H$, $\sigma$, $\nu$, $\gamma$, and the squared prediction error. The associated MSE curve is plotted in Figure~\ref{fig_NDR_new}(b), with $H=1$, $\gamma=1$, $\sigma=1.3$, and $\nu=2$. The combination  of a true Cauchy with a misspecified Student's $t$ distribution produces a scenario in which MSE improvements for inaccurate predictions are merely constant. In general, better behavior may be possible. For illustrative purposes, therefore, we shade the area below the curve in Figure~\ref{fig_NDR_new}(b).

Finally, for large penalty parameters $P$ (or, equivalently, small learning rates $H$), the ISD update for the Student's $t$ location model can be made to closely resemble the ESD update; hence, in the general case without concavity of the logarithmic density, we should likewise not expect MSE decreases larger than a constant.

\subsection{Contraction to noise-dominated region for the ESD update}
\label{App_NDRESD}

To obtain also a noise-dominated region contraction for the ESD update, we may write $\nabla(y_t|\theta_{t|t-1}) = \nabla(y_t|\theta^{\star}_{t}) -\mathcal{I}^{\star}_{t|t-1}(\theta_{t|t-1} - \theta^{\star}_{t})$ and substitute this in the final term on the right-hand side of~\eqref{Eq_Opt_ESD}. Expanding this term yields a cross-product that is hard to assess. However, we can use the general albeit loose bound $\|a + b\|_{P_t^{-1}}^2 \leq 2\|a\|_{P_t^{-1}}^2 + 2\|b\|_{P_t^{-1}}^2$ for arbitrary $K \times 1$ vectors $a,b$. We note that more generally, one could apply Young's inequality to the cross-term $2\langle P_t^{-1}a,b\rangle$ and fine-tune the exponent choice; the above is essentially a special case with exponent 2. Using $a = \nabla(y_t|\theta^{\star}_{t})$ and $b = -\mathcal{I}^{\star}_{t|t-1}(\theta_{t|t-1} - \theta^{\star}_{t})$, we obtain
\allowdisplaybreaks
\begin{align}
    \underset{y_t}{\mathbb{E}}\left[\big\|\nabla(y_t|\theta_{t|t-1})\big\|^2_{P_t^{-1}}\right] &=  \underset{y_t}{\mathbb{E}}\left[\big\| \nabla(y_t|\theta^{\star}_{t}) -\mathcal{I}^{\star}_{t|t-1}(\theta_{t|t-1} - \theta^{\star}_{t})\big\|^2_{P_t^{-1}}\right] \\
    &\leq 2\underset{y_t}{\mathbb{E}}\left[\big\|\nabla(y_t|\theta^{\star}_{t})\big\|^2_{P_t^{-1}}\right] + 2\underset{y_t}{\mathbb{E}}\left[\big\|\mathcal{I}^{\star}_{t|t-1}(\theta_{t|t-1} - \theta^{\star}_{t})\big\|^2_{P_t^{-1}}\right]\\
       &=2\underset{y_t}{\mathbb{E}}\left[\big\|\nabla(y_t|\theta^{\star}_{t})\big\|^2_{P_t^{-1}}\right] + 2\underset{y_t}{\mathbb{E}}\left[\big\|\theta_{t|t-1} - \theta^{\star}_{t}\big\|^2_{\mathcal{I}^{\star}_{t|t-1}P_t^{-1}\mathcal{I}^{\star}_{t|t-1}}\right].
\end{align}
Combining this with \eqref{Eq_ConExpESD} gives:
\begin{align}
    \underbrace{\underset{y_t}{\mathbb{E}}\left[\big\|\theta_{t|t}^\textnormal{ex} - \theta^\star_t\big\|_{P_t}^2\right]}_\text{MSE after update} \leq \underbrace{\big\|\theta_{t|t-1} - \theta^\star_t\big\|_{P_t}^2}_\text{SE before update} &-  2\underbrace{\underset{y_t}{\mathbb{E}}\left[\big\|\theta_{t|t-1} - \theta^{\star}_{t}\big\|^2_{\mathcal{I}^{\star}_{t|t-1} - \mathcal{I}^{\star}_{t|t-1}P_t^{-1}\mathcal{I}^{\star}_{t|t-1}}\right]}_\text{$\geq 0$, contractive force} \\& +  \underbrace{2\underset{y_t}{\mathbb{E}}\left[\big\|\nabla(y_t|\theta^{\star}_{t})\big\|^2_{P_t^{-1}}\right]}_\text{$\geq 0$, expansive force},
\end{align}
where positivity of the new contractive force can be guaranteed using $L_t$-Lipschitz continuity of the gradient combined with $\lambda_{\min}(P_t) \geq L_t \Rightarrow 1 -  L_t/\lambda_{\min}(P_t) \geq 0$. That is,
\allowdisplaybreaks
\begin{align}
    \underset{y_t}{\mathbb{E}}&\left[\|\theta_{t|t-1} - \theta^{\star}_{t}\|^2_{\mathcal{I}^{\star}_{t|t-1} - \mathcal{I}^{\star}_{t|t-1}P_t^{-1}\mathcal{I}^{\star}_{t|t-1}}\right] 
    \\&\quad\quad \quad= \underset{y_t}{\mathbb{E}}\left[\|(\mathcal{I}^{\star}_{t|t-1})^{1/2}(\theta_{t|t-1} - \theta^{\star}_{t})\|^2_{I_K- (\mathcal{I}^{\star}_{t|t-1})^{1/2}P_t^{-1}(\mathcal{I}^{\star}_{t|t-1})^{1/2}}\right]\\
    &\quad\quad \quad \geq \lambda_{\min}(I_K - (\mathcal{I}^{\star}_{t|t-1})^{1/2}P_t^{-1}(\mathcal{I}^{\star}_{t|t-1})^{1/2} )\underset{y_t}{\mathbb{E}}\left[\|(\mathcal{I}^{\star}_{t|t-1})^{1/2}(\theta_{t|t-1} - \theta^{\star}_{t})\|^2\right]\\
    &\quad\quad \quad\geq [1 - \lambda_{\max}(\mathcal{I}^{\star}_{t|t-1}P_t^{-1})] \underset{y_t}{\mathbb{E}}\left[\|(\mathcal{I}^{\star}_{t|t-1})^{1/2}(\theta_{t|t-1} - \theta^{\star}_{t})\|^2\right]\\
    &\quad\quad \quad\geq [1 -  \lambda_{\max}(\mathcal{I}^{\star}_{t|t-1})\lambda_{\max}(P_t^{-1})]\; \underset{y_t}{\mathbb{E}}\left[\|(\mathcal{I}^{\star}_{t|t-1})^{1/2}(\theta_{t|t-1} - \theta^{\star}_{t})\|^2\right]\\
     &\quad\quad \quad\geq [1 -  L_t/\lambda_{\min}(P_t)]\;\underset{y_t}{\mathbb{E}}\left[\|(\mathcal{I}^{\star}_{t|t-1})^{1/2}(\theta_{t|t-1} - \theta^{\star}_{t})\|^2\right]
     \\
    & \quad\quad \quad= [1 -  L_t/\lambda_{\min}(P_t)]\;\underset{y_t}{\mathbb{E}}\left[\|\theta_{t|t-1} - \theta^{\star}_{t}\|_{\mathcal{I}^{\star}_{t|t-1}}^2\right]\geq 0.
\end{align}

\clearpage
\setcounter{equation}{0}
\section{Details for simulations and empirical illustrations}
\label{Appendix C}

\subsection{Details for Section~\ref{sec_sim}: Overview of densities}
\label{app: DGPs}
\vspace*{1cm}
\begin{center}
\begin{sideways}
\begin{scriptsize}
    \begin{threeparttable}
\captionof{table}{Overview of data-generating processes in simulation studies.}
\label{tab:DGPs}

\begin{tabular}{l@{\hspace{0.1cm}}l@{\hspace{0.1cm}}c@{\hspace{0.1cm}}c@{\hspace{0.1cm}}c@{\hspace{0.1cm}}c}
  \toprule
\multicolumn{2}{l}{\textbf{DGP}} & \textbf{Link function}  & \textbf{Density} & \textbf{Score} & \textbf{Negative Hessian}
 \\
Type & Distribution &   & $p({y}_t|\theta_t)$ & $\frac{\mathrm{d}\log p(y|\theta_t)}{\mathrm{d}\theta_t}$ & $-\frac{\mathrm{d}^2\log p(y|\theta_t)}{\mathrm{d}\theta_t^2}$ 
\\
  \cmidrule(r{10pt}){1-2}  \cmidrule(lr){3-3} \cmidrule(lr){4-4} \cmidrule(lr){5-5} \cmidrule(lr){6-6} 

  Count & Poisson & $\lambda_t=\exp(\theta_t)$ &$\displaystyle\lambda_t^{y_t}\, \exp(-\lambda_t)/y_t! $ & $y_t-\lambda_t$ & $\lambda_t$ 
\\
Count  & Neg.\ bin.& $\lambda_t=\exp(\theta_t)$ & $\displaystyle \frac{\Gamma(\kappa+y_t)\left(\frac{\kappa}{\kappa+\lambda_t}\right)^\kappa\left(\frac{\lambda_t}{\kappa+\lambda_t}\right)^{y_t}}{\Gamma(\kappa)\Gamma(y_t+1)}$ &$\displaystyle y_t-\frac{\lambda_t(\kappa+y_t)}{\kappa+\lambda_t}$ & $\displaystyle \frac{\kappa \lambda_t(\kappa+y_t)}{(\kappa+\lambda_t)^2}$ 
\\
Intensity & Exponential & $\lambda_t=\exp(\theta_t)$ & $\displaystyle\lambda_t\, \exp(-\lambda_t y_t)$ &  $1-\lambda_t\,y_t$ & $y_t \lambda_t$ 
\\
Duration & Gamma & $\beta_t=\exp(\theta_t)$ & $\displaystyle
\frac{y_t^{\kappa-1}\exp(-y_t/\beta_t)}{\Gamma(\kappa)\beta_t^\kappa }
$ & $\displaystyle \frac{y_t}{\beta_t}-\kappa$& $\displaystyle \frac{y_t}{\beta_t}$ 
\\
Duration & Weibull  & $\beta_t=\exp(\theta_t)$ & $\displaystyle\frac{\kappa\, \left(y_t/\beta_t \right)^{\kappa-1} }{\beta_t \exp\{(y_t/\beta_t)^\kappa\} }
$& $\displaystyle \kappa\left(\frac{y_t}{\beta_t}\right)^\kappa-\kappa$& $\displaystyle \kappa^2 \left(\frac{y_t}{\beta_t}\right)^\kappa$ 
\\
Volatility & Gaussian & $\sigma^2_t=\exp(\theta_t)$ &$ \displaystyle
\frac{\exp\{-y_t^2/(2\sigma_t^2)\}}{ \{2\pi \sigma_t^2\}^{1/2} }$ & $\displaystyle \frac{y_t^2}{2\sigma_t^2}-\frac{1}{2}$
&$\displaystyle \frac{y_t^2}{2\sigma_t^2}$ 
\\
Volatility  & Student's $t$& $\sigma^2_t=\exp(\theta_t)$ &  $\displaystyle\frac{\Gamma\left(\frac{\nu+1}{2}\right)\left(1+\frac{y_t^2}{(\nu-2)\sigma_t^2}\right)^{-\frac{\nu+1}{2}}}{\sqrt{(\nu-2)\pi}\Gamma\left(\nu/2\right)\sigma_t}$ &
$\displaystyle \frac{\omega_t \, y_t^2}{2\sigma_t^2}-\frac{1}{2}$
& $\displaystyle \frac{\nu-2}{\nu+1}\,\frac{\omega_t^2\,y_t^2}{2\sigma_t^2}$
\\
&&&& $\displaystyle \omega_t:=\frac{\nu+1}{\nu-2+y_t^2/\sigma_t^2}$ & 
\\
Dependence & Gaussian & $\displaystyle\rho_t=\frac{1-\exp(-\theta_t)}{1+\exp(-\theta_t)}$ & $\displaystyle\frac{\exp\left\{-\frac{y_{1t}^2+y_{2t}^2-2\rho_t y_{1t}y_{2t} }{2(1-\rho_t^2)} \right\}}{2\pi \sqrt{1-\rho_t^2}}$ & $
\displaystyle \frac{\rho_t}{2}+
\frac{1}{2}\frac{ z_{1t}\,z_{2t}}{1-\rho_t^2} $
& $\displaystyle  \frac{1}{4}\frac{z_{1t}^2+z_{2t}^2}{1-\rho_t^2}- \frac{1-\rho_t^2}{4}$
\\
&& $\displaystyle$ && $z_{1t}:=y_{1t}-\rho_t y_{2t}$
\\
&&&& $z_{2t}:=y_{2t}-\rho_t y_{1t}$
\\
Dependence & Student's $t$ & $\displaystyle\rho_t=\frac{1-\exp(-\theta_t)}{1+\exp(-\theta_t)}$ & $\displaystyle\frac{\nu \left(1+\frac{y_{1t}^2+y_{2t}^2-2\rho_t y_{1t}y_{2t} }{(\nu-2)(1-\rho_t^2)}\right)^{-\frac{\nu+2}{2}}}{2\pi(\nu-2) \sqrt{1-\rho_t^2}}$ & $\displaystyle \frac{\rho_t}{2}+
\frac{\omega_t}{2}\frac{ z_{1t}\,z_{2t}}{1-\rho_t^2} $
& $\displaystyle  \frac{\omega_t}{4}\frac{z_{1t}^2+z_{2t}^2}{1-\rho_t^2}-\frac{1-\rho_t^2}{4}$
\\
&&&&& $-\frac{1}{2}\frac{\omega_t^2}{\nu+2}\frac{z_{1t}^2 \, z_{2t}^2}{(1-\rho_t^2)^2}$
\\
&& $\displaystyle$ && $z_{1t}:=y_{1t}-\rho_t y_{2t}$ &  \multirow{2}{*}{$\displaystyle \omega_t:=\frac{\nu+2}{\nu-2+\frac{y_{1t}^2+y_{2t}^2-2\rho_t y_{1t}y_{2t} }{1-\rho_t^2}}$}
\\
&&&& $z_{2t}:=y_{2t}-\rho_t y_{1t}$
\\
\\
  \bottomrule
\end{tabular}

\begin{tablenotes}
\item Note: The table contains nine distributions and link functions taken from~\citet{koopman2016predicting}. To aid the comparison (but at the expense of some consistency), we retain most of their parameter notation, even though some symbols are also used in the main text for different quantities.
\end{tablenotes}

\end{threeparttable}
\end{scriptsize}
\end{sideways}
\end{center}

\clearpage

\subsection{Details for Section~\ref{sec_sim}: True static (hyper-)parameters}

\label{app: DGPs2}

\begin{table}[ht]
\centering
\caption{Static parameters in the ISD filter as a DGP.} \label{tab:dgpparametervalues}
\begin{threeparttable}

\begin{tabular}{@{}llcccccl@{}}
\toprule
\textbf{Type}       & \textbf{Distribution} & $\omega^{0}$ & $\Phi^{0}$ & $H^0$ & shape ($\kappa^0$ or $\nu^0$) \\ \midrule
Count               & Poisson               & 0.00        & 0.97  &      0.10            &                \\
Count               & Negative binomial     & 0.00        & 0.97  &      0.10            & $\kappa^0 = 4$      \\ 
Intensity           & Exponential           & 0.00        & 0.97  &      0.10            &                \\ 
Duration            & Gamma                 & 0.00        & 0.97  &     0.10            & $\kappa^0 = 1.5$    \\
Duration            & Weibull               & 0.00        & 0.97  &    0.10            & $\kappa^0 = 1.2$    \\ 
Volatility          & Gaussian              & 0.00        & 0.97  &    0.10            &                \\
Volatility          & Student's $t$         & 0.00        & 0.97  &    0.10            & $\nu^0 = 6$       \\ 
Dependence              & Gaussian              & 0.00        & 0.97  &     0.10            &                \\
Dependence              & Student's $t$         & 0.00        & 0.97  &    0.10            & $\nu^0 = 6$       \\ \bottomrule
\end{tabular}
\end{threeparttable}
\end{table}

\subsection{Details for Section~\ref{sec_sim}: Closed-form ISD update step}
\label{app: DGPs3}

For five of nine distributions in Table~\ref{tab:DGPs}, the ISD update step~\eqref{ProParUpdate} can be solved in closed form using the principal branch of the Lambert $W$ function. This function, denoted by $W(z) := W_0(z)$, solves $w \exp(w) = z$ for $w$ in terms of $z$. For any $z\geq 0$, $W(z)$ then yields the unique nonnegative solution to $w \exp(w)= z$.

\begin{enumerate}
    \item  \textbf{Poisson distribution} with link $\lambda_t=\exp(\theta_t)$. The ISD first-order condition reads (for the score, see Table~\ref{tab:DGPs}):
\[
\theta_{t| t}=\theta_{t| t-1}+H_t(y_t-\mathrm{e}^{\theta_{t| t}}).
\]
Moving some terms to the left, we have
\[
\theta_{t| t}-\theta_{t| t-1}-H_t y_t =-H_t  \mathrm{e}^{\theta_{t| t}}.
\]
Multiply by $-\exp( -\theta_{t| t}+\theta_{t| t-1}+H_t y_t)$ to yield
\[
 (\underbrace{ -\theta_{t| t}+\theta_{t| t-1}+H_t y_t}_{=w})\exp( \underbrace{-\theta_{t| t}+\theta_{t| t-1}+H_t y_t }_{=w})
=\underbrace{H_t \exp(\theta_{t| t-1}+H_t y_t )}_{=z}.
\]
Setting $w= -\theta_{t| t}+\theta_{t| t-1}+H_t y_t=W(z)$ with $z=H_t \exp(\theta_{t| t-1}+H_t y_t)\geq 0$ and solving for $\theta_{t|t}$ gives
\begin{equation}
\theta_{t| t}
=
\theta_{t| t-1}+H_t y_t
-
W\! \left(
H_t\,\exp(\theta_{t| t-1}+H_t y_t)
\right).
\end{equation}

\item \textbf{Exponential distribution} with link $\lambda_t=\exp(\theta_t)$. The ISD first-order condition reads (for the score, see Table~\ref{tab:DGPs}):
\[
\theta_{t| t}=\theta_{t| t-1}+H_t(1-y_t \mathrm{e}^{\theta_{t| t}}).
\]
Moving some terms to the left, we have
\[
\theta_{t| t}-\theta_{t| t-1}-H_t=-H_t y_t \mathrm{e}^{\theta_{t| t}}.
\]
Multiply by $-\exp(-\theta_{t| t}+\theta_{t| t-1}+H_t)$ to yield
\[
(\underbrace{-\theta_{t| t}+\theta_{t| t-1}+H_t}_{=w}) \exp(\underbrace{-\theta_{t| t}+\theta_{t| t-1}+H_t}_{=w} )
=\underbrace{H_t y_t \exp(\theta_{t| t-1}+H_t}_{=z} ).
\]
Setting $w=-\theta_{t| t}+\theta_{t| t-1}+H_t=W(z)$ with $z=H_t y_t \exp(\theta_{t| t-1}+H_t)\geq 0$ and solving for $\theta_{t|t}$ gives
\begin{equation}
\theta_{t| t}
=
\theta_{t| t-1}+H_t
-
W\!\left(
H_t y_t\,\exp(\theta_{t| t-1}+H_t)
\right).
\end{equation}

\item \textbf{Gamma distribution} with link $\beta_t=\exp(\theta_t)$ and shape $\kappa>0$.
The ISD first-order condition reads (for the score, see Table~\ref{tab:DGPs}):
\[
\theta_{t| t}
=
\theta_{t| t-1}
+
H_t(y_t \mathrm{e}^{-\theta_{t| t}}-\kappa)
=
\theta_{t| t-1}-H_t\kappa
+
H_t y_t\,\mathrm{e}^{-\theta_{t| t}}.
\]
Moving some terms to the left, this can be written as
\[
\theta_{t| t}-\theta_{t| t-1}+H_t\kappa
=
H_t y_t\,\mathrm{e}^{-\theta_{t| t}}.
\]
Multiply by $\exp(\theta_{t| t}-\theta_{t| t-1}+ H_t\kappa )$ to yield
\[
( \underbrace{\theta_{t| t}-\theta_{t| t-1}+H_t\kappa}_{=w} ) \exp(\underbrace{\theta_{t| t}-\theta_{t| t-1}+ H_t\kappa }_{=w})
=
\underbrace{H_t y_t\,\exp(-\theta_{t| t-1}+H_t\kappa )}_{=z}.
\]
Setting $w=\theta_{t| t}-\theta_{t| t-1}+H_t\kappa=W(z)$ with $z = H_t y_t\,\exp(-\theta_{t| t-1}+H_t\kappa ) \geq 0$ and solving for $\theta_{t|t}$ gives
\begin{equation}
\theta_{t| t}
=
\theta_{t| t-1}-H_t\kappa
+
W\! \left(
H_t y_t\,\exp(-\theta_{t| t-1}+H_t\kappa)
\right).
\end{equation}

\item \textbf{Weibull distribution} with link $\beta_t=\exp(\theta_t)$ and shape $\kappa>0$. The ISD first-order condition reads (for the score, see Table~\ref{tab:DGPs}):
\[
\theta_{t| t}
=
\theta_{t| t-1}
+
H_t(\kappa y_t^{\kappa}\mathrm{e}^{-\kappa\theta_{t| t}}-\kappa)
=
\theta_{t| t-1}-H_t\kappa
+
H_t\kappa y_t^{\kappa}\,\mathrm{e}^{-\kappa\theta_{t| t}}.
\]
Moving some terms to the left, this can be written as
\[
\theta_{t| t}-\theta_{t| t-1}+H_t\kappa
=
H_t\kappa y_t^{\kappa}\,\mathrm{e}^{-\kappa\theta_{t| t}}.
\]
Multiply by $\kappa\exp(\kappa (\theta_{t|t}-\theta_{t|t-1}+H_t \kappa))$ to obtain
\[
\underbrace{\kappa(\theta_{t| t}-\theta_{t| t-1}+H_t\kappa)}_{=w}\exp(\underbrace{\kappa (\theta_{t|t}-\theta_{t|t-1}+H_t \kappa)}_{=w})
=
\underbrace{H_t\kappa^2 y_t^{\kappa}\,\exp(\kappa (-\theta_{t|t-1}+H_t \kappa))
}_{=z}.
\]
Setting $w=\kappa(\theta_{t| t}-\theta_{t| t-1}+H_t\kappa)=W(z)$ with $z = H_t\kappa^2 y_t^{\kappa}\,\exp(\kappa (-\theta_{t|t-1}+H_t \kappa))\geq 0$ and solving for $\theta_{t|t}$ gives
\begin{equation}
\theta_{t| t}
=
\theta_{t| t-1}-H_t\kappa
+
\frac{1}{\kappa}\,
W\!\left(
H_t\kappa^{2}y_t^{\kappa}\,
\exp(-\kappa\theta_{t| t-1}+H_t\kappa^{2})
\right).
\end{equation}

\item \textbf{Gaussian distribution} with link $\sigma^2_t=\exp(\theta_t)$. 
The ISD first-order condition reads (for the score, see Table~\ref{tab:DGPs}):
\[
\theta_{t|t}= \theta_{t|t-1}+\frac{H_t}{2}(y_t^2 \mathrm{e}^{-\theta_{t|t}}-1).
\]
Moving some terms to the left, this can be written as
\[
\theta_{t| t}-\theta_{t| t-1}+\frac{H_t}{2}
=
\frac{H_t}{2} y_t^2 \mathrm{e}^{-\theta_{t| t}}.
\]
Multiplying by $\exp(\theta_{t|t}-\theta_{t|t-1}+H_t/2)$, we find
\[
(\underbrace{\theta_{t| t}-\theta_{t| t-1}+H_t/2}_{=w})\exp(\underbrace{\theta_{t|t}-\theta_{t|t-1}+H_t/2}_{=w})
=
\underbrace{\frac{H_t}{2} y_t^2 \exp(-\theta_{t|t-1}+H_t/2)}_{=z}.
\]
Setting $w=\theta_{t|t}-\theta_{t|t-1}+H_t/2=W(z)$ with $z = \frac{H_t}{2} y_t^2 \exp(-\theta_{t|t-1}+H_t/2)\geq 0$ and solving for $\theta_{t|t}$ gives
\begin{equation}
\theta_{t| t}
=
\theta_{t| t-1}-\frac{H_t}{2}
+
W\!\left(
\frac{H_t}{2}y_t^2\,
\exp(-\theta_{t| t-1}+H_t/2)
\right).
\end{equation}
\end{enumerate}

\clearpage
\subsection{Details for Section~\ref{sec_sim}: Visual support for normality}
\label{app: DGPs4}

\begin{figure}[H]
\centering
\begin{subfigure}{1\textwidth}
\includegraphics[scale=0.3]{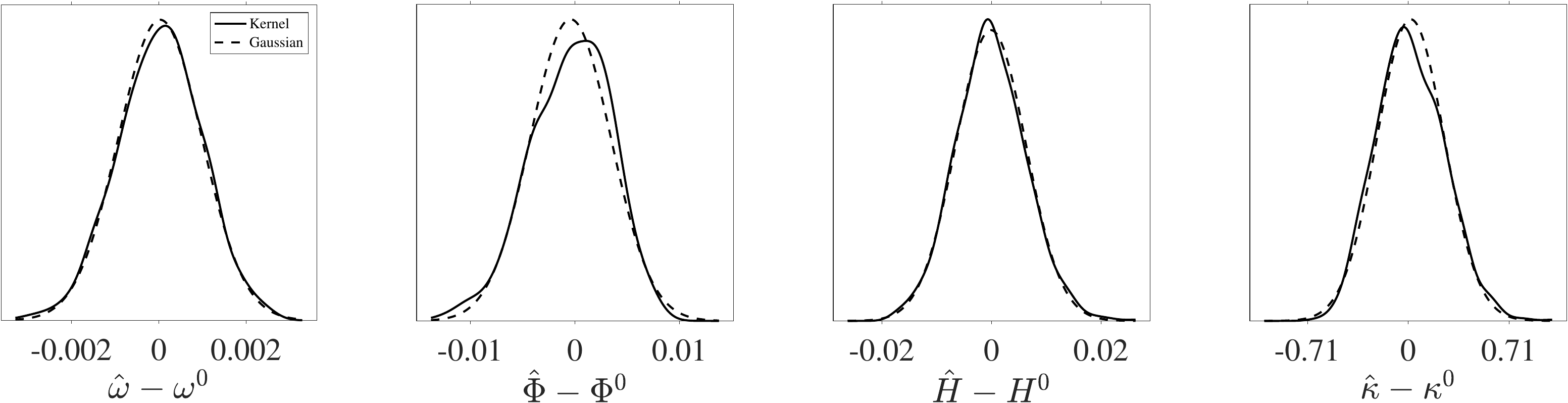}
\caption{Count: Negative Binomial}
\end{subfigure}

\begin{subfigure}{1\textwidth}
\includegraphics[scale=0.3]{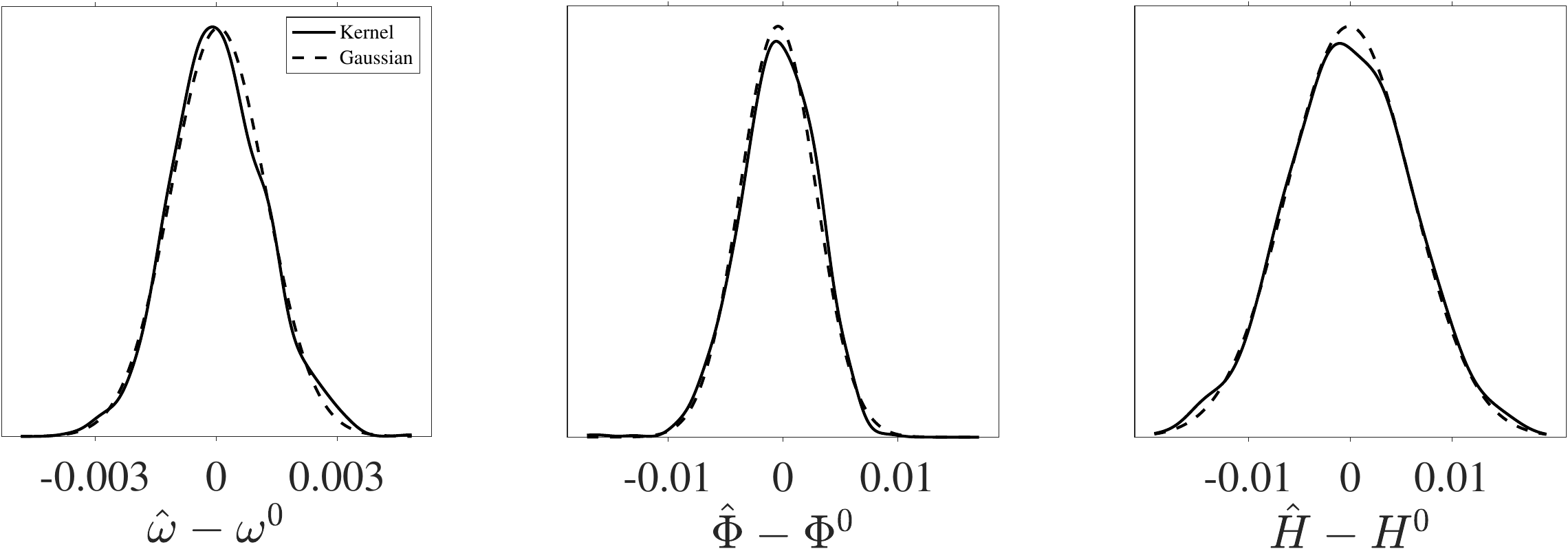}
\caption{Intensity: Exponential}
\end{subfigure}

\begin{subfigure}{1\textwidth}
\includegraphics[scale=0.3]{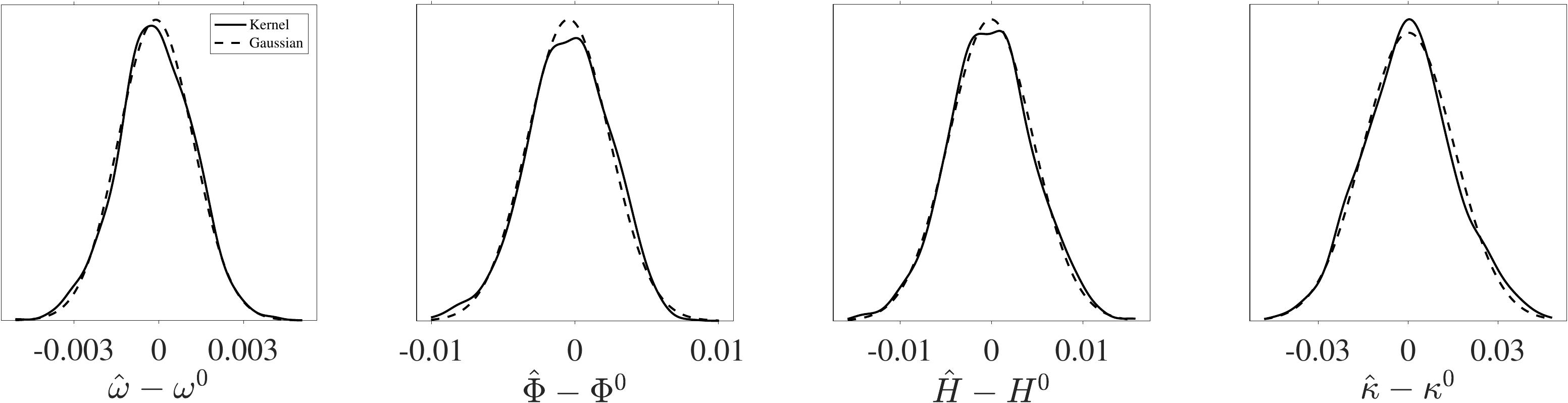}
\caption{Duration: Gamma}
\end{subfigure}

\begin{subfigure}{1\textwidth}
\includegraphics[scale=0.3]{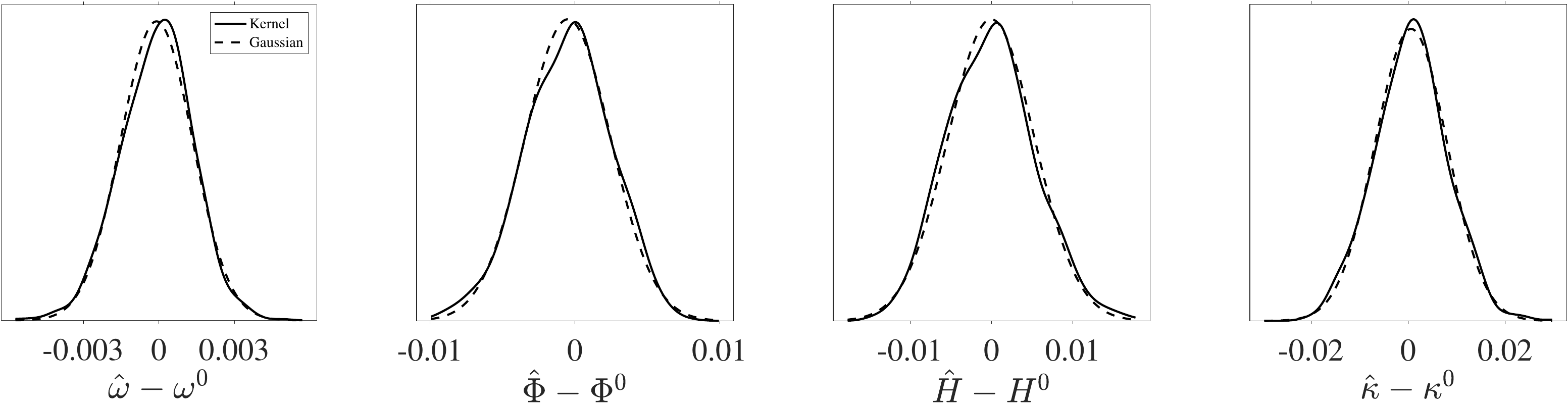}
\caption{Duration: Weibull}
\end{subfigure}
\caption{Distribution of estimation errors for different DGPs and sample size $T=16{,}000$.}
\label{fig1}
\end{figure}

\clearpage

\begin{figure}
\centering

\begin{subfigure}{1\textwidth}
\includegraphics[scale=0.3]{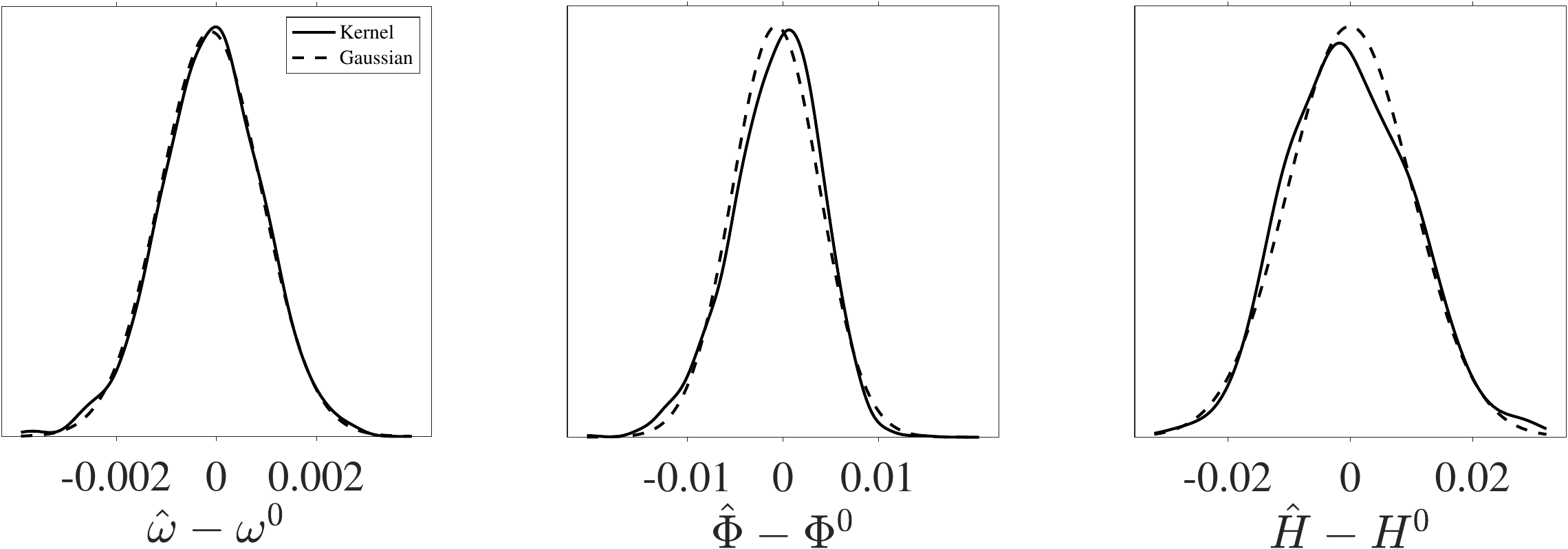}
\caption{Volatility: Gaussian}
\end{subfigure}

\begin{subfigure}{1\textwidth}
\includegraphics[scale=0.3]{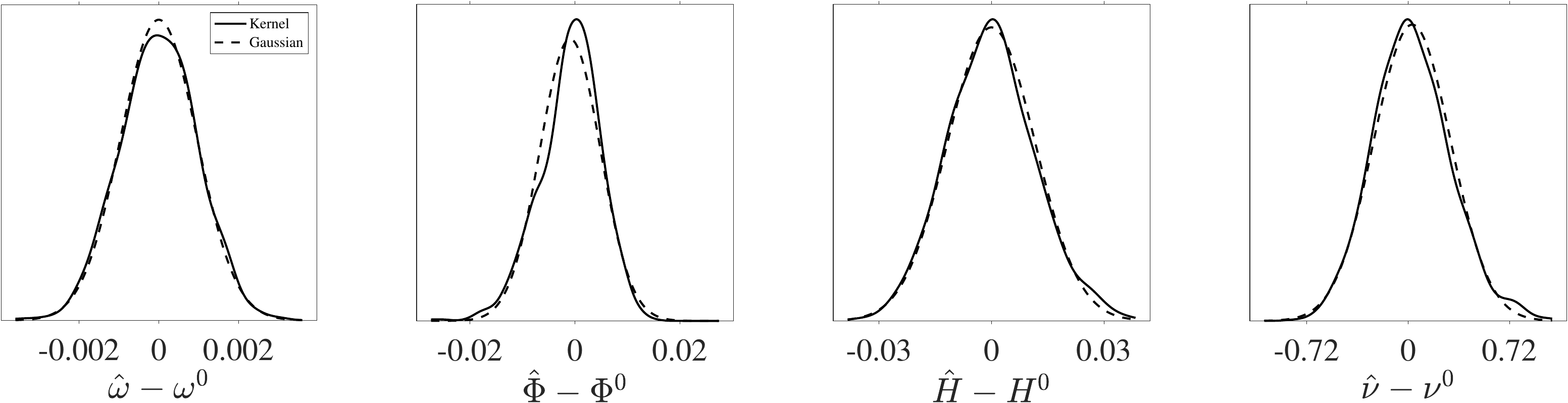}
\caption{Volatility: Student's $t$}
\end{subfigure}

\begin{subfigure}{1\textwidth}
\includegraphics[scale=0.3]{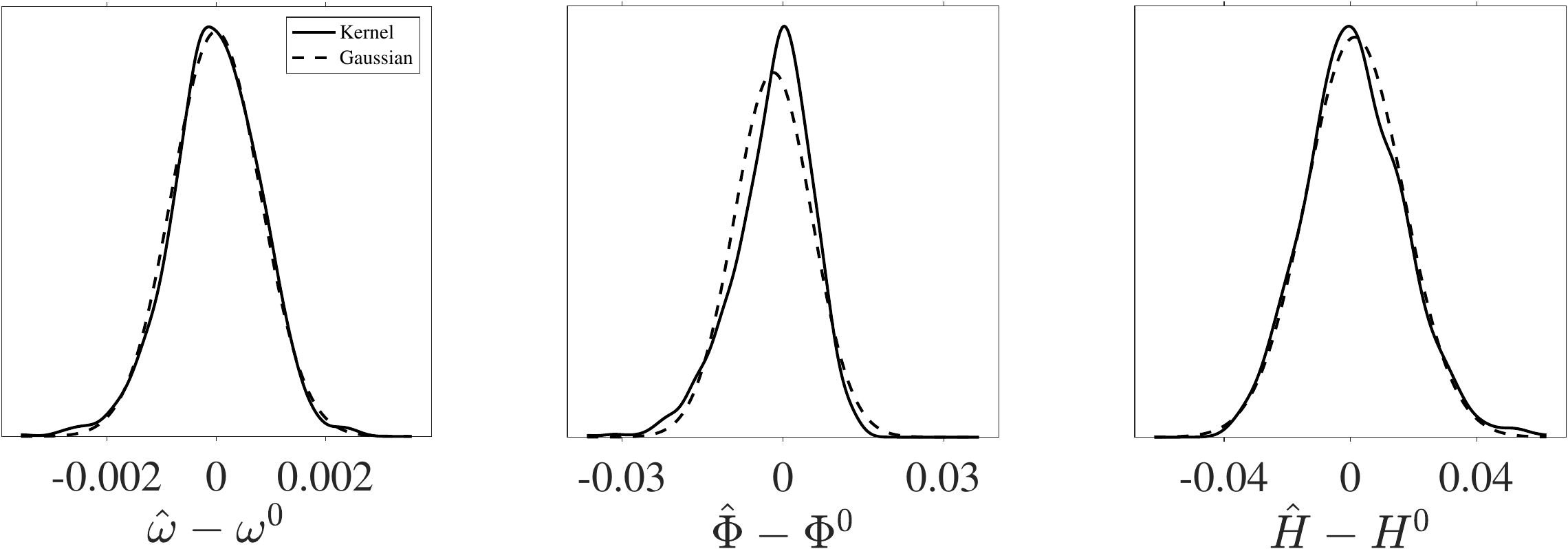}
\caption{Dependence: Gaussian}
\end{subfigure}

\begin{subfigure}{1\textwidth}
\includegraphics[scale=0.3]{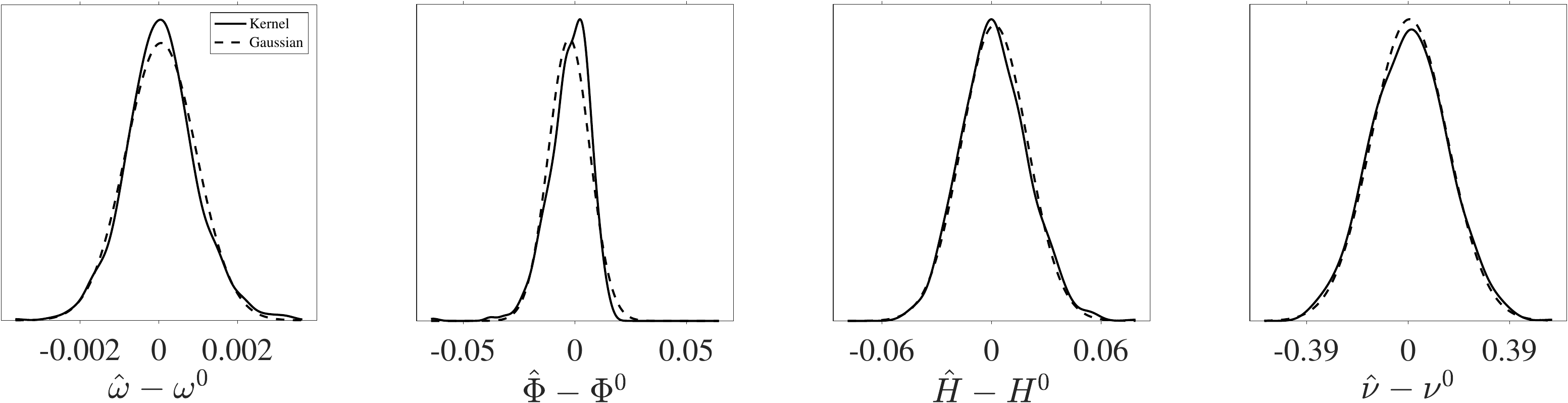}
\caption{Dependence: Student's $t$}
\end{subfigure}
\caption{Distribution of estimation errors for different DGPs and sample size $T=16{,}000$.}
\label{fig2}
\end{figure}

\clearpage

\subsection{Details for Section~\ref{sec: Network}: Dynamic network flows}
\label{appendix:gamma_flow_details}

This appendix derives the score and Hessian for the dynamic Gamma distribution in Section~\ref{sec: Network}; these are used in Newton's method for computing the ISD update. 

At time \(t\), let \(y_t\in\mathbb{R}_{>0}^n\) denote the vector of observed flows, stacked over all directed pairs \((i,j)\) with \(i,j=1,\ldots,N\) and \(i\neq j\), where \(n=N(N-1)\). Conditional on the state vector \(\theta_t=((\theta_{1,t})',(\theta_{2,t})')'\in\mathbb{R}^{2N}\), the elements of \(y_t\) are independent Gamma random variables with common shape parameter \(\psi_0>0\) and mean vector \(M(\theta_t)\in\mathbb{R}_{>0}^n\), where
\begin{equation}
    M(\theta_t)=\exp(\mu(\theta_t)),
\qquad
\mu(\theta_t)=\mu_0+Z\theta_t.
\end{equation}
Here \(\mu_0\in\mathbb{R}^n\) collects the baseline logarithmic means, the exponential is applied elementwise, and \(Z\) is the \(n\times 2N\) design matrix that maps sender and receiver effects into bilateral log-means. For each directed pair \((i,j)\),
\begin{equation}
    \mu_{ij,t}(\theta_t)= \mu_{0,ij}+\theta_{1,i,t}+\theta_{2,j,t},
\qquad
M_{ij,t}(\theta_t)=\exp(\mu_{0,ij}+\theta_{1,i,t}+\theta_{2,j,t}).
\end{equation}
Thus, each row of \(Z\) contains exactly two ones: one for the sender effect and one for the receiver effect. Under the mean-shape parametrization of the Gamma distribution,\footnote{The standard Gamma density is usually parameterized by shape \(\psi\) and scale \(s\), with density \(\Gamma(\psi)^{-1}s^{-\psi}y^{\psi-1}e^{-y/s}\). Writing the mean as \(M=\psi s\) gives \(s=M/\psi\), which yields the parametrization used here; see also \citet[Sec.~2.4]{murphy2012ml}.} the conditional density of \(y_t=(y_{1,t},\ldots,y_{n,t})'\) is
\begin{equation}
    p(y_t\mid \theta_t)
=
\prod_{i=1}^{n}
\frac{\psi_0^{\psi_0}}{\Gamma(\psi_0)}
\frac{y_{i,t}^{\psi_0-1}}{M_i(\theta_t)^{\psi_0}}
\exp\!\left(-\frac{\psi_0 y_{i,t}}{M_i(\theta_t)}\right).
\end{equation}
Hence,
\begin{equation}
\log p(y_t\mid \theta_t)
=
-\psi_0 \iota' (\mu_0+Z\theta_t)
-\psi_0 y_t' \exp\!\big(-(\mu_0+Z\theta_t)\big)+\text{constants},
\end{equation}
where we ignore constants that do not depend on \(\theta_t\), \(\iota\) is an \(n\times 1\) vector of ones, and the exponential is again applied elementwise. Differentiating with respect to \(\theta_t\) gives the score
\begin{equation}
\nabla(y_t\mid \theta_t)
=
\psi_0 Z'\left(y_t\oslash M(\theta_t)-\iota\right),
\end{equation}
where \(\oslash\) denotes elementwise division. Differentiating once more yields the Hessian
\begin{equation}
\frac{\partial^2 \log p(y_t\mid \theta_t)}{\partial \theta_t\partial \theta_t'}
=
-\psi_0 Z'\operatorname{diag}\!\left(y_t\oslash M(\theta_t)\right)Z.
\end{equation}

The conditional logarithmic density is therefore concave in \(\theta_t\), since \(\operatorname{diag}(y_t\oslash M(\theta_t))\) is positive semidefinite and the Hessian is negative semidefinite for all \(\theta_t\); hence, our guarantees for the ISD filter are applicable. However, the Hessian is not uniformly bounded over the full parameter space, as \(y_t\oslash M(\theta_t)\) can become arbitrarily large. Thus, although \(\log p(y_t\mid \theta_t)\) is concave, the score is not globally Lipschitz in \(\theta_t\); as such, no guarantees for the ESD filter are available.

To compute the ISD update in~\eqref{ProParUpdate}, we solve at each time \(t\) the regularized optimization problem~\eqref{RegLike}, that is, the conditional log density plus the quadratic penalty around the one-step-ahead prediction \(\theta_{t|t-1}\). In the present dynamic network model, this objective is smooth and concave in \(\theta\in \mathbb{R}^{2N}\), and after adding the penalty term it becomes strictly concave. Hence the update \(\theta_{t|t}\) is uniquely characterized by the first-order condition~\eqref{ProximalFOC} and can be computed efficiently by Newton's method.

Starting from the predictor \(\theta^{(0)}=\theta_{t|t-1}\), which is typically already close to the maximizer, Newton's method iteratively updates the current guess by solving a linear system based on the gradient and Hessian of the penalized objective. In each iteration, the score derived above is augmented by the gradient of the penalty term (i.e., \(-P_t(\theta-\theta_{t|t-1})\)), and the Hessian is augmented by the constant matrix \(-P_t\). Because the penalized objective is strictly concave, the Newton direction is well defined, and in practice only a few iterations are required. To ensure numerical stability, we combine Newton's method with backtracking line search, so that every step increases the regularized log density.

\subsection{Details for Section~\ref{Sec_Emp:linreg}: Dynamic linear regression}
\label{AppC}

Consider the linear regression model with dependent variable $y_t \in \mathbb{R}$ and independent variable $x_t \in \mathbb{R}^K$, that is,
\begin{equation}
   y_t \;=\; \beta_t'\,x_t\, +\, \varepsilon_t, \qquad \varepsilon_t \, \distas{\text{i.i.d.}}\,\textnormal{N}(0,\sigma^2),
   \label{Eq_App_LinReg}
\end{equation}
where $\beta_t$ is a $K\times1$ vector of time-varying parameters and $\varepsilon_t$ is an i.i.d.\ normally distributed innovation with variance $\sigma^2$. 

The log-likelihood contribution $\log p(y_t|\beta)$ is obviously twice continuously differentiable with respect to $\beta$ for all $y_t$, such that Assumption \ref{A4_Differentiability}b (differentiability) holds. In addition, the Hessian is equal to $-\frac{1}{\sigma^2}x_t x_t'$ and is therefore negative semi-definite. Combined with strong concavity of the penalty, this means that the regularized log-likelihood function $f(\beta|y_t, \beta_{t|t-1}, P_t) \; :=\; \log p(y_t|\beta) \, -\,  \frac{1}{2}\big\|\beta - \beta_{t|t-1}\big\|_{P_t}^2$ is strongly concave in $\beta$. Because $f(\beta|y_t, \beta_{t|t-1}, P_t)$ is finite-valued for any $\beta \in \mathbb{R}^K$, we have that it is thus strictly proper concave such that Assumption~\ref{A2_RL_StrictConcave} (strictly concave regularized log likelihood) holds. 

The first-order condition (FOC) of the ISD update at time $t$ associated with the model (\ref{Eq_App_LinReg}) takes the following form
\begin{equation}
    \beta_{t|t} = \beta_{t|t-1} + H_t \nabla(y_t | \beta_{t|t}, x_t),
    \label{Eq_App_LinFOC}
\end{equation}
where $H_t = P_t^{-1}$ is the learning-rate matrix and $\nabla(y_t | \beta_{t|t}, x_t)$ denotes the implicit score given by 
\begin{equation}
    \nabla(y_t | \beta_{t|t}, x_t) \;=\; \frac{y_t - \beta_{t|t}'\,x_t }{\sigma^2}\,x_t.
\end{equation}
Note that strong concavity of $f(\beta|y_t, \beta_{t|t-1}, P_t)$ and the unrestricted nature of the optimization (i.e., we maximize over $\mathbb{R}^K$) imply that if the FOC (\ref{Eq_App_LinFOC}) has a solution then it is the unique global maximizer. Solving the FOC will thus also directly verify Assumptions~ \ref{A1_RL_Existence} (existence) and \ref{A3_Boundary} (interior solution).

Collecting all terms containing $\beta_{t|t}$ on the left-hand side, we may write the FOC in (\ref{Eq_App_LinFOC}) as
\begin{equation}
    \left(I_K + H_t \frac{x_t x_t'}{\sigma^2}\right) \beta_{t|t} = \beta_{t|t-1} + H_t \frac{y_t x_t}{\sigma^2}.
\end{equation}
Now using the Sherman-Morrison identity, we left-multiply with $(I_K + H_t \frac{x_t x_t'}{\sigma^2})^{-1} = I_K - \frac{H_t x_t x_t'}{\sigma^2 + x_t'H_t x_t}$, which yields
\begin{equation}
   \beta_{t|t} = \left(I_K - \frac{H_t x_t x_t'}{\sigma^2 + x_t'H_t x_t}\right)\left(\beta_{t|t-1} + H_t \frac{y_t x_t}{\sigma^2}\right).
\end{equation}
Eliminating brackets and using the notation $\|x_t \|^2_{H_t} := x_t'H_t x_t$ then gives
\begin{equation}
   \beta_{t|t} = \beta_{t|t-1} + H_t \frac{y_t x_t}{\sigma^2} - \frac{H_t x_t x_t'}{\sigma^2 + \|x_t \|^2_{H_t}}\beta_{t|t-1} - \frac{H_t x_t x_t'}{\sigma^2 + \|x_t \|^2_{H_t}} H_t \frac{y_t x_t}{\sigma^2},
\end{equation}
where changing the ordering using the fact that $y_t$, $\sigma^2$, $x_t'\beta_{t|t-1}$ and $\|x_t \|^2_{H_t}$ are scalars and again using the definition of $\|x_t\|^2_{H_t}$, we get
\begin{equation}
   \beta_{t|t} = \beta_{t|t-1} + H_t \frac{y_t }{\sigma^2} x_t  - \frac{1}{\sigma^2 + \|x_t \|^2_{H_t}}H_t x_t'\beta_{t|t-1} x_t - \frac{\|x_t \|^2_{H_t}}{\sigma^2 + \|x_t \|^2_{H_t}}  H_t\frac{y_t }{\sigma^2} x_t.
\end{equation}
Multiplying the second and third term on the right-hand side with $\frac{\sigma^2 + \|x_t \|^2_{H_t}}{\sigma^2 + \|x_t \|^2_{H_t}}$ and $\frac{\sigma^2}{\sigma^2}$, respectively, allows us to combine the second through fourth terms as follows
\begin{equation}
   \beta_{t|t} = \beta_{t|t-1} + \frac{\sigma^2}{\sigma^2 + \|x_t \|^2_{H_t}} H_t \frac{y_t -x_t'\beta_{t|t-1}}{\sigma^2} x_t,
\end{equation}
where using the definition of the explicit gradient $\nabla(y_t | \beta_{t|t-1}, x_t)$ gives the final result
\begin{equation}
   \beta_{t|t} = \beta_{t|t-1} + \frac{\sigma^2}{\sigma^2 + \|x_t \|^2_{H_t}} H_t \nabla(y_t | \beta_{t|t-1}, x_t).
\end{equation}

\subsection{Details for Section~\ref{sec: GAR}: Dynamic quantile regression}
\label{App_QR}

The ISD update at time $t$ matches its ESD counterpart, as long as it does not overshoot the observation $y_t$. That is, we have
\begin{equation}
\label{App_QR_Propar}
    q_{t|t}(\tau) =
    \left\{ \begin{array}{l@{\hspace{1cm}}l}  
       \min\{y_t,q^\textnormal{ex}_{t|t}(\tau)\}, & y_t > q_{t|t-1}(\tau),\\
       \max\{y_t,q^\textnormal{ex}_{t|t}(\tau) \}, & y_t\leq q_{t|t-1}(\tau),
    \end{array} \right.
\end{equation}
where the ESD updated quantile $q^\textnormal{ex}_{t|t}(\tau)$ is given by the usual expression
\begin{equation}
\label{App_QR_caviar}
    q^\textnormal{ex}_{t|t}(\tau) = q_{t|t-1}(\tau) - 1[y_t < q_{t|t-1}(\tau)] \frac{H(1-\tau)}{\sigma} + 1[y_t > q_{t|t-1}(\tau)] \frac{H\tau}{\sigma}.
\end{equation}
To see that the ISD update prevents the crossing of the quantiles at different levels, we consider the different scenarios that can occur. Suppose that we have two predicted quantiles $q_{t|t-1}(\tau_1) < q_{t|t-1}(\tau_2)$ with $0<\tau_1 < \tau_2 < 1$. First, consider the case where the observation $y_t$ falls between the two quantiles, that is, $q_{t|t-1}(\tau_1) < y_t < q_{t|t-1}(\tau_2)$, then subsequently the $\tau_1$-level quantile is updated upwards, whereas the $\tau_2$-level quantile is updated downwards, see again \eqref{App_QR_caviar}. However, neither can surpass the observation $y_t$ as seen in \eqref{App_QR_Propar}, such that a crossing cannot occur. In this scenario, we could allow for one (or both) predicted quantiles to be exactly equal to $y_t$, in which case that updated quantile equals its predicted value and a crossing is still prevented.

Second, consider the case when $q_{t|t-1}(\tau_1) < q_{t|t-1}(\tau_2) < y_t$ such that both quantiles need to be adjusted upward. Combining \eqref{App_QR_Propar}--\eqref{App_QR_caviar}, we then have that
\begin{equation}
    q_{t|t}(\tau_1) = \min\left\{y_t,q_{t|t-1}(\tau) + \frac{H\tau_1}{\sigma}\right\} \leq \min\left\{y_t,q_{t|t-1}(\tau) + \frac{H\tau_2}{\sigma}\right\} = q_{t|t}(\tau_2),
\end{equation}
which directly follows from $H, \sigma > 0$ and $\tau_1 < \tau_2$.
Finally, consider the case when $y_t < q_{t|t-1}(\tau_1) < q_{t|t-1}(\tau_2)$ such that both quantiles need to be adjusted downwards. Using a similar argument as for the second case, we obtain
\begin{equation}
    q_{t|t}(\tau_1) = \max\left\{y_t,q_{t|t-1}(\tau) - \frac{H(1-\tau_1)}{\sigma}\right\} \leq \max\left\{y_t,q_{t|t-1}(\tau) - \frac{H(1-\tau_2)}{\sigma}\right\} = q_{t|t}(\tau_2).
\end{equation}
In sum, quantile crossings cannot occur in any of the three cases, i.e., for $y_t$ falling (i) between $q_{t|t-1}(\tau_1)$ and $q_{t|t-1}(\tau_2)$, (ii) above $q_{t|t-1}(\tau_2)$, or (iii) below $q_{t|t-1}(\tau_1)$.

\subsection{Details for Section~\ref{sec: Empirical illustration}: Student's $t$ location model}

\label{App T Bill}

\textbf{ESD update for Student's $t$ location model.}
The ESD update for the Student's \emph{t} location model (e.g.,\ \citealp[eq.\ 5]{harvey2014filtering})
is
\begin{align}
\label{ESD local level}
    \theta^\textnormal{ex}_{t|t}&= \theta^\textnormal{ex}_{t|t-1} + \underbrace{H\,\left[1+ \frac{1}{\nu}\left( \frac{y_t-\theta^\textnormal{ex}_{t|t-1}}{\sigma}\right)^2\right]^{-1}}_{:=w^\textnormal{ex}_{t|t}} \; (y_t-\theta^\textnormal{ex}_{t|t-1})
    \\
    &= (1-w^\textnormal{ex}_{t|t} )\;\theta^\textnormal{ex}_{t|t-1} + w^\textnormal{ex}_{t|t}\; y_t,
\end{align}
with learning rate $H>0$, scale parameter $\sigma>0$, and degrees of freedom $\nu>0$. 

As is standard (e.g., \citealp[Sec.\ 5]{blasques2022maximum}), the driving mechanism is taken to be $\nu \sigma^2/(\nu+1)$ times the score. Viewed differently, the factor $(\nu+1)/\nu \sigma^2$ that would have otherwise appeared is absorbed into the learning-rate parameter $H$. The advantage of this scaling is that the scaled Hessian (i.e., by $\nu \sigma^2/(\nu+1)$) falls in the simple range $[-1,1/8]$. This is consistent with equation~\eqref{Eq_tHesbounds} in Appendix~\ref{App_StabStudt} and \citet[Sec.\ 5]{blasques2022maximum}. Hence, the Lipschitz constant of the scaled score is unity.

The second equality above shows that $ \theta^\textnormal{ex}_{t|t}$ can be expressed as a weighted average of the prediction and the observation $y_t$, where the weight associated with $y_t$ is $w^\textnormal{ex}_{t|t}:=H [1+ \nu^{-1}(y_t-\theta^\textnormal{ex}_{t|t-1})^2/\sigma^2]^{-1}\geq 0$. Here, $w^\textnormal{ex}_{t|t}\leq H$ with equality holding in the limit as $\nu\to \infty$. The update is followed by the usual prediction step, given by $\theta^\textnormal{ex}_{t+1|t}=\omega+\Phi\theta^\textnormal{ex}_{t|t}$, where $|\Phi|\leq 1$. The combination of these two steps results in the standard prediction-to-prediction recursion in ESD models. The only distinction lies in the fact that we have the parameter combination $H \Phi$ in front of the score; up to a re-parametrization of $H$, this is immaterial.

Unfortunately, the ESD update~\eqref{ESD local level} does not guarantee that $\theta^\textnormal{ex}_{t|t}$ lies between the prediction $\theta^\textnormal{ex}_{t|t-1}$ and the observation $y_t$. Unless the learning rate $H$ is restricted to be less than one, the weight $w^\textnormal{ex}_{t|t}$ can exceed unity. In that case, the filter effectively goes short on the prediction and overshoots the observation. From a purely mechanical perspective, this behavior is not necessarily problematic. For example, \citet[p.\ 1117]{harvey2014filtering} note that their learning rate (corresponding to $H \Phi$ in our notation) can be estimated ``without unity imposed as an upper bound,'' and \citet[Table~1]{artemova2022score1} report learning rate estimates above one (e.g., $H\approx 2.2$ and $H \Phi\approx 1.6$) in their empirical analysis of Treasury bill spreads. Imposing the constraint $H\leq 1$ would appear to resolve the issue, but doing so would deteriorate model fit and dampen the filter's responsiveness.

\textbf{ISD update for Student's $t$ location model.}
Instead, here we introduce an ISD version of the Student's $t$ location model. The ISD update automatically ensures that, for any learning rate $H>0$, $\theta^\textnormal{im}_{t|t}$ lies in the interval between $ \min\{\theta^\textnormal{im}_{t|t-1},y_t\}$ and $\max\{\theta^\textnormal{im}_{t|t-1},y_t\}$. To see why, note that the Student's \emph{t} log density $\log p(y_t|\theta)$ is maximized at $\theta=y_t$, while the penalty increases with the distance $|\theta-\theta_{t|t-1}|$; hence, updating beyond $y_t$ cannot be optimal. The optimization problem corresponding to the ISD update reads
\begin{equation}
\label{ISD update Students t}
    \theta^\textnormal{im}_{t|t}:=\underset{\theta\in\mathbb{R}}{\text{arg max}} \left\{ \frac{\nu\sigma^2}{\nu+1} \log p(y_t|\theta) - \frac{P}{2}(\theta-\theta^\textnormal{im}_{t|t-1})^2 \right\},
\end{equation}
where $p(y_t|\theta)$ is the Student's $t$ density with location $\theta$ and the log-likelihood contribution is scaled by $\nu\sigma^2/(\nu+1)$ to align with the convention in the literature (e.g., \citealp{harvey2014filtering}; \citealp{blasques2022maximum}). Because the Hessian of the scaled log-likelihood contribution is at most $1/8$, the optimization problem~\eqref{ISD update Students t} is strongly concave if $P>1/8$; in this case, a single stationary point exists, which coincides with the global maximum. 

For an arbitrary $P=H^{-1}>0$, the first-order condition associated with~\eqref{ISD update Students t}
can be written as
\begin{equation}
\label{ISD local level}
    \theta^\textnormal{im}_{t|t}\;=\; \theta^\textnormal{im}_{t|t-1} \,+\, H\; \left[1+ \frac{1}{\nu}\left( \frac{y_t-\theta^\textnormal{im}_{t|t}}{\sigma}\right)^2\right]^{-1}\;(y_t-\theta^\textnormal{im}_{t|t}),
\end{equation}
which nearly matches the ESD update~\eqref{ESD local level}. The only difference is that the implicit update $\theta^\textnormal{im}_{t|t}$ appears on \emph{both} sides of the equation. To facilitate our analysis below, the update~\eqref{ISD local level} can be usefully rearranged as
\begin{equation}
   \left[1+ \frac{1}{\nu}\left( \frac{y_t-\theta^\textnormal{im}_{t|t}}{\sigma}\right)^2\right]\, (\theta^\textnormal{im}_{t|t}-\theta^\textnormal{im}_{t|t-1})\,-\,H\,(y_t-\theta^\textnormal{im}_{t|t})\;=\;0.
    \label{ISD local level2}
\end{equation}
We analyze this condition by supposing that the update $\theta^\textnormal{im}_{t|t}$ can be expressed as a weighted average of the prediction $\theta_{t|t-1}^\textnormal{im}$ and the observation $y_t$, i.e., we conjecture
\begin{align}
\label{ansatz}
\theta^\textnormal{im}_{t|t}\;&=\;(1-w^\textnormal{im}_{t|t})\,\theta^\textnormal{im}_{t|t-1}\,+\,w^\textnormal{im}_{t|t} \,y_t,
\\
&=\; \theta^\textnormal{im}_{t|t-1}\, +\, w^\textnormal{im}_{t|t} \, e_{t|t-1},
\label{ansatz2}
\end{align}
where $w^\textnormal{im}_{t|t} \in (0,1)$ is the weight (which is still to be determined) corresponding to the observation $y_t$, while $e_{t|t-1}:=y_t-\theta^\textnormal{im}_{t|t-1}$ is the prediction error. This hypothesis is natural, since we know that
\begin{equation}
\label{interval} 
  \theta_{t|t}^\textnormal{im} \in 
  \big[ \min\{\theta^\textnormal{im}_{t|t-1}, y_t\},
         \max\{\theta^\textnormal{im}_{t|t-1}, y_t\} \big].
\end{equation}
However, equation~\eqref{ansatz2} makes clear that $w_{t|t}^\textnormal{im}$ is not identified if  $e_{t|t-1}=0$ (i.e., if $y_t = \theta^\textnormal{im}_{t|t-1}$), in which case we obtain $ \theta_{t|t}^\textnormal{im}= \theta_{t|t-1}^\textnormal{im}$ regardless of $w_{t|t}^\text{im}$. Indeed, the interval in \eqref{interval} then collapses to a single point, which is the correct solution to the optimization problem~\eqref{ISD update Students t}. Without loss of generality, therefore, we focus on the case $e_{t|t-1}\neq 0$ and aim to find $w_{t|t}^\text{im}$.

Hypothesis~\eqref{ansatz} implies that  $\theta_{t|t}^\text{im} - \theta_{t|t-1}^{\text{im}}$ and $y_t - \theta_{t|t}^\text{im}$ are both linear in $w^\text{im}_{t|t}$, i.e., 
\[
\theta_{t|t}^\text{im} - \theta_{t|t-1}^{\text{im}}
\;=\; w_{t|t}^{\text{im}}\, e_{t|t-1},
\qquad
y_t - \theta_{t|t}^\text{im}
\;=\; (1 - w_{t|t}^\text{im}) \;e_{t|t-1}.
\]
Making these substitutions in equation~\eqref{ISD local level2}, we obtain
\begin{equation}
 \left[1 + \frac{1}{\nu}\frac{(1-w_{t|t}^\text{im})^2 e^2_{t|t-1}}{\sigma^2}\right]\,w^{\text{im}}_{t|t} \, e_{t|t-1}\,-\,H\,(1-w_{t|t}^\text{im})\, e_{t|t-1}
 \;=\; 0.
 \label{ISD local level3}
\end{equation}
Assuming $e_{t|t-1}\neq 0$, we may divide both sides of equation~\eqref{ISD local level3} by $e_{t|t-1}$ and rearrange the result to yield
\begin{equation}
    \frac{1}{\nu} \frac{\mathrm{e}2_{t|t-1}}{\sigma^2} \,(1-w^\textnormal{im}_{t|t})^2\, w^\textnormal{im}_{t|t}
    \,+\, w^\textnormal{im}_{t|t}\,-\,H\,(1-w^\textnormal{im}_{t|t})\;=\; 0,
   \label{cubic}
\end{equation}
which is a cubic equation in $w_{t|t}^\textnormal{im}$. If $H=0$, $w_{t|t}^\textnormal{im}=0$ becomes a solution. As $H\to \infty$, we require $w_{t|t}^\textnormal{im}\to 1$. For any fixed $0<H<\infty$, however, neither $w_{t|t}^\textnormal{im}=0$ nor $w_{t|t}^\textnormal{im}=1$ is a solution; this confirms the correctness of the open interval in our conjecture $w_{t|t}^\textnormal{im}\in (0,1)$.

The cubic equation~\eqref{cubic} allows at most three distinct solutions (at least one of them real valued), for which closed-form expressions are available in standard software packages. When $\nu\to \infty$ or $e_{t|t-1}\to 0$, the quadratic and cubic terms disappear and we get $w^\textnormal{im}_{t|t}\to H/(1+H)$.
More generally, $w^\textnormal{im}_{t|t}\leq H/(1+H)<1$ for any $\nu<\infty$, such that the update under the Student's~$t$ distribution is always more conservative than in the Gaussian case. That the weight is bounded above by $H/(1+H)$ also ensures that the update cannot surpass $y_t$. These desirable properties are not shared by the ESD update~\eqref{ESD local level}.

The possible multiplicity of solutions to the first-order condition~\eqref{ISD local level}, and hence to the cubic equation~\eqref{cubic}, stems from the fact that the Student's~$t$ distribution is not log-concave in the location parameter. As a result, there can be up to three stationary points: two local maxima and one local minimum. All three lie in the interval $[\min\{\theta^\text{im}_{t|t-1},y_t\},\max\{\theta^\text{im}_{t|t-1},y_t\}]$, with one local maximum located closer to $\theta^\text{im}_{t|t-1}$, the other closer to $y_t$, and the local minimum between them. When the observation deviates only moderately from the prediction, the local maximum near $y_t$ is dominant, meaning we effectively ``trust'' the data. When the observation is extreme, the local maximum near $\theta^\text{im}_{t|t-1}$ dominates, so $y_t$ is treated as an outlier. In the presence of multiple stationary points, we can compare the corresponding function values to determine the global maximum. In the unlikely event that the local maxima yield identical values, a tie-breaking rule can be applied.

{\small
\putbib[ref]
}
\end{bibunit}
\end{document}